\documentclass[floatfix,aps,preprint,superscriptaddress,nofootinbib,showpacs,amsmath,amssymb,prd,12pt]{revtex4}
\usepackage{graphicx}
\newcommand{\cha}[1]{\widetilde{\chi}^\pm_{#1}}
\newcommand{\neu}[1]{\widetilde{\chi}^0_{#1}}
\newcommand{\chap}[1]{\widetilde{\chi}^+_{#1}}
\newcommand{\cham}[1]{\widetilde{\chi}^-_{#1}}
\newcommand{\sq}{\widetilde{q}}
\newcommand{\alphas}{\alpha_s}
\newcommand{\msbar}{$\overline{\text{MS}}$}
\newcommand{\ps}{\displaystyle{\not{p}}}
\newcommand{\es}{\displaystyle{\not{\epsilon}^x}}
\newcommand{\uvcoeff}[1]{\frac{\alphas C_F}{4\pi}\frac{#1}{\epsilon}}

\begin{document}
\title{Threshold Resummation Effects in the
Associated Production of Chargino and Neutralino at Hadron
Colliders}
\author{Chong Sheng Li}
\email{csli@pku.edu.cn}
\affiliation{Department of Physics, Peking University, Beijing 100871, China}
\author{Zhao Li}
\email{zhli.phy@pku.edu.cn} \affiliation{Department of Physics,
Peking University, Beijing 100871, China}
\author{Robert J. Oakes}
\email{r-oakes@northwestern.edu}
\affiliation{Department of Physics and Astronomy, Northwestern University, Evanston, IL 60208-3112, USA}
\author{Li Lin Yang}
\email{llyang@pku.edu.cn}
\affiliation{Department of Physics, Peking University, Beijing 100871, China}


\begin{abstract}
We investigate the QCD effects in the associated production of the
chargino and the neutralino, $\tilde\chi^\pm_{1}$ and
$\tilde\chi^0_{2}$, in the Minimal Supersymmetric Standard Model
(MSSM) at both the Fermilab Tevatron and the CERN Large Hadron
Collider (LHC). We include the next-to-leading order (NLO) QCD
corrections (including supersymmetric QCD) and the threshold
resummation effects. Our results show that, compared to the NLO
predictions, the threshold resummation effects can increase the
total cross sections by $3.6\%$ and $3.9\%$ for the associated
production of $\tilde\chi^+_{1}\tilde\chi^0_{2}$ and
$\tilde\chi^-_{1}\tilde\chi^0_{2}$ at the LHC, respectively, and by
$4.7\%$ for those of $\tilde\chi^\pm_{1}\tilde\chi^0_{2}$ at the
Tevatron. In the invariant mass distributions the resummation
effects are significant for large invariant mass. The threshold
resummation reduces the dependence of the total cross sections at the LHC (Tevatron) on
the renormalization/factorization scales to $5\%$ ($4\%$) from up to $7\%$ ($11\%$)
at NLO.
\end{abstract}
\maketitle

\section{INTRODUCTION}
The search for new physics beyond the Standard Model (SM),
especially supersymmetry (SUSY), is one of the objectives at the
CERN Large Hadron Collider (LHC). Many calculations have been
carried out based on the Minimal Supersymmetric Standard Model
(MSSM), a version of SUSY. Phenomenologically SUSY predicts many new
particles; e.g., the superpartners of the SM particles.
Specifically, in the MSSM, there are squarks, gluino, sleptons,
charginos, neutralinos and more Higgs bosons in addition to the SM
particles. Besides squarks and gluinos, perhaps the most interesting new
particles are the four neutralinos $\neu{i} (i=1,2,3,4)$ and the two
charginos $\cha{j} (j=1,2)$, which are the mass eigenstates of the
superpartners of the Higgs and gauge bosons, since the lightest
chargino $\cha{1}$ and the two lightest neutralinos ($\neu{1}$,
$\neu{2}$) can be lighter than the squarks and the gluino in most of
the parameter space. In most of the MSSM parameter regions the
associated production of $\cha{1}\neu{2}$ is the main source of
trilepton events. In Refs.\cite{Barger:1998hp,Bityukov:1999sb} the
trilepton signal was investigated for the Fermilab Tevatron in the
Minimal Supergravity Model (mSUGRA) at leading order (LO). If
the leptonic decays of $\cha{1}\neu{2}$ are the dominant decay modes the
signal to background ratio can be quite large after suitable
cuts. Because the trileptonsignal is also quite sensitive to the SUSY parameters,
it is potentially also a sensitive probe of the SUSY
parameters. In fact, the trilepton signal from $\cha{1}\neu{2}$ is now
being searched for by the D0 Collaboration\cite{Hohlfeld:2006vk} at the Tevatron. So
far no excess has been observed above the expected SM background,
but the results have been used to constrain the masses. A plan
for searching for the trilepton signal from chargino and neutralino
has also been presented by the Compact Muon Solenoid (CMS)
Collaboration\cite{Boer:2006bg} at the LHC. Therefore, high
precision theoretical predictions for the associated production of
$\cha{1}\neu{2}$ are very important for the forthcoming experiments
at the LHC.

The next-to-leading order (NLO) SUSY QCD corrections to the process $pp\to\cha{1}\neu{2}$ in
mSUGRA was first investigated in
Refs.\cite{Plehn:1998nh,Beenakker:1999xh} where infrared
singularities were dealt with using the dipole subtraction
method\cite{Catani:1996vz}. Also, the NLO SUSY QCD and SUSY
electroweak (EW) corrections to this process in the general MSSM
were calculated in Ref.\cite{Hao:2006df}. In the following we further
investigate the QCD effects on this process, including the NLO
SUSY QCD corrections and, in addition, the next-to-leading-logarithmic (NLL)
threshold resummation effects in mSUGRA using the most recent SM
parameters\cite{Yao:2006px,CDF:2007bxa} at both the Tevatron and the LHC.

The paper is organized as follows: In Sec.~II we present the
analytic results at fixed order. In Sec.~III we briefly summarize
the threshold resummation formalism and derive the expressions for
the resummed cross sections. In Sec.~IV the numerical results are
presented and discussed. Sec.~V contains a brief summary of the conclusions. The
SUSY vertexes involved in our calculations are summarized in
Appendix \ref{vertexes}. The abbreviations for the
Passarino-Veltaman integrals are defined in Appendix \ref{BCD}. The
standard matrix elements and the explicit expressions for the form
factors are summarized in Appendix \ref{formfactor}.

\section{CALCULATIONS AT FIXED ORDER}
For hadron colliders the total cross section for the hadronic
process,
\begin{equation}
A+B\to\cha{1}+\neu{2}+X,
\end{equation}
can be factorized into the convolution of the parton distribution
functions and the parton cross section,
\begin{equation}
\sigma(S)=\sum_{a,b}\int~dx_a~dx_b~f_{a/A}(x_a,\mu_f)f_{b/B}(x_b,\mu_f)
\hat\sigma_{ab}(\hat{s}=x_a x_b S,\alphas), \label{ffhatsigma}
\end{equation}
where $\mu_f$ is the factorization scale, $f(x,\mu_f)$ is the parton
distribution function (PDF) and $\hat{s}$ is the parton center of
mass energy. A and B both refer to protons at the LHC and
proton and antiproton at the Tevatron, respectively. The parton
cross section $\hat{\sigma}$ is given by
\begin{equation}
\hat\sigma=\frac{1}{2\hat{s}}\int\overline{\sum}|{\cal M}|^2~dPS^{(n)},
\end{equation}
where $\overline{\sum}$ indicates the summation over final states
and the average over initial states and $\int dPS^{(n)}$ represents
the phase space integration.

For simplicity, in this section we only present the expressions for the subprocess
\begin{equation}
u+\bar{d}\to\chap{1}+\neu{2},
\label{LOPROC}
\end{equation}
The other processes are given by similar expressions.

\subsection{LEADING ORDER CALCULATION}
\begin{figure}[h]
\includegraphics[width=0.9\textwidth]{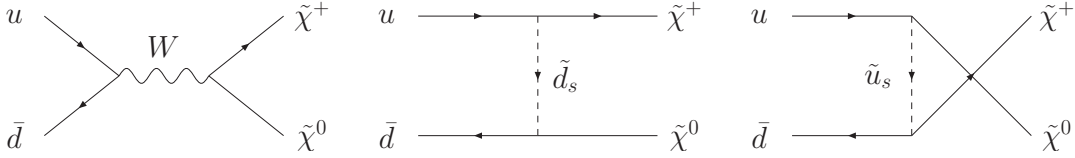}
\caption{The tree level Feynman diagrams for
chargino and neutralino associated production.
} \label{LOFD}
\end{figure}
The LO Feynman diagrams are shown in Fig.\ref{LOFD}. Considering the
light quarks as massless, at LO the production of $\chap{1}\neu{2}$
proceeds mainly via an s-channel exchange of a W boson, a t-channel
exchange of a down-type squark, and a u-channel exchange of an
up-type squark. The LO amplitude is
\begin{equation}
{\cal M}_0= {\cal M}_{s0}+\sum_{k=1}^2({\cal M}_{t0}^k+{\cal
M}_{u0}^k),
\end{equation}
with
\begin{equation}
{\cal M}_{s0}=
-\frac{D_L (A_R M_1^{RR}+A_L M_1^{RL})}{\hat{s}-M_W^2},
\end{equation}
\begin{equation}
{\cal M}_{t0}^k=
\frac{(M_2^{LR} a_{k2}^{\tilde{d}}+M_2^{LL} b_{k2}^{\tilde{d}}) C_U^k}{\hat{t}-M_{\tilde{d}_k}^2},
\end{equation}
and
\begin{equation}
{\cal M}_{u0}^k=
\frac{(M_3^{RL} a_{k2}^{\tilde{u}}+M_3^{RR} b_{k2}^{\tilde{u}}) C_V^k}{\hat{u}-M_{\tilde{u}_k}^2},
\end{equation}
where $D_L\equiv g_WV_{ud}/\sqrt{2}$, $A_L$, $A_R$,
$a_{k2}^{\tilde{d}}$, $b_{k2}^{\tilde{d}}$, $a_{k2}^{\tilde{u}}$,
$b_{k2}^{\tilde{u}}$, $C_U^k$ and $C_V^k$ are the coefficients
appearing in the SUSY couplings and their explicit expressions are
given in Appendix \ref{vertexes}. The standard matrix elements
$M_n^{ab}$ are given in Appendix \ref{formfactor}. The LO amplitude
and all of the NLO calculations in this paper are carried out in
t'Hooft-Feynman gauge. $\hat{s}$, $\hat{t}$ and $\hat{u}$ are the
Mandelstam variables defined as
\begin{equation}
\hat{s}=(p_1+p_2)^2,~~~~\hat{t}=(p_1-p_3)^2,~~~~~\hat{u}=(p_1-p_4)^2.
\end{equation}
In order to simplify the expressions, we further introduce the following modified Mandelstam variables:
\begin{equation}
\hat{t}'=\hat{t}-M_{\chap{1}}^2, ~~~~~~~~\hat{u}'=\hat{u}-M_{\chap{1}}^2.
\end{equation}

After the $D$-dimensional phase space integration, the LO parton
differential cross sections are given by
\begin{equation}
\begin{split}
\frac{d^2\hat{\sigma}^B}{d\hat{t}'d\hat{u}'}&=\frac{1}{16\pi\hat{s}^2\Gamma(1-\epsilon)}
\left(\frac{4\pi\mu_r^2\hat{s}}{\hat{t}'\hat{u}'-\hat{s}M_{\chap{1}}^2}\right)^\epsilon
\Theta(\hat{t}'\hat{u}'-\hat{s}M_{\chap{1}}^2)\Theta[\hat{s}-(M_{\chap{1}}+M_{\neu{2}})^2]
\\&
\times\delta(\hat{s}+\hat{t}+\hat{u}-M_{\chap{1}}^2-M_{\neu{2}}^2)
\overline{\sum}|{\cal M}_0|^2,
\end{split}
\end{equation}
where $\epsilon=(4-D)/2$ and the $\Theta$ function is the Heaviside
step function. The explicit expression for $\overline{\sum}|{\cal
M}_0|^2$ is
\begin{equation}
\begin{split}
\overline{\sum}&|{\cal M}_0|^2=
\frac{1}{6}\left\{
-\sum_{r,s=1}^2\frac{2\hat{s}M_{\chap{1}}M_{\neu{2}}a_{r2}^{\tilde{d}}a_{s2}^{\tilde{u}}C_U^rC_V^s}
{(\hat{t}-M_{\tilde{d}_r}^2)(\hat{u}-M_{\tilde{u}_s}^2)}
+\sum_{s=1}^2
\frac{[(a_{s2}^{\tilde{d}})^2+(b_{s2}^{\tilde{d}})^2](C_U^s)^2
(\hat{t}-M_{\chap{1}}^2)(\hat{t}-M_{\neu{2}}^2)}
{(\hat{t}-M_{\tilde{d}_s}^2)^2}
\right. \\& +
\sum_{s=1}^2
\frac{[(a_{s2}^{\tilde{u}})^2+(b_{s2}^{\tilde{u}})^2](C_V^s)^2
(\hat{u}-M_{\chap{1}}^2)(\hat{u}-M_{\neu{2}}^2)}
{(\hat{u}-M_{\tilde{u}_s}^2)^2}
+\frac{4D_L\hat{s}M_{\chap{1}}M_{\neu{2}}}{\hat{s}-M_W^2}
\sum_{s=1}^2\left[\frac{A_La_{s2}^{\tilde{d}}C_U^s}{\hat{t}-M_{\tilde{d}_s}^2}
-\frac{A_Ra_{s2}^{\tilde{u}}C_V^s}{\hat{u}-M_{\tilde{u}_s}^2}\right]
\\&
+\frac{4D_LA_R(\hat{t}-M_{\chap{1}}^2)(\hat{t}-M_{\neu{2}}^2)
}{\hat{s}-M_W^2}
\sum_{s=1}^2\frac{a_{s2}^{\tilde{d}}C_U^s}{(\hat{t}-M_{\tilde{d}_s}^2)}
-\frac{4D_LA_L(\hat{u}-M_{\chap{1}}^2)(\hat{u}-M_{\neu{2}}^2)
}{\hat{s}-M_W^2}
\sum_{s=1}^2\frac{a_{s2}^{\tilde{u}}C_V^s}{(\hat{u}-M_{\tilde{u}_s}^2)}
\\&+\frac{2C_U^1C_U^2(\hat{t}-M_{\chap{1}}^2)(\hat{t}-M_{\neu{2}}^2)
(a_{12}^{\tilde{d}}a_{22}^{\tilde{d}}+b_{12}^{\tilde{d}}b_{22}^{\tilde{d}})}
{(\hat{t}-M_{\tilde{d}_1}^2)(\hat{t}-M_{\tilde{d}_2}^2)}
 +\frac{2C_V^1C_V^2(\hat{u}-M_{\chap{1}}^2)(\hat{u}-M_{\neu{2}}^2)
(a_{12}^{\tilde{u}}a_{22}^{\tilde{u}}+b_{12}^{\tilde{u}}b_{22}^{\tilde{u}})}
{(\hat{u}-M_{\tilde{u}_1}^2)(\hat{u}-M_{\tilde{u}_2}^2)}
 \\& \left. +
\frac{4A_R^2D_L^2(\hat{t}-M_{\chap{1}}^2)(\hat{t}-M_{\neu{2}}^2)
+4A_LD_L^2(\hat{u}-M_{\chap{1}}^2)(\hat{u}-M_{\neu{2}}^2)
+8A_LA_RD_L^2\hat{s}M_{\chap{1}}M_{\neu{2}}}
{\hat{s}-M_W^2}
\right\}.
\end{split}
\end{equation}
The LO total cross section at the hadron colliders is obtained by
convoluting the parton cross section with the PDFs in the hadrons A and B:

\begin{equation}
\sigma^B=\int~dx_1dx_2~[f_{u/A}(x_1,\mu_f)f_{\bar{d}/B}(x_2,\mu_f)
+(A\leftrightarrow B)]\hat{\sigma}^B,
\end{equation}
where $\hat{\sigma}^B$ is the Born cross section
for $u\bar{d}\to\chap{1}\neu{2}$. Obviously, the  LO results 
are finite and free of  singularities.

\subsection{NEXT-TO-LEADING ORDER CALCULATION}
The NLO QCD (including SUSY QCD) corrections for the production of
$\cha{1}\neu{2}$ consist of the virtual corrections, generated by
loop diagrams of colored particles, and the real corrections with
the radiation of a real gluon or a massless (anti)quark. For both
virtual and real corrections, we will first give the results in the
dimensional regularization scheme (DREG)\cite{Hooft:1972fi}, in
which, to restore supersymmetry, we modify the Yukawa
coupling at the one loop level
\cite{Plehn:1998nh,Beenakker:1999xh,Beenakker:1996ch,Martin:1993yx}
\nocite{Hollik:1999xh}
:
\begin{equation}
g_W(q\widetilde{q}\widetilde{\chi})=g_W(1-\frac{\alphas}{6\pi}).
\label{MODYC}
\end{equation}
We will show the results in the dimensional reduction scheme
(DRED)\cite{Bern:2002zk} and compare the two schemes.

\subsubsection{VIRTUAL CORRECTIONS}
\begin{figure}[h]
\includegraphics[width=\textwidth]{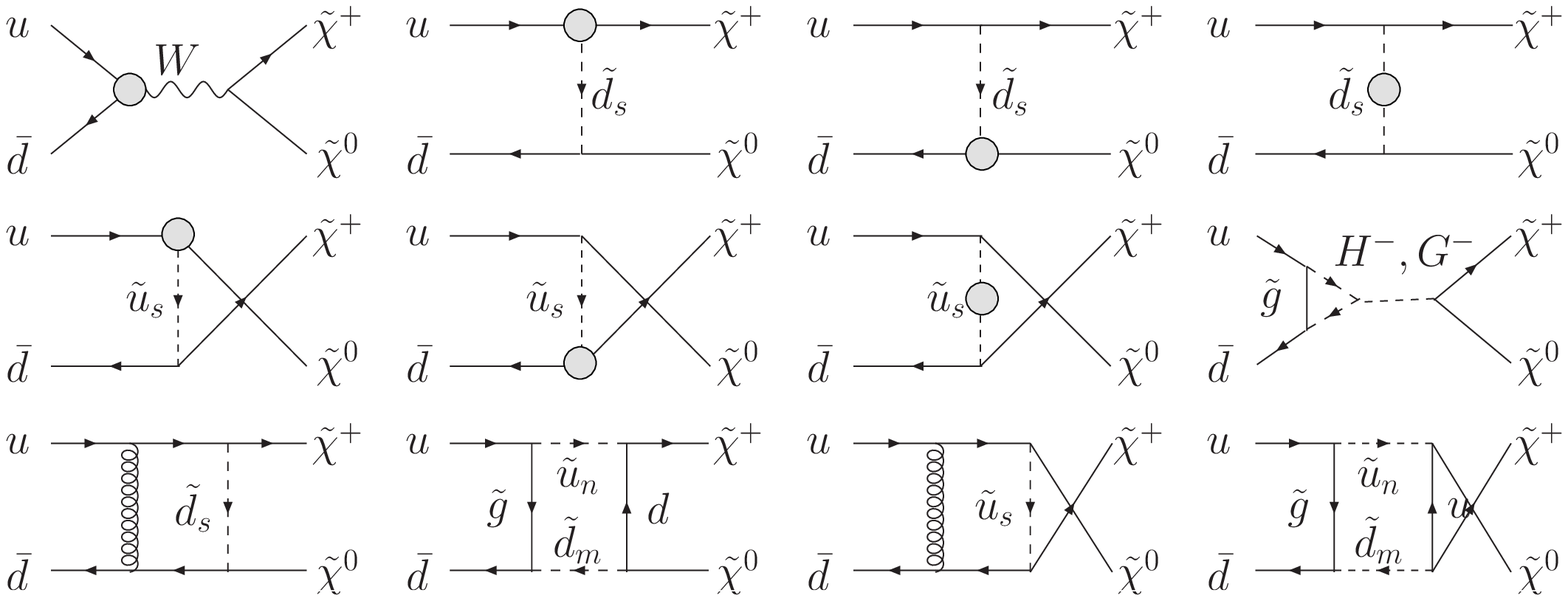}
\includegraphics[width=0.8\textwidth]{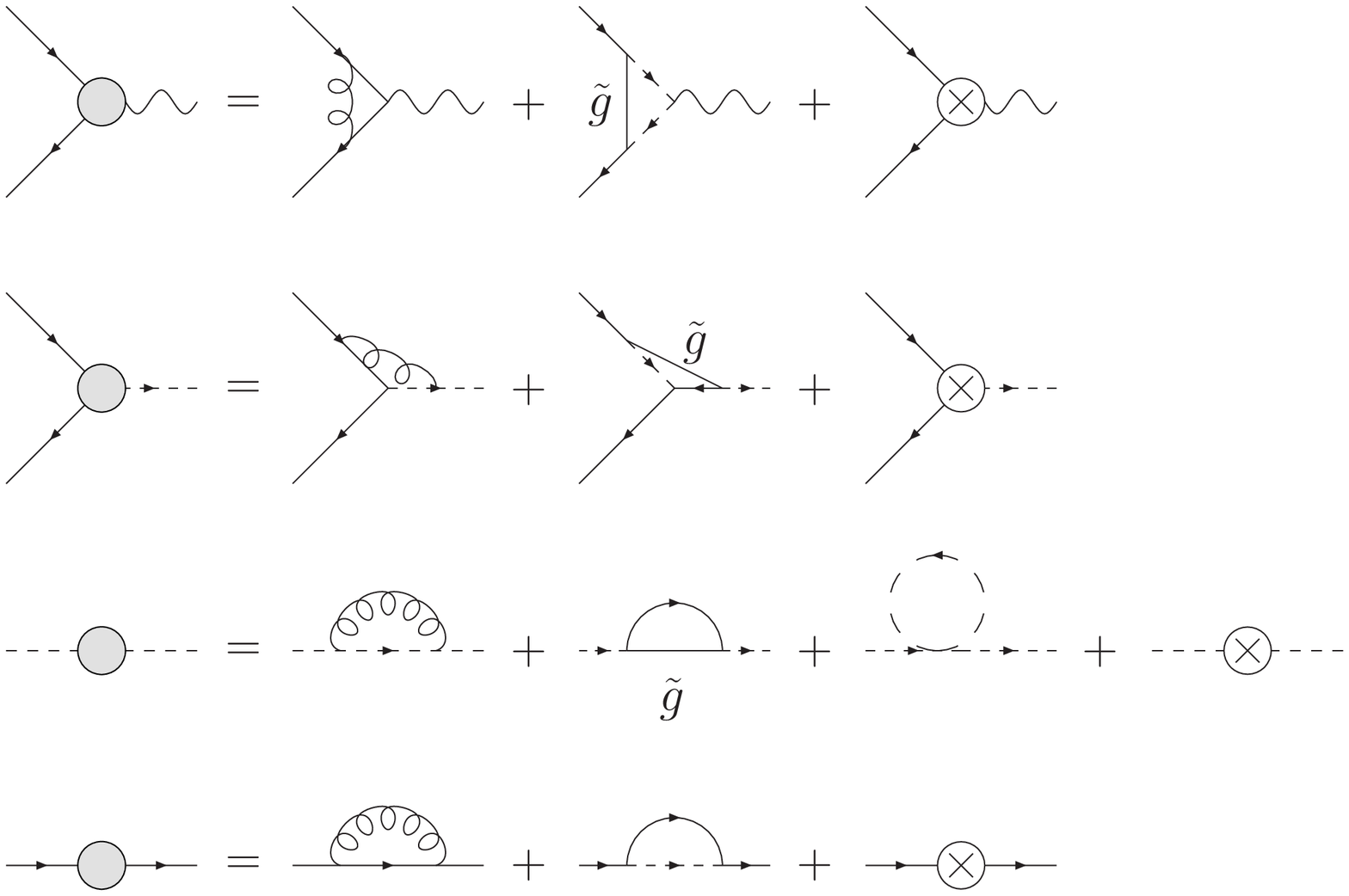}
\caption{The Feynman diagrams for the virtual corrections to
chargino and neutralino associated production.} \label{VFD}
\end{figure}

The Feynman diagrams for the virtual corrections are shown in
Fig.\ref{VFD}. In the calculations of the virtual corrections we
used the computer program package FormCalc\cite{Hahn:1998yk} to
generate the one loop amplitudes and the self energies. The
unrenormalized amplitudes for the virtual corrections are given by
\begin{equation}
{\cal M}_V=\sum_{n=1}^{24} \sum_{a,b=L,R}
(f^{ab}_{QCDVn}+f^{ab}_{SUSYVn}) M_n^{ab}. \label{VEQ}
\end{equation}
where the explicit expressions for the standard matrix elements $M_n^{ab}$
and the form factors $f^{ab}_{QCDVn}$ and $f^{ab}_{SUSYVn}$ are given in
Appendix \ref{formfactor}.
The ultraviolet (UV) divergence in the amplitude for the QCD
corrections can be expressed as
\begin{equation}
\begin{split}
{\cal M}_V\Big|_{UV}^{QCD}=&
\uvcoeff{1}\Big\{{\cal M}_{s0}
+\sum_{k=1}^2{\cal M}_{t0}^k
\frac{4\hat{t}-M_{\tilde{d}_k}^2}{\hat{t}-M_{\tilde{d}_k}^2}
+\sum_{k=1}^2{\cal M}_{u0}^k
\frac{4\hat{u}-M_{\tilde{u}_k}^2}{\hat{u}-M_{\tilde{u}_k}^2}
\Big\},
\end{split}\label{SUSYUV}
\end{equation}
and the UV divergence in the amplitude for the SUSY QCD corrections is
\begin{equation}
\begin{split}
{\cal M}_V\Big|_{UV}^{SUSY}=&
\uvcoeff{1}\Big\{{\cal M}_{s0}
+\sum_{k=1}^2{\cal M}_{t0}^k\frac{1}{\hat{t}-M_{\tilde{d}_k}^2}
\Big[-\sum_{r=1}^2{\mathcal S}^{\tilde{d}}_{kr}{\mathcal S}^{\tilde{d}}_{kr}M_{\tilde{d}_r}^2
+4M_{\tilde{g}}^2-2\hat{t}\Big]
\\ &
+\sum_{k=1}^2{\cal M}_{u0}^k\frac{1}{\hat{u}-M_{\tilde{u}_k}^2}
\Big[-\sum_{r=1}^2{\mathcal S}^{\tilde{u}}_{kr}{\mathcal S}^{\tilde{u}}_{kr}M_{\tilde{u}_r}^2
+4M_{\tilde{g}}^2-2\hat{u}\Big]
\\&
-\sum_{k=1}^2\sum_{r=1}^2\frac{C_U^r(M_2^{LR}a_{k2}^{\tilde{d}}+M_2^{LL}b_{k2}^{\tilde{d}})}
{(\hat{t}-M_{\tilde{d}_k}^2)(\hat{t}-M_{\tilde{d}_r}^2)}
\sum_{s=1}^2{\mathcal S}^{\tilde{d}}_{ks}{\mathcal S}^{\tilde{d}}_{rs}M_{\tilde{d}_s}^2
\\&
-\sum_{k=1}^2\sum_{r=1}^2\frac{C_V^r(M_3^{RL}a_{k2}^{\tilde{u}}+M_3^{RR}b_{k2}^{\tilde{u}})}
{(\hat{u}-M_{\tilde{u}_k}^2)(\hat{u}-M_{\tilde{u}_r}^2)}
\sum_{s=1}^2{\mathcal S}^{\tilde{u}}_{ks}{\mathcal S}^{\tilde{u}}_{rs}M_{\tilde{u}_s}^2
\Big\},
\end{split}\label{QCDUV}
\end{equation}
where $C_F=4/3$ and ${\mathcal S}^{\sq}_{ij}={R}^{\sq}_{i1}{R}^{\sq}_{j1}-{R}^{\sq}_{i2}{R}^{\sq}_{j2}$.
${R}^{\sq}$ is the $2\times2$ matrix shown below, and is defined to transform the squark $\sq$ current eigenstates to the mass eigenstates:
\begin{equation}
\left(\begin{array}{c}
\sq_1 \\
\sq_2
\end{array}\right)
={R}^{\sq}\left(\begin{array}{c}
\sq_L \\ \sq_R
\end{array}\right),~~~~~~
R^{\sq}=\left(
\begin{array}{cc}
\cos\theta_{\sq} & \sin\theta_{\sq}\\
-\sin\theta_{\sq}&\cos\theta_{\sq}
\end{array}\right),
\end{equation}
with $0\leq\theta_{\sq}<\pi$, by convention.
Correspondingly, the mass eigenstates $M_{\sq_1}$ and $M_{\sq_2}$
(with $M_{\sq_1}\leq M_{\sq_2}$) are given by
\begin{equation}
\left(\begin{array}{cc}
M_{\sq_1}^2 & 0 \\
0 & M_{\sq_2}^2
\end{array}
\right)=
R^{\sq}\hat{M}^2_{\sq}(R^{\sq})^{\dagger},
\end{equation}
\begin{equation}
\hat{M}_{\sq}^2=\left(
\begin{array}{cc}
M_{\sq_L}^2 & a_q M_q \\
a_q M_q & M_{\sq_R}^2
\end{array}
\right),
\end{equation}
with
\begin{equation}
M_{\sq_L}^2=M^2_{\widetilde{Q}}+M^2_q+M_Z^2\cos 2\beta(I_{3L}^q-e_q\sin^2\theta_W),
\end{equation}
\begin{equation}
M_{\sq_R}^2=M^2_{\widetilde{D}}+M^2_q+M_Z^2\cos 2\beta e_q \sin^2\theta_W,
\end{equation}
\begin{equation}
a_q=A_q-\mu\tan\beta.
\end{equation}
Here $\hat{M}^2_{\sq}$ is the squark mass matrix. $M_{\widetilde{Q},\widetilde{D}}$ and $A_q$ are soft SUSY breaking parameters and $\mu$ is the Higgsino mass parameter. $I_{3L}^q$  and $e_q$ are the third component of the weak isospin and the electric charge of the quark $q$, respectively.

In order to remove the UV divergences above we renormalized the
wave functions of the (s)quarks and the masses of squarks, adopting
the on-shell renormalization scheme \cite{Sirlin:1980nh}. And
the squark mixing matrix must also be renormalized.
\nocite{Marciano:1980pb,Sirlin:1981yz,Aoki:1982ed} Denoting
$M_{\sq_{s}0}$, $\sq_{s0}$ and $q_0$ as the bare squark mass, the
bare squark wave function, and the bare quark wave function,
respectively, the relevant renormalization constants are then
defined as
\begin{equation}
M_{\sq_{s}0}^2=M_{\sq_{s}}^2+\delta M_{\sq}^2,
\end{equation}
\begin{equation}
\sq_{s0}=(1+\frac{1}{2}\delta \widetilde{Z}_{ss}^{\sq})\sq_s+\frac{1}{2}\delta \widetilde{Z}^{\sq}_{sr}\sq_r,
\end{equation}
and
\begin{equation}
q_0=(1+\frac{1}{2}\delta Z^q_LP_L+\frac{1}{2}\delta Z^q_RP_R)q.
\end{equation}
After calculating the self energy diagrams in Fig.\ref{VFD},
we obtain the explicit expressions for the above renormalization constants:
\begin{equation}
\begin{split}
\delta M_{\sq_s}^2=&-\frac{\alphas C_F}{4\pi}
\Big\{A_0(M_{\sq_s}^2)+4M_{\sq_s}^2(B_0+B_1)(M_{\sq_s}^2,0,M_{\sq_s}^2)
+4A_0(M_{\tilde{g}}^2)+4M_{\sq_s}^2B_1(M_{\sq_s}^2,0,M_{\tilde{g}}^2)
\\& -\sum_{k=1}^2{\mathcal S}^{\sq}_{sk}{\mathcal S}^{\sq}_{ks}
A_0(M_{\sq_k}^2)\Big\},
\end{split}
\end{equation}
\begin{equation}
\delta \widetilde{Z}^{\sq}_{ss}=\frac{\alphas C_F}{\pi}\Big\{
[B_0+B_1+M_{\sq_s}^2(B'_0+B'_1)](M_{\sq_s}^2,0,M_{\sq_s}^2)
+(B_1+M_{\sq_s}^2B'_1)(M_{\sq_s}^2,0,M_{\tilde{g}}^2)\Big\},
\end{equation}
\begin{equation}
\delta\widetilde{Z}^{\sq}_{sr}=\frac{\alphas C_F}{2\pi(M_{\sq_s}^2-M_{\sq_r}^2)}
\sum_{k=1}^2{\mathcal S}^{\sq}_{sk}{\mathcal S}^{\sq}_{kr}A_0(M_{\sq_k}^2),
~~~~~~~~(s,r=1,2,~~s\neq r),
\end{equation}
\begin{equation}
\delta Z_L^q=-\frac{\alphas C_F}{2\pi}\sum_{k=1}^2(R_{k1}^{\sq})^2
(B_0+B_1)(0,M_{\sq_k}^2,M_{\tilde{g}}^2),
\end{equation}
and
\begin{equation}
\delta Z_R^q=-\frac{\alphas C_F}{2\pi}\sum_{k=1}^2(R_{k2}^{\sq})^2
(B_0+B_1)(0,M_{\sq_k}^2,M_{\tilde{g}}^2),
\end{equation}
where $B'_i=\partial B_i/\partial p^2$ and $A_0$ and $B_i$ are the
one-point and two-point integrals\cite{Denner:1991kt}, respectively.
Since we will factorize the collinear singularities into the
parton densities, as will be discussed below, the \msbar~scheme
for the renormalization of the initial quark wave functions should
be used here. However, the initial quark renormalization constants
have no finite terms, except in the SUSY QCD corrections which are
irrelevant for the PDF's. Therefore, the on-shell renormalization
scheme is equivalent to the \msbar~ scheme for initial quark
renormalization.

As for the renormalization of the squark mixing matrix, the
counterterm for the squark mixing matrix $R^{\sq}$ is defined as
\begin{equation}
R^{\sq}\rightarrow R^{\sq}+\delta R^{\sq},
\end{equation}
where the counterterm $\delta R^{\sq}$ can be fixed by requiring
that the counterterm $\delta R^{\sq}$ cancels the antisymmetric part
of the wave function corrections\cite{Denner:1990yz}
\nocite{Kniehl:1996bd,Eberl:2001eu}. The squark
mixing matrix $R^{\sq}$ counterterm can be written as
\begin{equation}
\delta R^{\sq}_{sr}=\frac{1}{4}\sum_{k=1}^2(\delta\widetilde{Z}^{\sq}_{sk}-\delta\widetilde{Z}^{\sq}_{ks})
R^{\sq}_{kr}.
\end{equation}
The corresponding counterterms for the virtual amplitudes are
given by
\begin{equation}
\begin{split}
{\cal M}_C=&\frac{1}{2}(\delta Z_L^u+\delta Z_L^d){\cal M}_{s0}
+\frac{1}{2}\delta Z_L^u\sum_{s=1}^2{\cal M}_{t0}^s
+\frac{1}{2}\sum_{s=1}^2
\frac{(M_2^{LR} a_{s2}^{\tilde{d}}\delta Z_L^d
+M_2^{LL} b_{s2}^{\tilde{d}}\delta Z_R^d
) C_U^s}{\hat{t}-M_{\tilde{d}_s}^2}
\\&
+\frac{1}{4}\sum_{s=1}^2\sum_{k=1}^2
(\frac{1}{\hat{t}-M_{\tilde{d}_s}^2}+\frac{1}{\hat{t}-M_{\tilde{d}_k}^2})
(M_2^{LR} a_{s2}^{\tilde{d}}+M_2^{LL} b_{s2}^{\tilde{d}}) C_U^k
(\delta\widetilde{Z}^{\tilde{d}}_{ks}+\delta\widetilde{Z}^{\tilde{d}}_{sk})
\\&
+\frac{1}{2}\delta Z_L^d\sum_{s=1}^2{\cal M}_{u0}^s
+\frac{1}{2}\sum_{s=1}^2
\frac{(M_3^{RL} a_{s2}^{\tilde{u}}\delta Z_L^u
+M_3^{RR} b_{s2}^{\tilde{u}}\delta Z_R^u
) C_V^s}{\hat{u}-M_{\tilde{u}_s}^2}
\\&
+\frac{1}{4}\sum_{s=1}^2\sum_{k=1}^2
(\frac{1}{\hat{u}-M_{\tilde{u}_s}^2}+\frac{1}{\hat{u}-M_{\tilde{u}_k}^2})
(M_3^{RL} a_{s2}^{\tilde{u}}+M_3^{RR} b_{s2}^{\tilde{u}}) C_V^k
(\delta\widetilde{Z}^{\tilde{u}}_{ks}+\delta\widetilde{Z}^{\tilde{u}}_{sk})
\\&
-\frac{1}{2}\sum_{s=1}^2\sum_{k=1}^2C_U^s(a_{k2}^{\tilde{d}}M_2^{LR}+b_{k2}^{\tilde{d}}M_2^{LL})
[\frac{\delta\widetilde{Z}_{sk}^{\tilde{d}}}{\hat{t}-M_{\tilde{d}_k}^2}
+\frac{\delta\widetilde{Z}_{ks}^{\tilde{d}}}{\hat{t}-M_{\tilde{d}_s}^2}]
\\&
-\frac{1}{2}\sum_{s=1}^2\sum_{k=1}^2C_V^s(b_{k2}^{\tilde{u}}M_3^{RR}+a_{k2}^{\tilde{u}}M_3^{RL})
[\frac{\delta\widetilde{Z}_{ks}^{\tilde{u}}}{\hat{u}-M_{\tilde{u}_s}^2}
+\frac{\delta\widetilde{Z}_{sk}^{\tilde{u}}}{\hat{u}-M_{\tilde{u}_k}^2}]
\\&
+\sum_{s=1}^2\frac{C_U^s\delta M_{\tilde{d}_s}^2}{(\hat{t}-M_{\tilde{d}_s}^2)^2}
(a_{s2}^{\tilde{d}}M_2^{LR}+b_{s2}^{\tilde{d}}M_2^{LL})
+\sum_{s=1}^2\frac{C_V^s\delta M_{\tilde{u}_s}^2}{(\hat{u}-M_{\tilde{u}_s}^2)^2}
(b_{s2}^{\tilde{u}}M_3^{RR}+a_{s2}^{\tilde{u}}M_3^{RL}),
\end{split}\label{CTAMP}
\end{equation}
and can be factorized: 
\begin{equation}
{\cal M}_C=\sum_{n=1}^{3} \sum_{a,b=L,R}
(f^{ab}_{QCDCn}+f^{ab}_{SUSYCn}) M_n^{ab}. \label{CEQ}
\end{equation}
The explicit expressions for the form factors $f^{ab}_{QCDCn}$
and $f^{ab}_{SUSYCn}$ are presented in Appendix \ref{formfactor}. In
Eq.(\ref{CEQ}), the UV divergences in QCD and SUSY QCD
corrections are given by
\begin{equation}
\begin{split}
{\cal M}_C\Big|_{UV}^{QCD}=&
-\uvcoeff{1}\Big\{{\cal M}_{s0}
+\sum_{k=1}^2{\cal M}_{t0}^k
\frac{\hat{t}+2M_{\tilde{d}_k}^2}{\hat{t}-M_{\tilde{d}_k}^2}
+\sum_{k=1}^2{\cal M}_{u0}^k
\frac{\hat{u}+2M_{\tilde{u}_k}^2}{\hat{u}-M_{\tilde{u}_k}^2}
\Big\},
\end{split}\label{QCDCTUV}
\end{equation}
and
\begin{equation}
\begin{split}
{\cal M}_C\Big|_{UV}^{SUSY}=&
\uvcoeff{1}\Big\{
-{\cal M}_{s0}
+\sum_{k=1}^2{\cal M}_{t0}^k\frac{1}{\hat{t}-M_{\tilde{d}_k}^2}
\Big[\sum_{r=1}^2{\mathcal S}^{\tilde{d}}_{kr}{\mathcal S}^{\tilde{d}}_{kr}M_{\tilde{d}_r}^2
-4M_{\tilde{g}}^2-\hat{t}+3M_{\tilde{d}_k}^2\Big]
\\ &
+\sum_{k=1}^2{\cal M}_{u0}^k\frac{1}{\hat{u}-M_{\tilde{u}_k}^2}
\Big[\sum_{r=1}^2{\mathcal S}^{\tilde{u}}_{kr}{\mathcal S}^{\tilde{u}}_{kr}M_{\tilde{u}_r}^2
-4M_{\tilde{g}}^2-\hat{u}+3M_{\tilde{u}_k}^2\Big]
\\&
+\sum_{k=1}^2\sum_{r=1}^2\frac{C_U^r(M_2^{LR}a_{k2}^{\tilde{d}}+M_2^{LL}b_{k2}^{\tilde{d}})}
{(\hat{t}-M_{\tilde{d}_k}^2)(\hat{t}-M_{\tilde{d}_r}^2)}
\sum_{s=1}^2{\mathcal S}^{\tilde{d}}_{ks}{\mathcal S}^{\tilde{d}}_{rs}M_{\tilde{d}_s}^2
\\&
+\sum_{k=1}^2\sum_{r=1}^2\frac{C_V^r(M_3^{RL}a_{k2}^{\tilde{u}}+M_3^{RR}b_{k2}^{\tilde{u}})}
{(\hat{u}-M_{\tilde{u}_k}^2)(\hat{u}-M_{\tilde{u}_r}^2)}
\sum_{s=1}^2{\mathcal S}^{\tilde{u}}_{ks}{\mathcal
S}^{\tilde{u}}_{rs}M_{\tilde{u}_s}^2 \Big\},
\end{split}\label{SUSYCTUV}
\end{equation}
respectively. From Eqs.(\ref{SUSYUV}),(\ref{QCDUV}), (\ref{QCDCTUV})
and (\ref{SUSYCTUV}), we obtain
\begin{equation}
({\cal M}_V+{\cal M}_C)\Big|_{UV}^{QCD}=
\uvcoeff{3}
\sum_{k=1}^2
({\cal M}_{t0}^k+{\cal M}_{u0}^k),
\label{DIV1}
\end{equation}
and
\begin{equation}
({\cal M}_V+{\cal M}_C)\Big|_{UV}^{SUSY}= -\uvcoeff{3} \sum_{k=1}^2
({\cal M}_{t0}^k+{\cal M}_{u0}^k).
\label{DIV2}
\end{equation}
The UV divergences above cancel, as they must.
The renormalized amplitude at one-loop order is UV convergent
\begin{equation}
({\cal M}_V+{\cal M}_C)\Big|_{UV}=0,
\label{DIV3}
\end{equation}
but it still contains infrared (IR) divergences:
\begin{equation}
{\cal M}_V\Big|_{IR}=
\frac{\alphas}{4\pi}\frac{\Gamma(1-\epsilon)}{\Gamma(1-2\epsilon)}
\left(\frac{4\pi\mu_r^2}{\hat{s}}\right)^\epsilon
\left(-\frac{2}{\epsilon^2}-\frac{4}{\epsilon}\right){\cal M}_0,
\end{equation}
\begin{equation}
{\cal M}_C\Big|_{IR}=
\frac{\alphas}{4\pi}\frac{\Gamma(1-\epsilon)}{\Gamma(1-2\epsilon)}
\left(\frac{4\pi\mu_r^2}{\hat{s}}\right)^\epsilon
\frac{1}{\epsilon}{\cal M}_0,
\end{equation}
and
\begin{equation}
({\cal M}_V+{\cal M}_C)\Big|_{IR}=
\frac{\alphas}{2\pi}\frac{\Gamma(1-\epsilon)}{\Gamma(1-2\epsilon)}
\left(\frac{4\pi\mu_r^2}{\hat{s}}\right)^\epsilon
\left(\frac{A_2^V}{\epsilon^2}+\frac{A_1^V}{\epsilon}\right){\cal
M}_0.
\end{equation}
Here $A_2^V=-C_F$ and $A_1^V=-\frac{3}{2}C_F$.

The ${\cal O}(\alphas)$ virtual corrections to the differential
cross section can be expressed as
\begin{equation}
\begin{split}
\frac{d^2\hat{\sigma}^V}{d\hat{t}'d\hat{u}'}&=\frac{1}{16\pi\hat{s}^2\Gamma(1-\epsilon)}
\left(\frac{4\pi\mu_r^2\hat{s}}{\hat{t}'\hat{u}'-\hat{s}M_{\chap{1}}^2}\right)^\epsilon
\Theta(\hat{t}'\hat{u}'-\hat{s}M_{\chap{1}}^2)\Theta[\hat{s}-(M_{\chap{1}}+M_{\neu{2}})^2]
\\&
\times\delta(\hat{s}+\hat{t}+\hat{u}-M_{\chap{1}}^2-M_{\neu{2}}^2)
2Re\Big[\overline{\sum}({\cal M}_C+{\cal M}_V)M_0^*\Big].
\end{split}\label{EQ44}
\end{equation}
The IR divergences in Eq.(\ref{EQ44}) include both the soft and collinear
divergences, which cancel after adding the real emission
corrections and absorbing divergences into the redefinition of
PDF's\cite{Altarelli:1979ub}, as will be discussed below.
\subsubsection{REAL CORRECTIONS}
\begin{figure}[h]
\includegraphics[width=\textwidth]{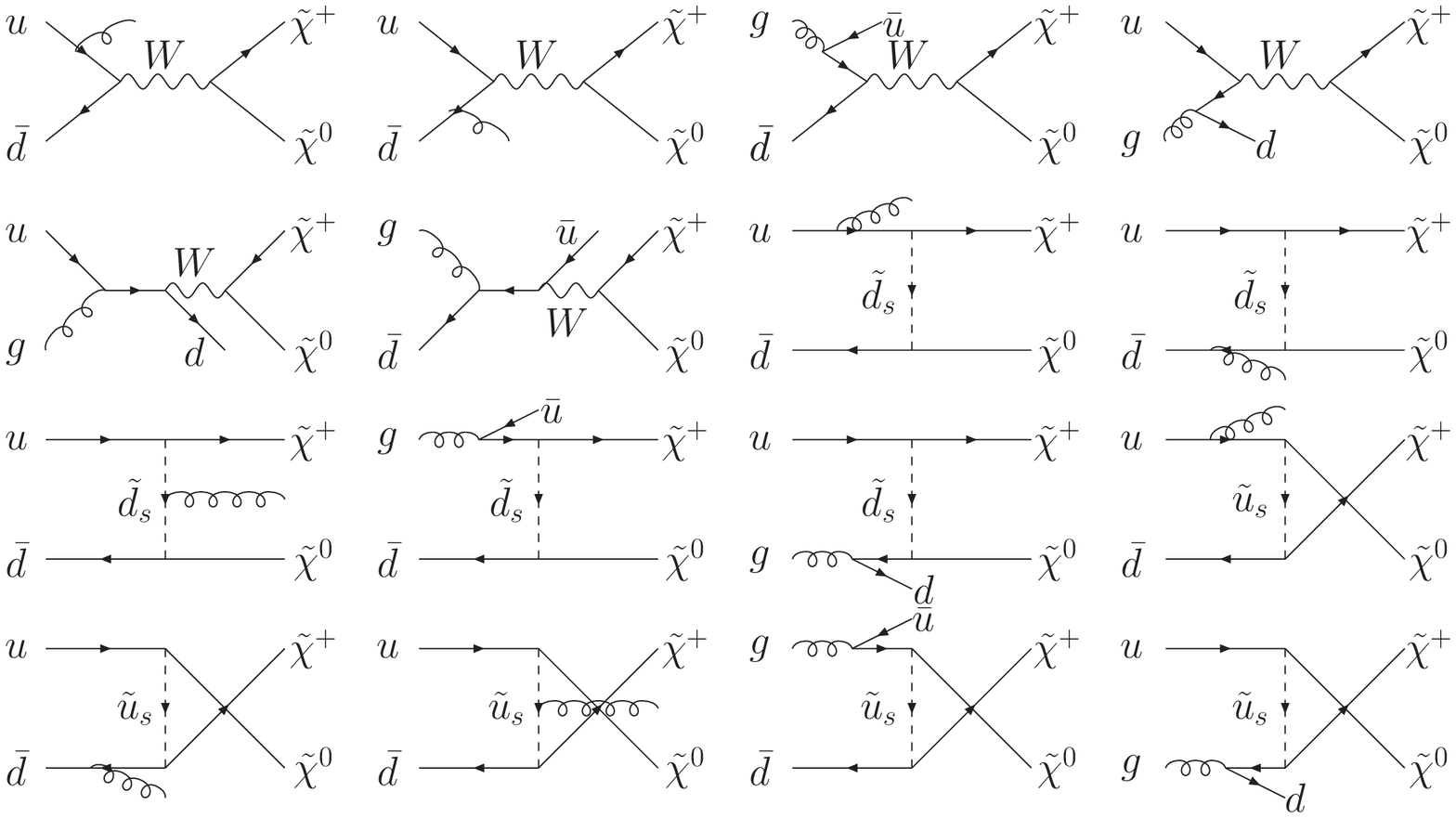}
\caption{The Feynman diagrams for the real corrections without
squark resonances  in chargino and neutralino associated production.
} \label{R1FD}
\end{figure}

\begin{figure}[h]
\includegraphics[width=\textwidth]{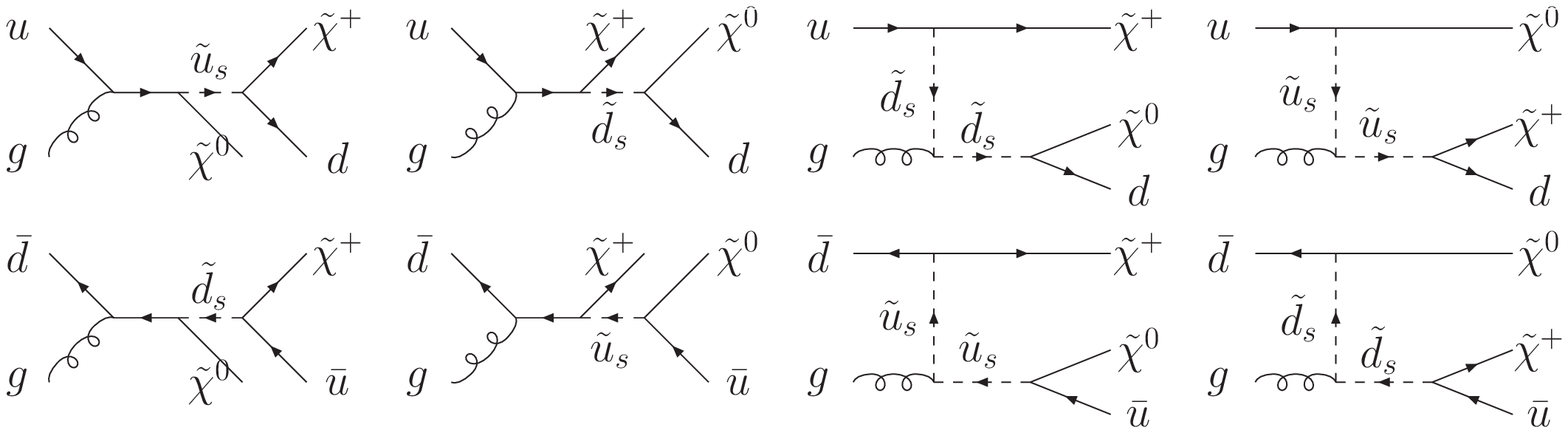}
\caption{The Feynman diagrams for the real corrections with squark
resonances in chargino and neutralino associated production.} \label{R2FD}
\end{figure}

The Feynman diagrams for the real emission corrections are shown in
Figs.\ref{R1FD} and \ref{R2FD}. 
\newpage
After calculating the relevant
Feynman diagrams the amplitudes for the real gluon emission process
\begin{equation}
u(p_1)+\bar{d}(p_2)\to\chap{1}(p_3)+\neu{2}(p_4)+g(p_5)
\label{realg}
\end{equation}
and the real massless (anti)quark emission processes
\begin{equation}
u(p_1)+g(p_2)\to\chap{1}(p_3)+\neu{2}(p_4)+d(p_5) \label{reald}
\end{equation}
and
\begin{equation}
\bar{d}(p_1)+g(p_2)\to\chap{1}(p_3)+\neu{2}(p_4)+\bar{u}(p_5)
\label{realubar}
\end{equation}
can be written as
\begin{equation}
{\cal M}_{RG}=\sum_{n=1}^{23} \sum_{a,b=L,R} f^{ab}_{RGn}
M_{Gn}^{ab} \label{RGEQ}
\end{equation}
and
\begin{equation}
{\cal M}_{RQ}=\sum_{n=1}^{21} \sum_{a,b=L,R} f^{ab}_{RQn}
M_{Qn}^{ab}, \label{RQEQ}
\end{equation}
respectively. The explicit expressions for the form factors
$f^{ab}_{RGn}$ and $f^{ab}_{RQn}$ and the standard matrix elements
$M_{Gn}^{ab}$ and $M_{Qn}^{ab}$ in Eqs.(\ref{RGEQ}) and (\ref{RQEQ})
are given in Appendix \ref{formfactor}.

The phase space integration for the real corrections will produce
soft and collinear singularities, which can be conveniently isolated
by slicing phase space into different regions using suitable
cutoffs. We used the two-cutoff phase space slicing
method\cite{Harris:2001sx}, which introduces two arbitrarily small
cutoffs, $\delta_s$ and $\delta_c$, to decompose the three-body
phase space into three regions.

The parton level cross section for real gluon emission
$\hat{\sigma}^R$ contains both the soft and the collinear
singularities and, in general, can be written as
\begin{equation}
\hat{\sigma}^R=\hat{\sigma}^S+\hat{\sigma}^{HC}+\hat{\sigma}^{\overline{HC}},
\end{equation}
where $\hat{\sigma}^S$ and $\hat{\sigma}^{HC}$ are the contributions
from the soft and the hard collinear regions, respectively, and
$\hat{\sigma}^{\overline{HC}}$ is the hard noncollinear par The
explicit forms are described below.

In the soft limit the energy of the emitted gluon is
small, i.e. $E_5\leq\delta_s\sqrt{\hat{s}}/2$, and the squared amplitude
$\overline\sum|{\cal M}_{RG}|^2$ can simply be factorized into the
squared Born amplitude times an eikonal factor $\Phi_{eik}$:
\begin{equation}
\overline\sum|{\cal M}_{RG}|^2\stackrel{soft}{\longrightarrow}(4\pi\alphas\mu_r^{2\epsilon})
\overline\sum|{\cal M}_0|^2\Phi_{eik},
\end{equation}
where the eikonal factor $\Phi_{eik}$ is given by
\begin{equation}
\Phi_{eik}=C_F\frac{\hat{s}}{(p_1\cdot p_5)(p_2\cdot p_5)}.
\end{equation}
The phase space in the soft limit also be factorizes: 
\begin{equation}
dPS^{(3)}(u\bar{d}\to\chap{1}\neu{2}g)\stackrel{soft}{\longrightarrow} dPS^{(2)}(u\bar{d}\to\chap{1}\neu{2})dS,
\end{equation}
where $dS$ is the integration over the phase space of the soft gluon
which is given by\cite{Harris:2001sx}
\begin{equation}
dS=\frac{1}{2(2\pi)^{3-2\epsilon}}\int_0^{\delta_s\sqrt{\hat{s}}/2}
dE_5 E_5^{1-2\epsilon}d\Omega_{2-2\epsilon}.
\end{equation}
The parton level cross section in the soft region can then be
expressed as
\begin{equation}
\hat{\sigma}^S=(4\pi\alphas\mu_r^{2\epsilon})\int dPS^{(2)}\overline\sum|{\cal M}_0|^2
\int dS\Phi_{eik}.
\label{soft}
\end{equation}
Using the approach in Ref.\cite{Harris:2001sx}, after integration over
the soft gluon phase space, Eq.(\ref{soft}) becomes
\begin{equation}
\hat{\sigma}^S=\hat{\sigma}^B
\left[\frac{\alphas}{2\pi}\frac{\Gamma(1-\epsilon)}{\Gamma(1-2\epsilon)}
\left(\frac{4\pi\mu_r^2}{\hat{s}}\right)^\epsilon\right]
(\frac{A_2^S}{\epsilon^2}+\frac{A_1^S}{\epsilon}+A^S_0),
\end{equation}
with
\begin{equation}
A_2^S=2C_F,~~A_1^S=-4C_F\ln\delta_s,~~A_0^S=4C_F\ln^2\delta_s.
\end{equation}

In the hard collinear region, $E_5>\delta_s\sqrt{\hat{s}}/2$ and
$-\delta_c\sqrt{s}<\hat{u}_{1,2}\equiv(p_{1,2}-p_5)^2<0$, the
emitted hard gluon is collinear to one of the partons. As a
consequence of the factorization theorems \cite{Bodwin:1984hc}
\nocite{Collins:1985ue} , the squared amplitude for the gluon
emission process (\ref{realg}) can be factorized into the product of
the squared Born amplitude and the Altarelli-Parisi splitting
function\cite{Altarelli:1977zs}\nocite{Ellis:1980wv,Bergmann:xxx,Mangano:1991jk,Kunszt:1992tn}
for $u(\bar{d})\to u(\bar{d})g$,
\begin{equation}
\overline\sum|{\cal M}_{RG}|^2\stackrel{collinear}{\longrightarrow}(4\pi\alphas\mu_r^{2\epsilon})
\overline\sum|{\cal M}_0|^2\Big(\frac{-2P_{uu}(z,\epsilon)}{z\hat{u}_1}
+\frac{-2P_{\bar{d}\bar{d}}(z,\epsilon)}{z\hat{u}_2}\Big).
\end{equation}
Here $z$ denotes the fraction of the momentum carried by parton $u(\bar{d})$
$u(\bar{d})$ with the emitted gluon carrying a fraction $(1-z)$ and $P_{ij}(z,\epsilon)$ are the
unregulated splitting functions in $D=4-2\epsilon$ dimensions for $0<z<1$,
which are related to the usual Altarelli-Parisi splitting kernels \cite{Altarelli:1977zs} as follows:
$P_{ij}(z,\epsilon)=P_{ij}(z)+\epsilon P'_{ij}(z)$. Explicitly
\begin{equation}
P_{uu}(z)=P_{\bar{d}\bar{d}}(z)=C_F\frac{1+z^2}{1-z}+C_F\frac{3}{2}\delta(1-z),
\end{equation}
and
\begin{equation}
P'_{uu}(z)=P'_{\bar{d}\bar{d}}(z)=-C_F(1-z)+C_F\frac{1}{2}\delta(1-z).
\end{equation}
The three-body phase space can also be factorized in the collinear limit
and, for example, in the limit $-\delta_c\hat{s}<\hat{u}_1<0$
it has the following form\cite{Harris:2001sx}:
\begin{equation}
dPS^{(3)}(u\bar{d}\to\chap{1}\neu{2}g)\stackrel{collinear}{\longrightarrow}dPS^{(2)}
(u\bar{d}\to\chap{1}\neu{2};\hat{s}'=z\hat{s})
\frac{(4\pi)^\epsilon}{16\pi^2\Gamma(1-\epsilon)}dzd\hat{u}_1
[-(1-z)\hat{u}_1]^{-\epsilon}.
\end{equation}
Here the two-body phase space is evaluated at a squared parton-parton energy  $z\hat{s}$.
The three-body cross section in the hard collinear region is then given by\cite{Harris:2001sx}
\begin{equation}
\begin{split}
d\sigma^{HC}=&\hat{\sigma}^B\Big[\frac{\alphas}{2\pi}\frac{\Gamma(1-\epsilon)}{\Gamma(1-2\epsilon)}
\Big(\frac{4\pi\mu_r^2}{\hat{s}}\Big)^\epsilon\Big]
\Big(-\frac{1}{\epsilon}\Big)\delta_c^{-\epsilon}[P_{uu}(z,\epsilon)f_{u/A}(x_1/z)f_{\bar{d}/B}(x_2)
\\& +P_{\bar{d}\bar{d}}(z,\epsilon)f_{\bar{d}/A}(x_1/z)f_{u/B}(x_2)
+(A\leftrightarrow B)]\frac{dz}{z}
\Big(\frac{1-z}{z}\Big)^{-\epsilon}dx_1dx_2,
\end{split}
\end{equation}
where $f(x)$ is a bare PDF.

After subtracting the soft and collinear region of the phase space,
the remaining hard noncollinear part $\hat{\sigma}^{\overline{HC}}$
is finite and can be numerically computed using Monte-Carlo
integration techniques\cite{Hahn:2004fe}. The result and can be written in the
form
\begin{equation}
d\hat{\sigma}^{\overline{HC}}=\frac{1}{2\hat{s}}\overline{\sum} |{\cal M}_{RG}|^2d\overline{PS}^{(3)},
\end{equation}
where $d\overline{PS}^{(3)}$ is the hard noncollinear region of the three-body phase space.

In addition to real gluon emission, other real
emission corrections to the inclusive cross section for
$A+B\to\cha{1}\neu{2}$ at NLO involve the processes with an
additional massless (anti)quark in the final state. Since the
contributions from real massless (anti)quark emission contain
initial state collinear singularities, we also need to use the
two-cutoff phase space slicing method \cite{Harris:2001sx} to
isolate these collinear divergences. But we only split the phase
space into two regions since there are no soft divergences.
Consequently, using the approach in Ref.\cite{Harris:2001sx}, the
cross sections for the processes with an additional massless
(anti)quark in the final state can be expressed as
\begin{equation}
\begin{split}
d\sigma^{add}=&
\sum_{(\alpha=u,\bar{d})}\hat{\sigma}^{\overline{C}}(g\alpha\to\chap{1}\neu{2}+X)
[f_{g/A}(x_1)f_{\alpha/B}(x_2)+ (A\leftrightarrow B)]dx_1dx_2
\\&+d\hat{\sigma}^B\Big[\frac{\alphas}{2\pi}\frac{\Gamma(1-\epsilon)}{\Gamma(1-2\epsilon)}
\Big(\frac{4\pi\mu_r^2}{\hat{s}}\Big)^\epsilon\Big]\Big(-\frac{1}{\epsilon}\Big)
\delta_c^{-\epsilon}
[P_{ug}(z,\epsilon)f_{g/A}(x_1/z)f_{\bar{d}/B}(x_2)
\\&+f_{u/A}(x_1)P_{\bar{d}g}(z,\epsilon)f_{g/B}(x_2/z)
+(A\leftrightarrow B)]
\frac{dz}{z}\Big(\frac{1-z}{z}\Big)^{-\epsilon}dx_1dx_2,
\end{split}
\label{addsigma}
\end{equation}
where
\begin{equation}
P_{ug}(z)=P_{\bar{d}g}(z)=\frac{1}{2}[z^2+(1-z)^2],
\end{equation}
\begin{equation}
P'_{ug}(z)=P'_{\bar{d}g}(z)=-z(1-z).
\end{equation}
The first term in Eq.(\ref{addsigma}) represents the noncollinear
cross section. The parton cross section
$\hat{\sigma}^{\overline{C}}$ can be written in the form
\begin{equation}
\hat{\sigma}^{\overline{C}}=\int\frac{1}{2\hat{s}}\overline{\sum} |{\cal M}_{RQ}|^2d\overline{PS}^{(3)},
\end{equation}
where $d\overline{PS}^{(3)}$ is the three-body phase space in the
noncollinear region. The second term in Eq.(\ref{addsigma})
represents the collinear singular cross sections.

\subsubsection{MASS FACTORIZATION AND NLO TOTAL CROSS SECTIONS}
As mentioned above, after adding the renormalized virtual
corrections and the real corrections the parton level cross
sections still contain collinear divergences. These can be
absorbed into a redefinition of the PDF's at NLO, using
mass factorization\cite{Altarelli:1977zs}. In
practice this means that first we convolute the parton cross sections
with the bare PDF's $f_{\alpha/H}(x)~(H=A,B)$ and then use the
renormalized PDF's $f_{\alpha/H}(x, \mu_f)$ to replace
$f_{\alpha/H}(x)$. In the \msbar~convention the scale-dependent PDF's
$f_{\alpha/H}(x,\mu_f)$ are given by \cite{Harris:2001sx}
\begin{equation}
f_{\alpha/H}(x,\mu_f)=f_{\alpha/H}(x)+\sum_{\beta}\left(-\frac{1}{\epsilon}\right)
\left[\frac{\alphas}{2\pi}\frac{\Gamma(1-\epsilon)}{\Gamma(1-2\epsilon)}
\left(\frac{4\pi\mu_r^2}{\mu_f^2}\right)^\epsilon\right]
\int^1_x\frac{dz}{z}P_{\alpha\beta}^+(z)f_{\beta/H}(x/z).
\end{equation}
This replacement produces a collinear singular counterterm
which, when combined with the hard collinear contributions,gives,
as in Ref.\cite{Harris:2001sx}, the ${\cal O}(\alphas)$
expression for the remaining collinear contribution:
\begin{eqnarray}
d\sigma^{coll}&&=
\left[\frac{\alphas}{2\pi}\frac{\Gamma(1-\epsilon)}{\Gamma(1-2\epsilon)}
\left(\frac{4\pi\mu_r^2}{\hat{s}}\right)^\epsilon\right] \left\{
\tilde{f}_{u/A}(x_1,\mu_f)f_{\bar{d}/B}(x_2,\mu_f) +
f_{u/A}(x_1,\mu_f)\tilde{f}_{\bar{d}/B}(x_2,\mu_f) \right. \nonumber
\\ && \left. +\left[\frac{A_1^{sc}}{\epsilon}+A_0^{sc}\right]
f_{u/A}(x_1,\mu_f)f_{\bar{d}/B}(x_2,\mu_f) +(A\leftrightarrow
B)\right\} \hat\sigma^B dx_1dx_2, \label{sigmacoll}
\end{eqnarray}
where
\begin{equation}
A_1^{sc}=C_F(4\ln\delta_s+3),~~
A_0^{sc}=A_1^{sc}\ln(\frac{\hat{s}}{\mu_f^2}),
\end{equation}
\begin{equation}
\tilde{f}_{\alpha(=u,\bar{d})/H}(x,\mu_f)=\sum_{\beta=g,\alpha}\int^{1-\delta_s\delta_{\alpha\beta}}_x
\frac{dy}{y}f_{\beta/H}(x/y,\mu_f)\widetilde{P}_{\alpha\beta}(y),
\end{equation}
with
\begin{equation}
\widetilde{P}_{\alpha\beta}(y)=P_{\alpha\beta}(y)\ln(\delta_c\frac{1-y}{y}\frac{\hat{s}}{\mu_f^2})
-P'_{\alpha\beta}(y).
\end{equation}

Finally, the NLO total cross section for $A+B\to\chap{1}\neu{2}$ in
the \msbar~ factorization scheme is
\begin{eqnarray}
{\sigma}^{NLO}&&=\sum_{\alpha,\beta=u,\bar{d}}\int dx_1dx_2\left\{
\left[f_{\alpha/A}(x_1, \mu_f)f_{\beta/B}(x_2, \mu_f)\right]
(\hat{\sigma}^B+\hat{\sigma}^V+\hat{\sigma}^S+\hat{\sigma}^{\overline{HC}})\right\}+\sigma^{coll}
\nonumber \\ && +\sum_{\alpha=u,\bar{d}} \int dx_1 dx_2\left[
f_{g/A}(x_1,\mu_f)f_{\alpha/B}(x_2,\mu_f)+ (A\leftrightarrow B)
\right] \hat{\sigma}^{\overline{C}}(g\alpha\to\chap{1}\neu{2}X) .
\label{totalcs}
\end{eqnarray}
Note that the expression above contains no singularities for
$2A_2^V+A_2^S=0$ and $2A_1^V+A_1^S+A_1^{sc}=0$.

\subsubsection{ON-SHELL SUBTRACTION}
In the massless (anti)quark corrections there is resonance
production of squarks, which actually
corresponds to squark and gaugino production at the LO followed by
squark decay to a gaugino and a quark, as shown in Fig.\ref{R2FD}. We used the method in
Ref.\cite{Beenakker:1996ch} to subtract their contributions. For
example, consider a representative process
\begin{equation}
u+g\to\tilde{u}+\neu{2},~~~~\tilde{u}\to\chap{1}+d,
\end{equation}
which is shown as the first Feynman diagram in Fig.\ref{R2FD}.
Using the Breit-Wigner propagator $1/(p^2-m^2+im\Gamma)$, the
squared resonance matrix elements can be expressed as
\begin{equation}
|{\cal M}|^2=\frac{f(Q^2)}{(Q^2-M_{\tilde{u}}^2)^2+M_{\tilde{u}}^2\Gamma_{\tilde{u}}^2},
\end{equation}
where $Q^2=(p_{\chap{1}}+p_d)^2$. After subtracting the
contributions due to resonance production the squared matrix
element is
\begin{equation}
|{\cal M}|^2=\frac{f(Q^2)}{(Q^2-M_{\tilde{u}}^2)^2+M_{\tilde{u}}^2\Gamma_{\tilde{u}}^2}
-\frac{f(M_{\tilde{u}}^2)}{(Q^2-M_{\tilde{u}}^2)^2+M_{\tilde{u}}^2\Gamma_{\tilde{u}}^2}.
\end{equation}
This subtracted result avoids double counting and makes
the numerical calculation more stable since the resonance peaks are
subtracted before the phase space integration. The dependence on the squark widths
will be discussed in Sec.~IV.

\subsubsection{NLO TOTAL CROSS SECTIONS IN BOTH DREG AND DRED SCHEMES}
In our calculations we used the DREG scheme. However, this scheme is
not appropriate for SUSY models because it violates
supersymmetry. To restore supersymmetry we modified the
Yukawa coupling at the one loop level as shown in Eq.(\ref{MODYC}).

The real corrections and NLO total cross sections in the DREG scheme have been given above.
Next we show the corresponding results in the DRED scheme. The
contributions from soft gluon emission remain the same, but in addition
to the the modified Yukawa couplings those from hard collinear gluon
emission and massless (anti)quark emission are also different. These
differences arise from the splitting functions and the PDF's.

First, note the LO amplitude in the DREG scheme with modified Yukawa
couplings (DREGM) is different from that in the DRED scheme: 
\begin{equation}
{\cal M}_0^{DREGM}-{\cal M}_0^{DRED}=-\frac{\alphas C_F}{4\pi}
\sum_{k=1}^2({\cal M}_{t0}^k+{\cal M}_{u0}^k). \label{DIFF1}
\end{equation}
Here, and below, the LO amplitudes and cross sections in the right
hand side of equations are all in 4-dimensions, and their Yukawa
couplings are not modified.
Calculating the virtual corrections in the DRED scheme one
finds that $\delta Z_L^f$, $\delta Z_R^f$,
$\delta\widetilde{Z}_{ij}^{\sq}$ and $\delta M_{\sq}^2$ remain
the same as in the DREG scheme. Thus 
\begin{equation}
{\cal M}_C^{DREGM}-{\cal M}_C^{DRED}=0. \label{DIFF2}
\end{equation}
However, the unrenormalized amplitudes ${\cal M}_V$ differ:
\begin{equation}
{\cal M}_V^{DREGM}-{\cal M}_V^{DRED}=-\frac{\alphas C_F}{4\pi}{\cal
M}_{s0}.\label{DIFF3}
\end{equation}
From Eqs.(\ref{DIFF1}), (\ref{DIFF2}) and (\ref{DIFF3}), one finds
the following relations:
\begin{equation}
({\cal M}_0+{\cal M}_V+{\cal M}_C)^{DREGM} -({\cal M}_0+{\cal
M}_V+{\cal M}_C)^{DRED} =-\frac{\alphas C_F}{4\pi}{\cal M}_0,
\end{equation}
\begin{equation}
(\hat{\sigma}^B+\hat{\sigma}^V)^{DREGM}
-(\hat{\sigma}^B+\hat{\sigma}^V)^{DRED} =-\frac{\alphas
C_F}{2\pi}\hat{\sigma}^B+{\cal O}(\alphas^2). \label{formdiff}
\end{equation}

Second, note the splitting functions in the DRED scheme have no dependence on
$\epsilon$:
\begin{equation}
P_{ij}(z,\epsilon)^{DRED}=P_{ij}(z).
\label{DREDP}
\end{equation}
From Eqs. (\ref{sigmacoll}) and (\ref{DREDP}), one finds
\begin{equation}
\begin{split}
(\sigma^{coll})^{DREGM}-(\sigma^{coll})^{DRED}&=
-\frac{\alphas}{2\pi}\Big\{
\sum_{\alpha}\int_{x_1}^{1-\delta_s\delta_{u\alpha}}\frac{dy}{y}
f_{\alpha/A}(x_1/y,\mu_f)P'_{u\alpha}(y)f_{\bar{d}/B}(x_2,\mu_f)
\\&+\sum_{\alpha}\int_{x_2}^{1-\delta_s\delta_{\bar{d}\alpha}}
\frac{dy}{y}f_{\alpha/B}(x_2/y,\mu_f)
P'_{\bar{d}\alpha}(y)f_{u/A}(x_1,\mu_f)
\\&+(A\leftrightarrow B)\Big\}\hat{\sigma}^B dx_1dx_2
+{\cal O}(\alphas^2).
\end{split}\label{colldiff}
\end{equation}

Third, note the PDF's in the DRED and DREG schemes are
related\cite{Kamal:1995as}:
\begin{equation}
f_{\alpha/A,B}(x,\mu_f)^{DREG}=f_{\alpha/A,B}(x,\mu_f)^{DRED}
+\frac{\alphas}{2\pi}\sum_{\beta}\int_{x}^1\frac{dy}{y}P'_{\alpha\beta}(x/y)f_{\beta/A,B}(y,\mu_f)^{DRED}.
\end{equation}
Substituting into the formula for the Born cross section we obtain
an additional difference at ${\cal O}(\alphas)$  arising
from the PDF's:
\begin{equation}
\begin{split}
(\sigma^B)^{DREGM}-(\sigma^B)^{DRED}=& \frac{\alphas}{2\pi}\Big\{
\sum_{\alpha}\int^1_{x_1}\frac{dy}{y}f_{\alpha/A}(x_1/y,\mu_f)^{DRED}
P'_{u\alpha}(y)f_{\bar{d}/B}(x_2,\mu_f)^{DRED}
\\&+\sum_{\alpha}\int^1_{x_2}\frac{dy}{y}f_{\alpha/B}(x_2/y,\mu_f)^{DRED}
P'_{\bar{d}\alpha}(y)f_{u/A}(x_1,\mu_f)^{DRED}
\\&+(A\leftrightarrow B)\Big\}\hat{\sigma}^B dx_1dx_2.
\end{split}\label{pdfdiff}
\end{equation}

Finally note that Eqs. (\ref{colldiff}) and (\ref{pdfdiff}) are very similar except for the limits on
the integral over y.
Substituting Eqs. (\ref{formdiff}), (\ref{colldiff}) and (\ref{pdfdiff}) into Eq. (\ref{totalcs}),
we obtain the following relations for the NLO total cross sections in two schemes:
\begin{equation}
\begin{split}
(\sigma^{NLO})^{DREGM}-(\sigma^{NLO})^{DRED}=&
\frac{\alphas}{2\pi}\Big\{
\sum_{\alpha}\int^1_{1-\delta_s\delta_{u\alpha}}
\frac{dy}{y}f_{\alpha/A}(x_1/y,\mu_f)P'_{u\alpha}(y)f_{\bar{d}/B}(x_2,\mu_f)
\\&+\sum_{\alpha}\int^1_{1-\delta_s\delta_{\bar{d}\alpha}}\frac{dy}{y}
f_{\alpha/B}(x_2/y,\mu_f)P'_{\bar{d}\alpha}(y)
f_{u/A}(x_1,\mu_f)\\&+ (A\leftrightarrow B)\Big\}\hat{\sigma}^B
dx_1dx_2 -\frac{\alphas C_F}{2\pi}\sigma^B +{\cal O}(\alphas^2).
\end{split}
\end{equation}
Using the explicit expressions, including the $\epsilon$ dependece, for the splitting functions $P'$,
one finds
\begin{equation}
(\sigma^{NLO})^{DREGM}-(\sigma^{NLO})^{DRED}={\cal O}(\alphas^2).
\end{equation}
Therefore, the NLO total cross sections in the two schemes are the same at NLO.

\section{THRESHOLD RESUMMATION}
Here we briefly summarize the basic formalism for threshold
resummation Refs.
\cite{
Contopanagos:1996nh,
Kidonakis:1996aq,
Kidonakis:2006bu,
Arnold:1990yk,
Eynck:2003fn,
Catani:1989ne,
Catani:1996yz,
Kulesza:2002rh}.
\nocite{Kidonakis:1999ze}
\nocite{Kidonakis:1997gm,Laenen:1998qw}
\nocite{Bozzi:2007qr}
\nocite{Catani:1992ua}
\nocite{Catani:1999hs}
Using pair inclusive (PIM) kinematics the invariant mass differential
cross section can be written as
\begin{equation}
\omega(S,Q^2)= \sum_{a,b}\int^1_\tau~dz\int~dx_a dx_b f_{a/A}(x_a,
\mu_f) f_{b/B}(x_b, \mu_f) \delta(z-\frac{Q^2}{x_a x_b S})
\hat{\omega}_{ab}(z,Q^2),
\end{equation}
where
\begin{equation}
\tau=\frac{Q^2}{S},~~~~
\omega=\frac{d\sigma}{dQ^2},~~~~
\hat{\omega}_{ab}=\frac{d\hat{\sigma}_{ab}}{dQ^2},~~~~
z=\frac{Q^2}{\hat{s}},
\end{equation}
and $Q^2$ is the invariant mass of the chargino and neutralino. The
differential cross section $\hat{\omega}_{ab}$ contains large
logarithmic terms $\alphas^n[\ln^m(1-z)/(1-z)]_+$, which come from the
incomplete cancelation between real gluon emission and
virtual gluon corrections. In the region
$z\approx 1$ ($Q^2\approx\hat{s}$) these large logarithms have to
be resummed to all orders in $\alphas$.

In order to calculate the hard-scattering function $\hat{\omega}_{ab}$
we consider the IR regularized cross section for parton-parton scattering
which factorizes:
\begin{equation}
\omega_{ab}= \int^1_\tau~dz\int~dx_a dx_b \phi_{a/a}(x_a, \mu_f)
\phi_{b/b}(x_b, \mu_f) \delta(z-\frac{Q^2}{x_a x_b S})
\hat{\omega}_{ab}(z,Q^2), \label{DIFFCSEQ}
\end{equation}
where $\phi_{a/a}$ and $\phi_{b/b}$ are the flavor diagonal parton
distributions in partons. Using a Mellin transformation with respect to $\tau$ the
convolution in Eq.(\ref{DIFFCSEQ}) can be simplified as the product
\begin{equation}
\tilde{\omega}_{ab}(N)=\tilde{\phi}_{a/a}(N)
\tilde{\phi}_{b/b}(N)\tilde{\hat{\omega}}_{ab}(N), \label{PHIFAC}
\end{equation}
where
\begin{equation}
\tilde\omega(N)=\int~d\tau~\tau^{N-1}\omega(\tau),
\end{equation}
\begin{equation}
\tilde\phi_{a/a}(N)=\int~dx~x^{N-1}\phi_{a/a}(x),
\end{equation}
\begin{equation}
\tilde\phi_{b/b}(N)=\int~dx~x^{N-1}\phi_{b/b}(x),
\end{equation}
and
\begin{equation}
\tilde{\hat{\omega}}=\int~dz~z^{N-1}\hat{\omega}(z).
\end{equation}
Here the large logarithmic terms in $\tilde{\hat{\omega}}_{ab}$
turn out to be $\alphas^n\ln^mN$ and $z\to 1$ corresponds
to $N \to \infty$. The next step is to resum the logarithms of
$N$.

In order to separate the soft gluon effects from the short distance
hard scattering we can factorize the differential cross section
into the form
\begin{equation}
\tilde{\omega}_{ab}(N)= \tilde{\psi}_{a/a}(N)\tilde{\psi}_{b/b}(N)
\sum_{IJ}H_{IJ}^{ab}\tilde{S}_{JI}^{ab}(\frac{Q}{N\mu_f}),\label{PSIFAC}
\end{equation}
where $I,J$ are color indices, $H_{IJ}$ describes the short
distance hard scattering and $\tilde{S}_{JI}$ is a soft gluon
function associated with noncollinear soft gluons. The explicit
definitions of $H_{IJ}$ and $\tilde{S}_{JI}$ can be found in
Ref.\cite{Contopanagos:1996nh}. The $\psi$'s are the center-of-mass
parton distribution functions in which the universal collinear
singularities associated with the initial partons are absorbed.

From Eq.(\ref{PHIFAC}) and Eq.(\ref{PSIFAC}) we have
\begin{equation}
\tilde{\hat\omega}_{ab}=\frac{\tilde\psi_{a/a}\tilde\psi_{b/b}}
{\tilde\phi_{a/a}\tilde\phi_{b/b}}Tr[H\tilde{S}].
\end{equation}

\begin{figure}[h]
\includegraphics[width=0.3\textwidth]{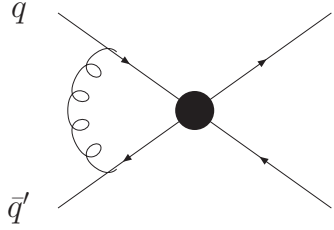}
\caption{The one-loop eikonal diagram for
chargino and neutralino associated production.
} \label{EKFD}
\end{figure}

After resumming the terms with the $N$ dependence
we obtain the exponentiated differential cross section in the
space of moments \cite{Contopanagos:1996nh,Kidonakis:1996aq}:
\begin{eqnarray}
\tilde{\hat\omega}_{ab}^{EXP}= && \exp\left[\sum_i E^{(f_i)}(N)
\right] \exp\left[\sum_i
2\int^Q_{\mu_f}\frac{d\mu^\prime}{\mu^\prime}
\gamma_i(\alphas(\mu^{\prime 2}))\right] \nonumber\\ && \times
\left.\exp\left[\sum_i 2d_{\alphas}\int^Q_{\mu_r}
\frac{d\mu^\prime}{\mu^\prime} \beta(\alphas(\mu^{\prime 2}))\right]
Tr \right\{ H^{ab}(\alphas(\mu_r^2)) \nonumber\\ && \times
\bar{P}\exp\left[\int^{Q/N}_Q
\frac{d\mu^\prime}{\mu^\prime}(\Gamma_S^{ab})^\dagger
(\alphas(\mu^{\prime 2}))\right] \tilde{S}^{ab}(1,\alphas(Q^2/N^2))
\nonumber\\ && \left.\times P \exp\left[\int^{Q/N}_Q
\frac{d\mu^\prime}{\mu^\prime}\Gamma_S^{ab}(\alphas(\mu^{\prime
2}))\right]\right\}, \label{EXPEQ}
\end{eqnarray}
where $d_{\alphas}$ is a constant and its definition is given in
Ref.\cite{Kidonakis:2006bu}. $P$ and $\bar{P}$ denote path ordering in
the same sense as the integration variable $\mu^\prime$ and in the opposite sense, respectively.
The first exponent in Eq.(\ref{EXPEQ}) resums the
collinear and soft gluon emission from initial partons in the hard
scattering and is given in the modified minimal subtraction (\msbar)
scheme by
\begin{equation}
E^{(f_i)}(N)=-\int^1_0dz\frac{z^{N-1}-1}{1-z}\left\{\int^{\mu_f^2}_{(1-z)^2Q^2}
\frac{d\mu^{\prime 2}}{\mu^{\prime 2}} A^{(f_i)}[\alphas(\mu^{\prime
2})] +\frac{1}{2}\nu^{(f_i)}[\alphas((1-z)^2Q^2)] \right\},
\end{equation}
with
\begin{equation}
A^{(f_i)}(\alphas)=C_f\left(\frac{\alphas}{\pi}
+\frac{1}{2}K\left(\frac{\alphas}{\pi}\right)^2\right),
\end{equation}
\begin{equation}
K=C_A\left(\frac{67}{18}-\frac{\pi^2}{6}\right)-\frac{5}{9}n_f,
\end{equation}
\begin{equation}
\nu^{(f_i)}=2C_f(\alphas/\pi)[1-\ln(2v_{(f_i)})],
\end{equation}
\begin{equation}
v_{(f_i)}=(\hat\beta_i \cdot \hat{n})^2/|\hat{n}|^2,
\end{equation}
\begin{equation}
\hat\beta_i=p_i\sqrt{2/\hat{s}},
\end{equation}
where $\hat\beta_i$ is the particle velocity, $\hat{n}$ is the axial
gauge vector, $N_c$ is the number of colors, and $n_f$ is the flavor
number of light quarks. $C_f=C_F=(N_c^2-1)/(2N_c)$ for initial
quarks and $C_f=C_A=N_c$ for initial gluons. In Eq.(\ref{EXPEQ})
$\gamma_i$ is the anomalous dimension of $\psi$ and is given at one
loop by $\gamma_q=3C_F\alphas/4\pi$ for quarks and
$\gamma_g=\beta_0\alphas/\pi$ for gluons. The $\beta$ function is
defined as
\begin{equation}
\beta(\alphas)=
\frac{1}{2}\mu \frac{d\ln g}{d\mu}=-\sum_{n=0}^\infty\beta_n(\frac{\alphas}{\pi})^{(n+2)},
\end{equation}
with
\begin{equation}
\beta_0=(11C_A-2n_f)/12,
\end{equation}
\begin{equation}
\beta_1=(17C_A^2-5C_An_f-3C_Fn_f)/24.
\end{equation}
$\Gamma_S$ in Eq.(\ref{EXPEQ}) is the soft anomalous dimension
matrix\cite{Kidonakis:1996aq}, which can be derived from the eikonal
diagrams as shown in Fig.\ref{EKFD}, and is given by
\begin{equation}
\Gamma_S=\frac{\alphas C_F}{2\pi}[-2\ln 2-\ln(v_{(u)}
v_{(\bar{d})})+2-2\pi i].
\end{equation}

Next Eq.(\ref{EXPEQ}) at NLL can be written in the simplified form
\begin{equation}
\tilde{\hat\omega}_{ab}^{NLL}=
\tilde{\hat\omega}_{0}C_{ab}(\alphas)\exp\left[X(N,\alphas)\right],
\label{RESEXP}
\end{equation}
with
\begin{equation}
\tilde{\hat\omega}_0=\int~d\tau~\tau^{N-1}\hat\omega_0,
\end{equation}
\begin{equation}
X(N,\alphas)=g_1(\lambda)\ln\bar{N}+g_2(\lambda),
\end{equation}
\begin{equation}
g_1(\lambda)=\frac{C_F}{\beta_0\lambda}[2\lambda+(1-2\lambda)\ln(1-2\lambda)],
\end{equation}
\begin{eqnarray}
g_2(\lambda)=&&\frac{C_F\beta_1}{\beta_0^3}[2\lambda+\ln(1-2\lambda)
+\frac{1}{2}\ln^2(1-2\lambda)]
-\frac{C_FK}{2\beta_0^2}[2\lambda+\ln(1-2\lambda)]
\nonumber \\ &&
+\frac{C_F}{\beta_0}[2\lambda+\ln(1-2\lambda)]\ln\frac{Q^2}{\mu^2_r}
-\frac{C_F}{\beta_0}2\lambda\ln\frac{Q^2}{\mu^2_f},
\end{eqnarray}
where $\hat\omega_0=d\hat\sigma_0/dQ^2$ is the Born
differential cross section, $\bar{N}=N\exp(\gamma_E)$, $\gamma_E$ is
the Euler constant, and $\lambda=\beta_0\alphas\ln\bar{N}/\pi$.

The function $C_{ab}$ can be expanded as
\begin{equation}
C_{ab}(\alphas)=1+\sum_{n=1}^\infty\left(\frac{\alphas}{\pi}\right)^nC_{ab}^{\prime(n)}.
\end{equation}
By matching the moments of the NLO cross section\cite{Arnold:1990yk}
we obtain the function $C_{ab}$ in Eq.(\ref{RESEXP}) at the NLO. The
contributions to the first term $C_{ab}^{(1)}\equiv
C_{ab}^{\prime(1)}\alphas/\pi$ in the expansion above come from the
constant terms in the moments of the differential cross section,
which are primarily the coefficients of the $\delta(1-z)$ terms in the
differential cross sections. The other terms come from the Mellin
transformations of the logarithms depending on $z$.

As shown in Eqs.(\ref{DIV1}), (\ref{DIV2}) and (\ref{DIV3}) in
Sec.~II, the divergences from QCD and SUSY QCD corrections
cancel each other. Therefore, combining the contributions
from real gluon corrections and PDF renormalization, the
QCD contributions to $C_{ab}^{(1)}$ are given by
\begin{eqnarray}
C_{ab}^{(1)}=&&\frac{2Re( \widetilde{\cal M}^{QCD}_V{\cal M}_0^*)}{|{\cal M}_0|^2}
+\frac{\alphas C_F}{2\pi}\left\{\frac{2}{\epsilon^2}+\frac{1}{\epsilon}
(2\ln\frac{4\pi \mu^2_r}{Q^2}-2\gamma_E+3)
\right. \nonumber \\ && \left.
+(\ln\frac{4\pi\mu^2_r}{Q^2}-\gamma_E)^2+3\ln\frac{4\pi\mu^2_r}{Q^2}+\frac{\pi^2}{6}-3\gamma_E+3\ln\frac{Q^2}{\mu^2_f}
\right. \nonumber \\ && \left.
+4\ln^2\bar{N}-4\ln\frac{Q^2}{\mu^2_f}\ln\bar{N}
+\frac{4\ln\bar{N}-2\ln\frac{Q^2}{\mu^2_f}}{N}\right\},
\end{eqnarray}
with
\begin{equation}
\widetilde{\cal
M}_V^{QCD}=\sum_{n=1}^{24}\sum_{a,b=L,R}\left[f^{ab}_{QCDVn}+f^{ab}_{QCDCn}
+\left.\left.\right(f^{ab}_{SUSYVn}\Big|_{UV}
+f^{ab}_{SUSYCn}\Big|_{UV} \right)\right]M_n^{ab}.
\end{equation}
Here the terms of order ${\cal O}(\ln\bar{N}/N)$ and
${\cal O}(\ln(Q^2/\mu^2)/N)$ are included.

In order to more completely include the behavior of the full towers
of logarithms\cite{Eynck:2003fn} Eq.(\ref{RESEXP}) is modified:
\begin{equation}
\tilde{\hat\omega}_{ab}^{NLL}=
\tilde{\hat\omega}_{0}\exp[C_{ab}^{(1)}(\alphas)]\exp\left[X(N,\alphas)\right].
\end{equation}

To obtain the physical cross section we perform the inverse
Mellin transformation,
\begin{equation}
\omega(\tau)=\frac{1}{2\pi
i}\int^{C+i\infty}_{C-i\infty}dN\tau^{-N}\tilde{\omega}(N).
\label{omegatau}
\end{equation}
where the "minimal prescription" is used\cite{Catani:1996yz}.

To improve the convergence of the integration in Eq.(\ref{omegatau})
we adopt the methods in Ref.\cite{Kulesza:2002rh}.
First, we rotate the contour by an angle $\phi$ with respect to the
real axis and parameterize it in the form
\begin{equation}
N=C+z\exp^{\pm i \phi},
\end{equation}
where the upper (lower) sign applies to the upper (lower) half plane $0\le z\le\infty$ ($\infty\ge z\ge 0$).
Then we rewrite the inverse transformation convolution:
\begin{equation}
\omega(\tau)=\frac{1}{2\pi i}\int_{C_N} dN \tau^{-N}\sum_{a,b}[(N-1)\tilde{f}_a(N)][(N-1)\tilde{f}_b(N)]
\frac{\tilde{\hat\omega}_{ab}(N)}{(N-1)^2},
\end{equation}
where $C_N$ represents the modified contour. The inverse Mellin
transformation of $\tilde{\hat\omega}_{ab}/(N-1)^2$,
\begin{equation}
{\cal Y}_{ab}(z)=\frac{1}{2\pi i}\int_{C_N} dN
z^{-N}\frac{\tilde{\hat\omega}_{ab}(N)}{(N-1)^2},
\end{equation}
is well behaved near the region $z\approx1$ due to the suppression by the factor
$1/(N-1)^2$. The inverse Mellin transformation of
$(N-1)\tilde{f}_i(N)$ is then
\begin{equation}
\frac{1}{2\pi i}\int_{C_N} dN x^{-N}(N-1)\tilde{f}_i(N)=-\frac{d}{dx}[xf_i(x)]={\cal F}_i(x).
\end{equation}
Finally,
\begin{equation}
\omega(\tau)=\sum_{a,b}\int^\infty_\tau\frac{dz}{z}\int^1_{\tau/z}\frac{dx}{x}{\cal
F}_a(x){\cal F}_b(\frac{\tau}{xz}){\cal Y}_{ab}(z).
\end{equation}
And after integrating over the invariant mass $Q^2$ in the differential
cross section and inserting the terms ignored in the Mellin
transformation we obtain the resummed total cross section
\begin{equation}
\sigma^{RES}=\sigma^{NLO}+\left[\sigma^{NLL}-\left.\sigma^{NLL}\right|_{\alphas=0}
-\alphas\left(\frac{\partial\sigma^{NLL}}{\partial\alphas}\right)_{\alphas=0}\right].
\end{equation}

\section{NUMERICAL RESULTS AND DISCUSSIONS}
In the numerical calculations the following SM input parameters were
chosen\cite{Yao:2006px,CDF:2007bxa}:
\begin{equation}
\begin{array}{lll}
M_t=170.9\text{GeV\cite{CDF:2007bxa}},& \alpha(M_Z)^{-1}=127.918,&
\alphas(M_Z)=0.1176, \\ M_W=80.403\text{GeV}, &
M_Z=91.1876\text{GeV}. & \\
\end{array}
\end{equation}
The masses of the light quarks were neglected. The running QCD
coupling $\alphas$ was evaluated at the two-loop
level\cite{Gorishnii:1990zu}
\nocite{Gorishnii:1991zr,Djouadi:1995gt,Spira:1997dg}
and the CTEQ6.5M PDF's
\cite{Whalley:2005nh}\nocite{Tung:2006tb}
were used
to calculate the various cross sections, either at LO or at NLO.
As for the renormalization and factorization scales, we chose
$\mu_r=\mu_f=Q_{\cha{1}\neu{2}}\equiv\sqrt{(p_{\cha{1}}+p_{\neu{2}})^2}$,
unless specified otherwise.

Using the program package SPheno\cite{Porod:2003um} the MSSM
spectrum, including the widths of the squarks,
was calculated in the mSUGRA scenario in which there are
five input parameters: the ratio of Higgs-field vacuum
expectation values (VEV's) $\tan\beta$, the common scalar mass $m_0$,
the common gaugino mass $m_{1/2}$, the trilinear coupling $A_0$, and
the sign of the Higgs mixing parameter $\mu$.
The value of $A_0$ does not significantly affect our numerical
results so we put $A_0=0$ and, based on the
analysis in the literature \cite{Barger:1998hp,Chattopadhyay:1995ae}, focused on
$\mu>0$.

\newpage
\begin{table}[ht]
\caption{The dependence of the chargino mass $M_{\cha{1}}$ and the neutralino mass $M_{\neu{2}}$ on the top quark mass $M_t$.
The masses were calculated using SPheno\cite{Porod:2003um} 
for $\tan\beta=5$, $m_0=200$GeV, and
$m_{1/2}=250$GeV.}
\label{RGEDEP}
\begin{ruledtabular}
\begin{tabular}{ccc}
$M_t$/GeV & $M_{\cha{1}}$/GeV & $M_{\neu{2}}$/GeV \\
 170.9 & 188.506  & 189.915  \\
 176.1 & 190.609  & 191.779  \\
 180.1 & 191.978  & 193.003  \\
 \end{tabular}
 \end{ruledtabular}
 \end{table}
Table \ref{RGEDEP} shows the dependence of the chargino mass
$M_{\cha{1}}$ and the neutralino mass $M_{\neu{2}}$ on top quark
mass $M_t$. We see that the chargino and
neutralino masses depend slightly on top quark mass due to the fact
that they are calculated using the SUSY renormalization group
evolution (RGE). The explicit expressions for the total cross
sections for the associated production of a chargino and a neutralino
are independent of $M_t$ as shown in Sec.~II and Sec.~III. Thus the top
quark mass $M_t$ only enters in the SUSY RGE.

\begin{table}[ht]
\caption{The NLO total cross sections for different squark widths
$\Gamma(\sq)$ using and not using on-shell subtraction, respectively.
} \label{WIDTHDEP}
\begin{ruledtabular}
\begin{tabular}{ccc}
$\Gamma(\sq)/\Gamma_0(\sq)$ & $\sigma'_{NLO}(pp\to\chap{1}\neu{2})/pb$
& $\sigma_{NLO}(pp\to\chap{1}\neu{2})/pb$\\
2   &  0.734 & 0.650\\
1   &  0.825 & 0.653\\
0.5  & 1.006 & 0.661\\
\end{tabular}
\end{ruledtabular}
\end{table}

Table \ref{WIDTHDEP} shows the NLO total cross sections for
$\chap{1}\neu{2}$ production at the LHC, using on-shell subtraction,
($\sigma_{NLO}(pp\to\chap{1}\neu{2})$), or not, 
($\sigma'_{NLO}(pp\to\chap{1}\neu{2})$),
for different squark widths $\Gamma(\sq)$, assuming $\tan\beta=5$,
$m_0=200$GeV and $m_{1/2}=250$GeV. The squark widths, which were
calculated using SPheno\cite{Porod:2003um}, are
$\Gamma_0(\sq)$.
Table \ref{WIDTHDEP}  shows that the variation in
$\sigma'_{NLO}(pp\to\chap{1}\neu{2})$ is about 26\% while the
variation in $\sigma_{NLO}(pp\to\chap{1}\neu{2})$ is only about 2\%.
Obviously, using on-shell subtraction reduces the dependence on the
squark widths.

To present the resummation effects we defined the
following quantities:
\begin{equation}
\delta K=\frac{\sigma_{RES}-\sigma_{NLO}}{\sigma_{NLO}},
\end{equation}
\begin{equation}
\delta K_d=\frac{d\sigma_{RES}-d\sigma_{NLO}}{d\sigma_{NLO}}.
\end{equation}
These represent the threshold resummation effects relative to the
NLO cross sections.

We present the numerical results for both $\chap{1}\neu{2}$ and
$\cham{1}\neu{2}$ production at the LHC, but show only those for 
$\chap{1}\neu{2}$ production at the Tevatron since these cross
sections are different at the LHC but the same
at the Tevatron.

In Fig.\ref{CUTDEP} we chose $\chap{1}\neu{2}$ production at
the LHC as an example to show that it is reasonable to use the
two-cutoff phase space slicing method in the NLO calculations, i.e.
the dependence of the NLO predictions on the arbitrary cutoffs
$\delta_s$ and $\delta_c$ is indeed very weak, as shown in
Ref.\cite{Harris:2001sx}. Here $\sigma_{other}$ includes the
contributions from the Born cross section and the virtual
corrections, which are cutoff independent. Both the soft plus hard
collinear contributions and the hard noncollinear contributions
depend strongly on the cutoffs. However, these two contributions in
($\sigma_{soft}+\sigma_{hard coll}+\sigma_{virtual}$ and
$\sigma_{hard non-coll}$) nearly cancel,
especially for the cutoff $\delta_s$ between $5\times 10^{-5}$ and
$10^{-3}$, where the final results for $\sigma_{NLO}$ are almost
independent of the cutoffs and very near $7.1$pb. Therefore, we will
take $\delta_s=10^{-4}$ and $\delta_c=\delta_s/100$ in the numerical
calculations below.

Using the same parameters we reproduced the results in
Ref.\cite{Beenakker:1999xh} as shown in Fig.\ref{PLEHN}, which
provides a check on our calculations. However, our results are not
exactly the same as the results in Ref.\cite{Beenakker:1999xh}
because the masses calculated using SPheno\cite{Porod:2003um} are
different from those in Ref.\cite{Beenakker:1999xh}.

Fig.\ref{TBDEP} shows the total cross sections as a function of
$\tan\beta$, assuming $m_{1/2}=150$GeV, for $m_0=200$GeV and
$1000$GeV. The general shapes of the
cross sections are similar. The main difference is that the absolute
values of the total cross sections are different. 
$\chap{1}\neu{2}$ production at the LHC has the largest cross
section. In general, the total cross sections at the LHC are a few
pb while those at the Tevatron are hundreds of fb. Fig.\ref{TBDEP}
also shows that the total cross sections for large
$\tan\beta$($>10$) are almost independent of $\tan\beta$ while those
for small $\tan\beta$($<10$) decrease with the increasing
$\tan\beta$ especially for $m_0=200$GeV. We note that the contributions from
the resummation effects do not change the shapes of the curves very
much.

With the same parameters as in Fig.\ref{TBDEP} the resummation
effects $\delta K$ are presented in Fig.\ref{RELTBDEP} as a function
of $\tan\beta$ for $m_0=200$GeV and $1000$GeV. Note
that $\delta K$ is almost independent of $\tan\beta$ for large
$\tan\beta$($>10$) and there are larger resummation effects for
$m_0=1000$GeV than for $m_0=200$GeV. However, $\delta K$
at the LHC decreases with the increasing $\tan\beta$ for
$m_0=200$GeV and $\tan\beta<10$. Fig.\ref{RELTBDEP} shows that
$\delta K$ at the LHC for $\cham{1}\neu{2}$ production is larger
than that for $\chap{1}\neu{2}$ production and $\delta K$ at the
Tevatron is larger than at the LHC. For $m_{0}=1000$GeV the
resummation effects can reach about $4\%$ at the LHC and about
$4.7\%$ at the Tevatron.

Fig.\ref{MHFDEP} shows the total cross sections as a function of
$m_{1/2}$ assuming $m_0=200$GeV and $\tan\beta=5$. As $m_{1/2}$
varies from $150$GeV to $250$GeV
$M_{\neu{2}}$ increases from $101$GeV to $190$GeV and
$M_{\widetilde{u}_1}$ increases from $406$GeV to $599$GeV,
respectively. And the total cross sections decrease rapidly with the
increasing of $m_{1/2}$. For example, when $m_{1/2}>240$GeV the
total cross sections are less than $1$pb and $100$fb at the LHC and
Tevatron, respectively. Note that the total cross section for
$\chap{1}\neu{2}$ production at the LHC is the largest while 
$\chap{1}\neu{2}$ production at the Tevatron is the smallest.

With the same parameters as in Fig.\ref{MHFDEP} the resummation
effects $\delta K$ are shown in Fig.\ref{RELMHFDEP} as a function of
$m_{1/2}$. At the LHC the resummation effects
increase with the decreasing of $m_{1/2}$, reaching $3.3\%$ for
$m_{1/2}=150$GeV. At the Tevatron $\delta K$ increases with the
increasing $m_{1/2}$, reaching $4.9\%$ for $m_{1/2}=250$GeV. We also
find that the smallest resummation effects $\delta K$ at the
Tevatron for $m_{1/2}=150$GeV are about $3.8\%$, which is larger
that at the LHC for all values of $m_{1/2}$.

Fig.\ref{M0DEP} shows the total cross sections as a function of
$m_0$ assuming $m_{1/2}=150$GeV and $\tan\beta=5$. As $m_0$ varies
from $100$GeV to $1000$GeV,
$M_{\cha{1}}$ increases from $96$GeV to $116$GeV and $M_{\neu{2}}$
increases from $100$GeV to $117$GeV, respectively. The
total cross sections decrease with the increasing $m_0$ for
$m_0>300$GeV. However, note that the total cross section for
$\chap{1}\neu{2}$ production at the LHC is independent of $m_0$ for
$m_0<300$GeV.

With the same parameters as in Fig.\ref{M0DEP} the resummation
effects $\delta K$ are presented in Fig.\ref{RELM0DEP} as a function
of $m_{0}$. The resummation effects increase at both the LHC and
the Tevatron as $m_0$ increases. For $m_0=1000$GeV the resummation
effects reach $3.9\%$ at the LHC and $4.7\%$ at the Tevatron. The
total cross sections increase rapidly with the increasing $m_0$ for
$m_0<500$GeV while they become independent of $m_0$ when
$m_0>900$GeV.

Figs.\ref{LHCSCALEDEP} and \ref{TEVSCALEDEP} show the total cross
section for $\chap{1}\neu{2}$ production at the LHC and the
Tevatron, respectively, as functions of the renormalization scale
$\mu_{r}$ and the factorization scale $\mu_f$, and for
$\mu_r=\mu_f$, assuming $m_{1/2}=250$GeV,
$m_0=200$GeV and $\tan\beta=5$. The $\mu_r$ dependence in the LO cross sections
at both colliders is increased by the NLO corrections
and the $\mu_r$ dependence is
slightly decreased by the resummation effects. The $\mu_f$ dependence in the
LO cross sections at the LHC (Tevatron) is decreased by
the NLO corrections and is further increased (decreased) by the
resummation effects. However,  setting
$\mu_{f}=\mu_{r}=\mu_{scale}$, the resummation effects reduce the
scale dependence at NLO.
In fact, from
Fig.\ref{LHCSCALEDEP} it can be seen that the
renormalization/factorization scale dependence of the total cross
sections at the LHC(Tevatron) is reduced to $5\%$ ($4\%$) with the threshold
resummation from up to $7\%$ ($11\%$) at NLO.

Figs.\ref{INV200} and \ref{INV1000} present the differential cross
sections as a function of the invariant mass $Q_{\cha{1}\neu{2}}$
assuming $m_{1/2}=150$GeV and $\tan\beta=5$ for $m_0=200$GeV and
$1000$GeV, respectively. We see that the maximum in the differential
cross section occurs at about $Q_{\cha{1}\neu{2}}=230$GeV and
$280$GeV for $m_0=200$GeV and $1000$GeV, respectively, and the
differential cross sections decrease rapidly with the increasing
$Q_{\cha{1}\neu{2}}$. The NLO corrections change the shapes of the
differential cross sections, especially for
$Q_{\cha{1}\neu{2}}<300$GeV. The threshold resummation effects
enhance the NLO differential cross sections more at moderate values
of $Q_{\cha{1}\neu{2}}$ and much less so at low or high values of
$Q_{\cha{1}\neu{2}}$.

Fig.\ref{RELINV} shows $\delta K_d$ as a function of the invariant
mass. In general, after slightly decreasing, $\delta K_d$ increases
more rapidly for $m_0=200$GeV than that for $m_0=1000$GeV. 
The resummation effects are significant for large invariant mass.
For example, for $m_0=200$GeV, $\delta K_d$ is larger than $18\%$
and $35\%$ at the LHC and Tevatron for $Q_{\cha{1}\neu{2}}>5000$GeV
and $Q_{\cha{1}\neu{2}}>1200$GeV, respectively. However, in general,
$\delta K$ is only a few percent as shown in Figs.\ref{RELTBDEP},
\ref{RELMHFDEP} and \ref{RELM0DEP}.


\section{CONCLUSION}
In conclusion, we have calculated the QCD effects in the associated
production of $\cha{1}\neu{2}$ in the MSSM within the mSUGRA scenario
at both the Tevatron and the LHC, including the NLO SUSY QCD
corrections and the NLL threshold resummation effects. Our results
show that, compared to the NLO predictions, the threshold
resummation effects can increase the total cross sections by $3.6\%$
and $3.9\%$ for the associated production of $\chap{1}\neu{2}$ and
$\cham{1}\neu{2}$ at the LHC, respectively, and $4.7\%$ for the
associated production of $\cha{1}\neu{2}$ at the Tevatron. In the
invariant mass distributions the resummation effects are significant
for large invariant mass. The renormalization/factorization scale
dependence of the total cross sections at the LHC (Tevatron) is reduced to
$5\%$ ($4\%$) with threshold resummation from up to $7\%$ ($11\%$) at NLO.

\begin{acknowledgments}
This work was supported in part by the National Natural Science Foundation of China,
under grants No. 10421503, No. 10575001 and No. 10635030,
and the Key Grant Project of Chinese Ministry of Education under grant No. 305001
and the Specialized Research Fund for the Doctoral Program of Higher Education, and
the U.S. Department of Energy, Division of High Energy Physics, under Grant No.
DE-FG02-91-ER4086.
\end{acknowledgments}

\appendix
\section{}\label{vertexes}
In this appendix we summarize\cite{Gunion:1984yn}\nocite{Kraml:1999qd,Bartl:2003pd} the SUSY vertexes involved
in our calculations.
\begin{enumerate}
\item The chargino-neutralino-W vertex is
\begin{equation}
{\mathcal L}_{\widetilde{\chi}^+\widetilde{\chi}^0W}=
-\overline{\chap{i}}\gamma^\mu(\hat A_L^{ij}P_L+\hat A_R^{ij}P_R)\neu{j}W^+_\mu
-\overline{\neu{j}}\gamma^\mu(\hat A_L^{ij}P_L+\hat A_R^{ij}P_R)\chap{i}W^-_\mu,
\end{equation}
with
\begin{equation}
\hat A_L^{ij}=g_W(\frac{V_{i2}Z_{j4}}{\sqrt{2}}-V_{i1}Z_{j2}),
\end{equation}
\begin{equation}
\hat A_R^{ij}=-g_W(\frac{U_{i2}Z_{j3}}{\sqrt{2}}+U_{i1}Z_{j2}),
\end{equation}
where $g_W=e/\sin\theta_W$, $P_L=(1-\gamma_5)/2$ and $P_R=(1+\gamma_5)/2$.
$\theta_W$ is the weak mixing angle.
$Z$ is the neutralino mixing matrix while $V$ and $U$ are the chargino mixing matrixes.
The chargino index is $i(=1,2)$ and the neutralino index is $j(=1,2,3,4)$.
Also we define $A_L=\hat A_L^{12}$ and $A_R=\hat A_R^{12}$.
\begin{figure}[ht!]
\includegraphics[width=\textwidth]{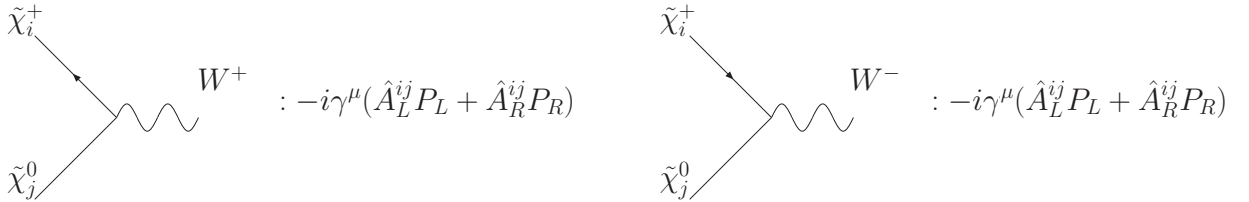}
\caption{The Feynman rules for the chargino-neutralino-W vertex.}
\end{figure}

\item The chargino-quark-squark vertex is
\begin{equation}
{\mathcal L}_{\widetilde{\chi}^+q\widetilde{q}}=
-\hat C_U^{si}\overline{u}P_R\chap{i}\widetilde{d}_s
-\hat C_U^{si}\overline{\chap{i}}P_Lu\widetilde{d}_s^*
-\hat C_V^{si}\overline{d}P_R\widetilde{\chi}^{+c}_{i}\widetilde{u}_s
-\hat C_V^{si}\overline{\chap{i}}^cP_Ld\widetilde{u}_s^*,
\end{equation}
with
\begin{equation}
\hat C_U^{si}=g_WV_{ud}U_{i1}R_{s1}^{\widetilde{d}},
\end{equation}
\begin{equation}
\hat C_V^{si}=g_WV_{ud}V_{i1}R_{s1}^{\widetilde{u}},
\end{equation}
where $V_{ud}$ is the ($u$,$d$) component of the CKM matrix and $R^{\sq}$ is
the squark mixing matrix.
$s(=1,2)$ is the index of the relevant squarks in the mass eigenstates
and $i(=1,2)$ is the chargino index.
We define $\displaystyle{C_U^s=\hat C_U^{s1}}$
and $\displaystyle{C_V^s=\hat C_V^{s1}}$.
\begin{figure}[ht!]
\includegraphics[width=\textwidth]{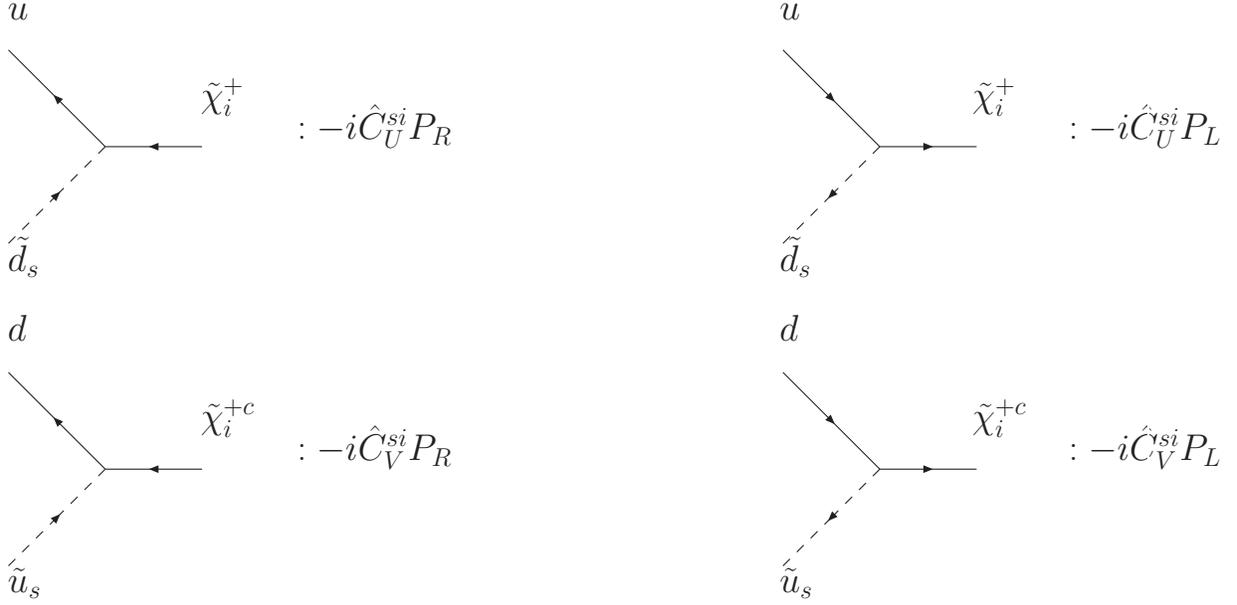}
\caption{The Feynman rules for the chargino-quark-squark vertex.}
\end{figure}

\item The neutralino-quark-squark vertex is
\begin{equation}
{\mathcal L}_{\widetilde{\chi}^0 q\widetilde{q}}=
-\overline{\neu{j}}
(a_{sj}^{\sq}P_L+b_{sj}^{\sq}P_R)q\sq_s^*
-\overline{q}
(a_{sj}^{\sq}P_R+b_{sj}^{\sq}P_L)\neu{j}\sq_s,
\end{equation}
where
\begin{equation}
a_{sj}^{\sq}=\sqrt{2}g_WR_{s1}^{\sq}[(e_q-I^q_{3L})\tan\theta_WZ_{j1}+I^q_{3L}Z_{j2}],
\end{equation}
and
\begin{equation}
b_{sj}^{\sq}=-\sqrt{2}g_We_q\tan\theta_WR_{s2}^{\sq}Z_{j1}.
\end{equation}
$e_q$ and $I^q_{3L}$ is the electric charge and the third component of the weak isospin of the
left-handed quark $q$.
\begin{figure}[ht!]
\includegraphics[width=\textwidth]{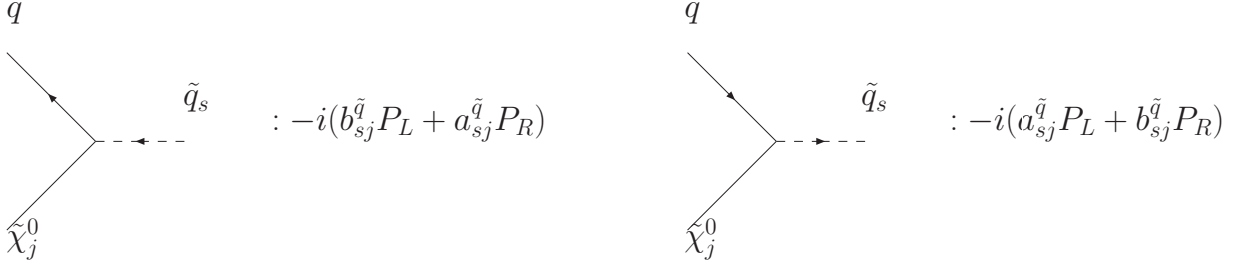}
\caption{The Feynman rules for the neutralino-quark-squark vertex.}
\end{figure}

\item The squark-Higgs vertex is
\begin{equation}
{\mathcal L}_{\sq\sq'H(G)}=
-D_H^{mn}\tilde{u}_{m}\tilde{d}_{n}^*H^-
-D_H^{mn}\tilde{u}_{m}^*\tilde{d}_{n}H^+
-D_G^{mn}\tilde{u}_{m}\tilde{d}_{n}^*G^-
-D_G^{mn}\tilde{u}_{m}^*\tilde{d}_{n}G^+
,
\end{equation}
where
\begin{equation}
D_H^{mn}=g_WV_{ud}R_{m1}^{\tilde{u}}R_{n1}^{\tilde{d}}\sin(2\beta)M_W/\sqrt{2},
\end{equation}
and
\begin{equation}
D_G^{mn}=-g_WV_{ud}R_{m1}^{\tilde{u}}R_{n1}^{\tilde{d}}\cos(2\beta)M_W/\sqrt{2}.
\end{equation}
$m$ and $n$ are the indices of the relevant squarks in the mass eigenstates.
\begin{figure}[ht!]
\includegraphics[width=0.8\textwidth]{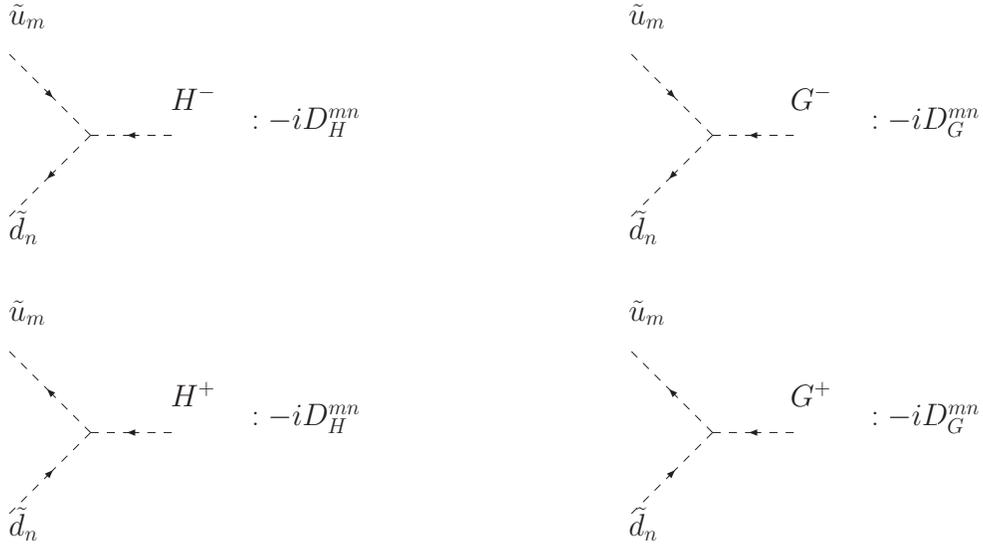}
\caption{The Feynman rules for the squark-Higgs vertex.}
\end{figure}

\item The squark-W vertex is
\begin{equation}
{\mathcal L}_{\sq\sq'W}=
iD_W^{mn}[\tilde{d}_{n}^*(\partial^\mu\tilde{u}_{m})
-(\partial^\mu\tilde{d}_n^*)\tilde{u}_m]W^-_\mu
+iD_W^{mn}[\tilde{u}_{m}^*(\partial^\mu\tilde{d}_{n})
-(\partial^\mu\tilde{u}_m^*)\tilde{d}_n]W^+_\mu,
\end{equation}
where
\begin{equation}
D_W^{mn}=g_WV_{ud}R_{m1}^{\tilde{u}}R_{n1}^{\tilde{d}}/\sqrt{2}.
\end{equation}
\begin{figure}[ht!]
\includegraphics[width=\textwidth]{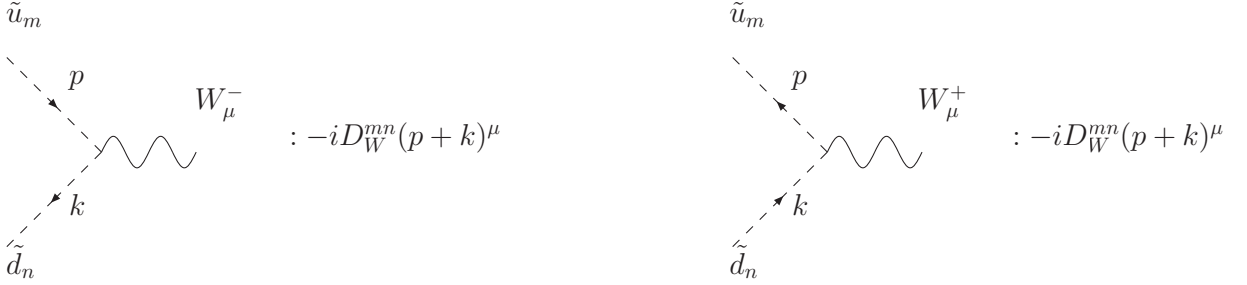}
\caption{The Feynman rules for the squark-W vertex. $p$ and $k$ are the four-momenta of $\tilde{u}_m$ and $\tilde{d}_n$
in direction of the charge flow, respectively.}
\end{figure}

\item The chargino-neutralino-Higgs vertex is
\begin{equation}
\begin{split}
{\mathcal L}_{\widetilde{\chi}^+\widetilde{\chi}^0H(G)}=&
-\overline{\chap{i}}[(\hat{C}_R^H)_{ij}P_L+(\hat{C}_L^H)_{ij}P_R]\neu{j}H^+
-\overline{\chap{i}}[(\hat{C}_R^G)_{ij}P_L+(\hat{C}_L^G)_{ij}P_R]\neu{j}G^+
\\&
-\overline{\neu{j}}[(\hat{C}_L^H)_{ij}P_L+(\hat{C}_R^H)_{ij}P_R]\chap{i}H^-
-\overline{\neu{j}}[(\hat{C}_L^G)_{ij}P_L+(\hat{C}_R^G)_{ij}P_R]\chap{i}G^-,
\end{split}
\end{equation}
where
\begin{equation}
(\hat{C}_L^H)_{ij}=g_W\cos\beta(V_{i1}Z_{j4}+\frac{V_{i2}}{\sqrt{2}}(\tan\theta_WZ_{j1}+Z_{j2})),
\end{equation}
\begin{equation}
(\hat{C}_R^H)_{ij}=g_W\sin\beta(U_{i1}Z_{j3}-\frac{U_{i2}}{\sqrt{2}}(\tan\theta_WZ_{j1}+Z_{j2})),
\end{equation}
\begin{equation}
(\hat{C}_L^G)_{ij}=g_W\sin\beta(V_{i1}Z_{j4}+\frac{V_{i2}}{\sqrt{2}}(\tan\theta_WZ_{j1}+Z_{j2})),
\end{equation}
and
\begin{equation}
(\hat{C}_R^G)_{ij}=-g_W\cos\beta(U_{i1}Z_{j3}-\frac{U_{i2}}{\sqrt{2}}(\tan\theta_WZ_{j1}+Z_{j2})).
\end{equation}
We define $C_L^H=(\hat{C}_L^H)_{12}$,
$C_R^H=(\hat{C}_R^H)_{12}$,
$C_L^G=(\hat{C}_L^G)_{12}$,
and
$C_R^G=(\hat{C}_R^G)_{12}$.
\begin{figure}[ht!]
\includegraphics[width=\textwidth]{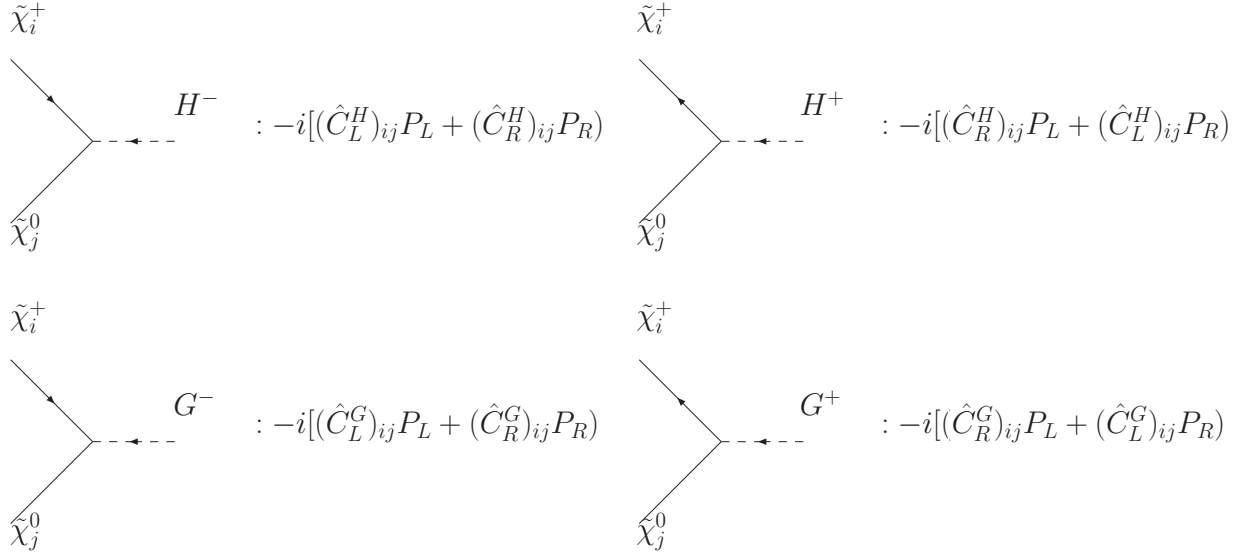}
\caption{The Feynman rules for the chargino-neutralino-Higgs vertex.}
\end{figure}

\item The SUSY QCD sector of the four-squark vertex is
\begin{equation}
{\mathcal L}_{\sq\sq\sq\sq}=
-\frac{1}{2}g_S^2 T^a_{rs}T^a_{tu}{\mathcal S}^\alpha_{ij}{\mathcal S}^\beta_{kl}
\sq^{\alpha*}_{jr}\sq^\alpha_{is}\sq^{\beta*}_{lt}\sq^\beta_{ku}
\end{equation}
where
\begin{equation}
{\mathcal S}^\alpha_{ij}={R}^\alpha_{i1}{R}^\alpha_{j1}-{R}^\alpha_{i2}{R}^\alpha_{j2}.
\end{equation}
$\alpha$ and $\beta$ represent the flavors of the relevant squarks. Here $i$, $j$, $k$ and $l$ are
the relevant squark indices. $r$, $s$, $t$ and $u$ are the color indices of the relevant squarks.
\begin{figure}[ht!]
\includegraphics[width=0.7\textwidth]{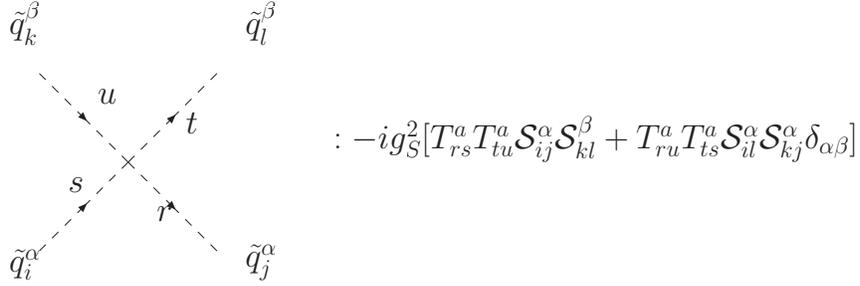}
\caption{The Feynman rules for the SUSY QCD interaction of the four-squark vertex.}
\end{figure}

\item The squark-gluon vertex is
\begin{equation}
{\mathcal L}_{\sq\sq g}=ig_ST^a_{rs}\delta_{ij}G^a_\mu[\sq^*_{j,r}(\partial^\mu\sq_{i,s})-(\partial^\mu\sq^*_{j,r})\sq_{i,s}].
\end{equation}
Here $i$ and $j$ are the indices of the relevant squarks in the mass eigenstates.
$r$ and $s$ are the color indices of the relevant squarks.
\begin{figure}[ht!]
\includegraphics[width=0.4\textwidth]{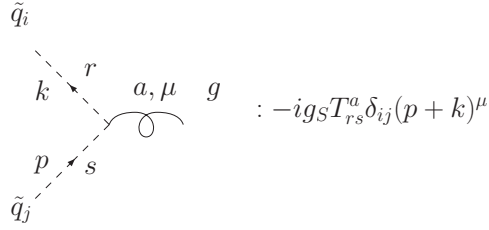}
\caption{The Feynman rules for the squark-gluon vertex. $p$ and $k$ are the relevant four-momenta
of squark $\sq$ in direction of the charge flow.}
\end{figure}

\item The quark-squark-gluino vertex is
\begin{equation}
{\mathcal L}_{q\sq\tilde{g}}=
-\sqrt{2}g_ST^a_{rs}[\overline{q}_r(R^{\sq}_{i1}P_R-R^{\sq}_{i2}P_L)\tilde{g}^a\sq_{i,s}
+\overline{\tilde{g}}^a(R^{\sq}_{i1}P_L-R^{\sq}_{i2}P_R)q_r\sq^*_{i,s}]
.\end{equation}
Here $i$ is the mass eigenstate index and $s$ is the color index of the squark.
\begin{figure}[ht!]
\includegraphics[width=\textwidth]{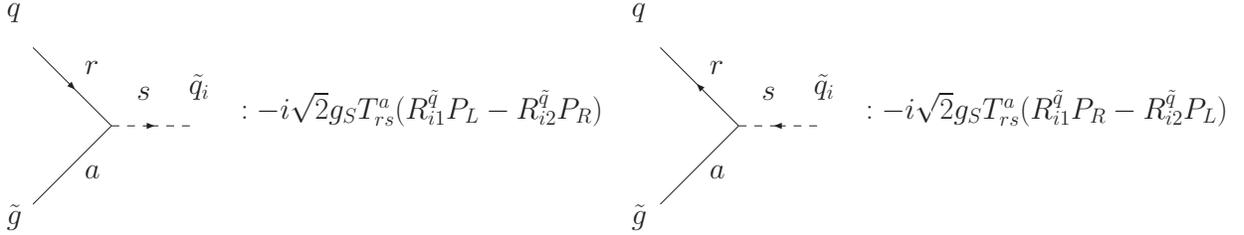}
\caption{The Feynman rules for the quark-squark-gluino vertex.}
\end{figure}

\end{enumerate}

\section{}\label{BCD}
In this appendix, for simplicity, we
introduce the following abbreviations for the Passarino-Veltman
two-point integrals $B_i$, three-point integrals $C_{i(j)}$, and
four-point integrals $D_{i(j)}$, which are defined as in
Ref.\cite{Denner:1991kt}
except that we use internal masses squared as arguments:

$\displaystyle{
B_i^a=B_i(M_{\chap{1}}^2,0,M_{\tilde{d}_s}^2),
}$

$\displaystyle{
B_i^b=B_i(M_{\neu{2}}^2,0,M_{\tilde{d}_s}^2),
}$

$\displaystyle{
B_i^c=B_i(\hat{t},0,M_{\tilde{d}_s}^2),
}$

$\displaystyle{
B_i^d=B_i(M_{\chap{1}}^2,0,M_{\tilde{u}_s}^2),
}$

$\displaystyle{
B_i^e=B_i(\hat{u},0,M_{\tilde{u}_s}^2),
}$

$\displaystyle{
B_i^f=B_i(M_{\neu{2}}^2,0,M_{\tilde{u}_s}^2),
}$

$\displaystyle{
B_i^g=B_i(0,M_{\tilde{g}}^2,M_{\tilde{u}_s}^2),
}$

$\displaystyle{
B_i^h=B_i(M_{\chap{1}}^2,0,M_{\tilde{d}_m}^2),
}$

$\displaystyle{
B_i^i=B_i(\hat{t},0,M_{\tilde{g}}^2),
}$

$\displaystyle{
B_i^j=B_i(0,M_{\tilde{g}}^2,M_{\tilde{d}_s}^2),
}$

$\displaystyle{
B_i^k=B_i(\hat{u},M_{\tilde{g}}^2,0),
}$

$\displaystyle{
B_i^l=B_i(\hat{s},0,0),
}$

$\displaystyle{
B_i^m=B_i(M_{\neu{2}}^2,0,M_{\tilde{u}_m}^2),
}$

$\displaystyle{
B_i^n=B_i(\hat{u},0,M_{\tilde{g}}^2),
}$

$\displaystyle{
B_i^p=B_i(0,M_{\tilde{u}_k}^2,M_{\tilde{g}}^2),
}$

$\displaystyle{
B_i^q=B_i(0,M_{\tilde{d}_k}^2,M_{\tilde{g}}^2),
}$

$\displaystyle{
B_i^r=B_i(M_{\tilde{d}_s}^2,0,M_{\tilde{g}}^2),
}$

$\displaystyle{
B_i^s=B_i(M_{\tilde{u}_s}^2,0,M_{\tilde{g}}^2),
}$

$\displaystyle{
B_i^t=B_i(M_{\tilde{d}_s}^2,0,M_{\tilde{d}_s}^2),
}$

$\displaystyle{
B_i^u=B_i(M_{\tilde{u}_s}^2,0,M_{\tilde{u}_s}^2),
}$

$\displaystyle{
C_{i(j)}^a=C_{i(j)}(0,\hat{s},0,0,0,0),
}$

$\displaystyle{
C_{i(j)}^b=C_{i(j)}(0,M_{\chap{1}}^2,\hat{t},0,0,M_{\tilde{d}_s}^2),
}$

$\displaystyle{
C_{i(j)}^c=C_{i(j)}(0,M_{\neu{2}}^2,\hat{t},0,0,M_{\tilde{d}_s}^2),
}$

$\displaystyle{
C_{i(j)}^d=C_{i(j)}(\hat{s},M_{\neu{2}}^2,M_{\chap{1}}^2,0,0,M_{\tilde{u}_s}^2),
}$

$\displaystyle{
C_{i(j)}^e=C_{i(j)}(\hat{s},M_{\chap{1}}^2,M_{\neu{2}}^2,0,0,M_{\tilde{d}_s}^2),
}$

$\displaystyle{
C_{i(j)}^f=C_{i(j)}(0,M_{\chap{1}}^2,\hat{u},0,0,M_{\tilde{u}_s}^2),
}$

$\displaystyle{
C_{i(j)}^g=C_{i(j)}(0,M_{\neu{2}}^2,\hat{u},0,0,M_{\tilde{u}_s}^2),
}$

$\displaystyle{
C_{i(j)}^h=C_{i(j)}(\hat{t},0,M_{\chap{1}}^2,0,M_{\tilde{g}}^2,M_{\tilde{u}_m}^2),
}$

$\displaystyle{
C_{i(j)}^i=C_{i(j)}(0,\hat{s},0,M_{\tilde{g}}^2,M_{\tilde{u}_s}^2,M_{\tilde{d}_m}^2),
}$

$\displaystyle{
C_{i(j)}^j=C_{i(j)}(\hat{u},M_{\chap{1}}^2,0,M_{\tilde{g}}^2,0,M_{\tilde{d}_m}^2),
}$

$\displaystyle{
C_{i(j)}^k=C_{i(j)}(\hat{u},M_{\chap{1}}^2,0,M_{\tilde{g}}^2,0,M_{\tilde{d}_m}^2),
}$

$\displaystyle{
C_{i(j)}^l=C_{i(j)}(\hat{u},M_{\neu{2}}^2,0,M_{\tilde{g}}^2,0,M_{\tilde{u}_m}^2),
}$

$\displaystyle{
C_{i(j)}^m=C_{i(j)}(\hat{t},M_{\neu{2}}^2,0,M_{\tilde{g}}^2,0,M_{\tilde{d}_m}^2),
}$

$\displaystyle{
D_{i(j)}^a=D_{i(j)}(0,\hat{s},M_{\chap{1}}^2,\hat{t},0,M_{\neu{2}}^2,0,0,0,M_{\tilde{d}_s}^2),
}$

$\displaystyle{
D_{i(j)}^b=D_{i(j)}(0,\hat{s},M_{\neu{2}}^2,\hat{u},0,M_{\chap{1}}^2,0,0,0,M_{\tilde{u}_s}^2),
}$

$\displaystyle{
D_{i(j)}^c=D_{i(j)}(\hat{u},M_{\neu{2}}^2,\hat{s},0,0,M_{\chap{1}}^2,M_{\tilde{g}}^2,
0,M_{\tilde{u}_s}^2,M_{\tilde{d}_m}^2),
}$

$\displaystyle{
D_{i(j)}^d=D_{i(j)}(\hat{t},0,\hat{s},M_{\neu{2}}^2,M_{\chap{1}}^2,0,0,M_{\tilde{g}}^2,
M_{\tilde{u}_s}^2,M_{\tilde{d}_m}^2),
}$

Many of the above functions contain soft and/or collinear
singularities, but all the Passarino-Veltman integrals can be
reduced\cite{Denner:2005nn} to the scalar functions $B_0$, $C_0$ and
$D_0$. And the explicit expressions for these singular scalar
functions have been calculated previously in a different 
context\cite{Jin:2002nu}. The remaining IR finite functions can be
calculated by LoopTools\cite{Hahn:1998yk}.

\section{}\label{formfactor}
In this appendix we collect the explicit expressions for the nonzero
form factors in Eqs.(\ref{VEQ})-(\ref{RQEQ}).
The standard matrix elements in Eqs.(\ref{VEQ})-(\ref{CEQ}) for the subprocess
\begin{equation}
u(p_1)+\bar{d}(p_2)\to\chap{1}(p_3)+\neu{2}(p_4),
\end{equation}
are defined as follows:

$M_1^{ab}=\bar{v}_2P_a\gamma^\mu u_1 \bar{u}_4P_b\gamma_\mu v_3,$

$M_2^{ab}=\bar{v}_1P_av_3\bar{v}_2P_bv_4,$

$M_3^{ab}=\bar{v}_2P_av_3\bar{u}_4P_bu_1,$

$M_4^{ab}=\bar{v}_2P_av_4\bar{u}_3P_bu_1,$

$M_5^{ab}=\bar{v}_2P_av_3\bar{u}_4P_b\ps_2u_1,$

$M_6^{ab}=\bar{v}_2P_av_3\bar{u}_4P_b\ps_2\ps_3u_1,$

$M_{7}^{ab}=\bar{v}_2P_av_3\bar{u}_4P_b\ps_3u_1,$

$M_{8}^{ab}=\bar{v}_2P_av_4\bar{u}_3P_b\ps_2u_1,$

$M_{9}^{ab}=\bar{u}_4P_au_1\bar{v}_2P_b\ps_1v_3,$

$M_{10}^{ab}=\bar{u}_3P_au_1\bar{v}_2P_b\ps_1v_4,$

$M_{11}^{ab}=\bar{v}_2P_a\ps_1\gamma^\mu v_3\bar{u}_4P_b\ps_2\gamma_\mu u_1,$

$M_{12}^{ab}=\bar{u}_4P_a\gamma^\mu u_1\bar{v}_2P_b\ps_1\gamma_\mu v_3,$

$M_{13}^{ab}=\bar{u}_3P_a\gamma^\mu u_1\bar{v}_2 P_b \ps_1 \gamma_\mu v_4,$

$M_{14}^{ab}=\bar{v}_2 P_a \ps_1 \gamma^\mu v_4 \bar{u}_3 P_b \ps_2 \gamma_\mu u_1,$

$M_{15}^{ab}=\bar{v}_2 P_a \gamma^\mu v_3 \bar{u}_4 P_b \ps_2 \gamma_\mu u_1,$

$M_{16}^{ab}=\bar{v}_2 P_a \gamma^\mu v_3 \bar{u}_4 P_b \gamma_\mu u_1,$

$M_{17}^{ab}=\bar{v}_2 P_a \gamma^\mu v_4 \bar{u}_3 P_b \gamma_\mu u_1,$

$M_{18}^{ab}=\bar{v}_2 P_a \gamma^\mu v_4 \bar{u}_3 P_b \ps_2\gamma_\mu u_1,$

$M_{19}^{ab}=\bar{v}_2 P_a \gamma^\mu\gamma^\nu v_3 \bar{u}_4 P_b \gamma_\mu\gamma_\nu u_1,$

$M_{20}^{ab}=\bar{v}_2 P_a \gamma^\mu\gamma^\nu v_4 \bar{u}_3 P_b \gamma_\mu\gamma_\nu u_1,$

$M_{21}^{ab}=\bar{v}_2P_a u_1 \bar{u}_4 P_b v_3,$

$M_{22}^{ab}=\bar{v}_2 P_a u_1 \bar{u}_4 P_b \ps_1 v_3,$

$M_{23}^{ab}=\bar{u}_4 P_a v_3 \bar{v}_2 P_b \ps_3 u_1,$

$M_{24}^{ab}=\bar{v}_2 P_a \ps_3 u_1 \bar{u}_4 P_b \ps_1 v_3,$

where $a$ and $b$ are the left-hand index $L$ or right-hand index $R$, while $u_i=u(p_i)$
and $v_i=v(p_i)$ are the spinors of the particle with momentum $p_i$.

The nonzero form factors in Eqs.(\ref{VEQ})-(\ref{CEQ}) are the following:

$\displaystyle{
f_{QCDV1}^{RR}=
\frac{A_R
D_L \alpha _s}{3 \pi
\left(\hat{s}-M_W^2\right)}
\left(-2 B^l_{0}+2 \hat{s} C^a_{0}+4
C^a_{00}+2 \hat{s} C^a_{1}+2 \hat{s}
C^a_{2}+1\right)
}$

$\displaystyle{
f_{QCDV1}^{RL}=
\frac{A_L
D_L \alpha _s}{3 \pi
\left(\hat{s}-M_W^2\right)}
\left(-2 B^l_{0}+2 \hat{s} C^a_{0}+4
C^a_{00}+2 \hat{s} C^a_{1}+2 \hat{s}
C^a_{2}+1\right)
}$

$\displaystyle{
f_{QCDV2}^{LR}=
\sum_{s=1}^{2}
\frac{a^{\tilde{d}}_{s2} C_U^s \alpha _s
}{3 \pi  (\hat{t}-M_{\tilde{d}_s}^2)^2}
\left\{(\hat{t}-M_{\tilde{d}_s}^2) B^b_{0}+4 \hat{t}
B^c_{0}+4 \hat{t}
B^c_{1}+(\hat{t}-M_{\tilde{d}_s}^2) \left[(\hat{s}+4
\hat{t}+\hat{u}
\right.\right.
}$

$\displaystyle{
\hspace{36pt}\left.\left.
-M_{\chap{1}}^2-2 M_{\neu{2}}^2)
C^c_{2}-2 (\hat{s}+\hat{u}-M_{\chap{1}}^2)
(C^c_{0}+C^c_{1})\right]+A_0(M_{\tilde{d}_s}^2
)\right\}
}$

$\displaystyle{
f_{QCDV2}^{LL}=
\sum_{s=1}^{2}
\frac{b^{\tilde{d}}_{s2} C_U^s \alpha _s
}{3 \pi  (\hat{t}-M_{\tilde{d}_s}^2)^2}
\left\{(\hat{t}-M_{\tilde{d}_s}^2) B^b_{0}+4 \hat{t}
B^c_{0}+4 \hat{t}
B^c_{1}+(\hat{t}-M_{\tilde{d}_s}^2) \left[(\hat{s}+4
\hat{t}+\hat{u}
\right.\right.
}$

$\displaystyle{
\hspace{36pt}\left.\left.
-M_{\chap{1}}^2-2 M_{\neu{2}}^2)
C^c_{2}-2 (\hat{s}+\hat{u}-M_{\chap{1}}^2)
(C^c_{0}+C^c_{1})\right]+A_0(M_{\tilde{d}_s}^2
)\right\}
}$

$\displaystyle{
f_{QCDV3}^{RR}=
\sum_{s=1}^{2}
\frac{b^{\tilde{u}}_{s2} C_V^s \alpha _s
}{3 \pi (\hat{u}-M_{\tilde{u}_s}^2)^2}
\left((\hat{u}-M_{\tilde{u}_s}^2) B^d_{0}+4 \hat{u}
B^e_{0}+4 \hat{u}
B^e_{1}+(\hat{u}-M_{\tilde{u}_s}^2) \left(B^f_{0}
\right.\right.
}$

$\displaystyle{
\hspace{36pt}\left.\left.
+2
(\hat{u}-M_{\tilde{u}_s}^2) C^d_{0}+2
(\hat{u}-M_{\chap{1}}^2) C^f_{0}-(\hat{s}+\hat{t})
C^g_{0}+\hat{u} C^g_{0}+M_{\chap{1}}^2
C^g_{0}-M_{\neu{2}}^2 C^g_{0}
\right.\right.
}$

$\displaystyle{
\hspace{36pt}\left.\left.
+2
(\hat{u}-M_{\chap{1}}^2) C^f_{1}-2 \hat{s}
C^g_{1}-2 \hat{t} C^g_{1}+2 M_{\chap{1}}^2
C^g_{1}+(3 \hat{u}-M_{\chap{1}}^2)
C^f_{2}+\hat{s} C^g_{2}+\hat{t} C^g_{2}
\right.\right.
}$

$\displaystyle{
\hspace{36pt}\left.\left.
+4
\hat{u} C^g_{2}-M_{\chap{1}}^2 C^g_{2}-2
M_{\neu{2}}^2 C^g_{2}-2 \hat{s} \hat{u}
D^b_{0}+2 \hat{s} M_{\tilde{u}_s}^2 D^b_{0}-2
\hat{s} \hat{u} D^b_{1}+2 \hat{s} M_{\tilde{u}_s}^2
D^b_{1}
\right.\right.
}$

$\displaystyle{
\hspace{36pt}\left.\left.
+2 \hat{t} \hat{u} D^b_{12}-2 \hat{u}
M_{\neu{2}}^2 D^b_{12}-2 \hat{t} M_{\tilde{u}_s}^2
D^b_{12}+2 M_{\neu{2}}^2 M_{\tilde{u}_s}^2 D^b_{12}+2
\hat{t} \hat{u} D^b_{13}-2 \hat{u} M_{\neu{2}}^2
D^b_{13}
\right.\right.
}$

$\displaystyle{
\hspace{36pt}\left.\left.
-2 \hat{t} M_{\tilde{u}_s}^2 D^b_{13}+2
M_{\neu{2}}^2 M_{\tilde{u}_s}^2 D^b_{13}+2 \hat{u}^2
D^b_{2}+2 \hat{t} \hat{u} D^b_{2}-2 \hat{u}
M_{\chap{1}}^2 D^b_{2}-2 \hat{u} M_{\neu{2}}^2
D^b_{2}
\right.\right.
}$

$\displaystyle{
\hspace{36pt}\left.\left.
-2 \hat{t} M_{\tilde{u}_s}^2 D^b_{2}-2
\hat{u} M_{\tilde{u}_s}^2 D^b_{2}+2 M_{\chap{1}}^2
M_{\tilde{u}_s}^2 D^b_{2}+2 M_{\neu{2}}^2 M_{\tilde{u}_s}^2
D^b_{2}+2 \hat{t} \hat{u} D^b_{23}-2 \hat{u}
M_{\chap{1}}^2 D^b_{23}
\right.\right.
}$

$\displaystyle{
\hspace{36pt}\left.\left.
-2 \hat{u} M_{\neu{2}}^2
D^b_{23}-2 \hat{t} M_{\tilde{u}_s}^2 D^b_{23}+2
M_{\chap{1}}^2 M_{\tilde{u}_s}^2 D^b_{23}+2 M_{\neu{2}}^2
M_{\tilde{u}_s}^2 D^b_{23}+2 \hat{u}^2 D^b_{3}-2
\hat{s} \hat{u} D^b_{3}
\right.\right.
}$

$\displaystyle{
\hspace{36pt}\left.\left.
+2 \hat{t} \hat{u}
D^b_{3}-2 \hat{u} M_{\chap{1}}^2 D^b_{3}-2
\hat{u} M_{\neu{2}}^2 D^b_{3}+2 \hat{s}
M_{\tilde{u}_s}^2 D^b_{3}-2 \hat{t} M_{\tilde{u}_s}^2
D^b_{3}-2 \hat{u} M_{\tilde{u}_s}^2 D^b_{3}
\right.\right.
}$

$\displaystyle{
\hspace{36pt}\left.\left.
+2
M_{\chap{1}}^2 M_{\tilde{u}_s}^2 D^b_{3}+2 M_{\neu{2}}^2
M_{\tilde{u}_s}^2 D^b_{3}-2
(-\hat{t}+M_{\chap{1}}^2+M_{\neu{2}}^2)
(\hat{u}-M_{\tilde{u}_s}^2)
D^b_{33}\right)+A_0(M_{\tilde{u}_s}^2)\right)
}$

$\displaystyle{
f_{QCDV3}^{RL}=
\sum_{s=1}^{2}
\frac{a^{\tilde{u}}_{s2} C_V^s \alpha _s
}{3 \pi (\hat{u}-M_{\tilde{u}_s}^2)^2}
\left((\hat{u}-M_{\tilde{u}_s}^2) B^d_{0}+4 \hat{u}
B^e_{0}+4 \hat{u}
B^e_{1}+(\hat{u}-M_{\tilde{u}_s}^2) \left(B^f_{0}
\right.\right.
}$

$\displaystyle{
\hspace{36pt}\left.\left.
+2
(\hat{u}-M_{\tilde{u}_s}^2) C^d_{0}+2
(\hat{u}-M_{\chap{1}}^2) C^f_{0}-(\hat{s}+\hat{t})
C^g_{0}+\hat{u} C^g_{0}+M_{\chap{1}}^2
C^g_{0}-M_{\neu{2}}^2 C^g_{0}
\right.\right.
}$

$\displaystyle{
\hspace{36pt}\left.\left.
+2
(\hat{u}-M_{\chap{1}}^2) C^f_{1}-2 \hat{s}
C^g_{1}-2 \hat{t} C^g_{1}+2 M_{\chap{1}}^2
C^g_{1}+(3 \hat{u}-M_{\chap{1}}^2)
C^f_{2}+\hat{s} C^g_{2}+\hat{t} C^g_{2}
\right.\right.
}$

$\displaystyle{
\hspace{36pt}\left.\left.
+4
\hat{u} C^g_{2}-M_{\chap{1}}^2 C^g_{2}-2
M_{\neu{2}}^2 C^g_{2}-2 \hat{s} \hat{u}
D^b_{0}+2 \hat{s} M_{\tilde{u}_s}^2 D^b_{0}-2
\hat{s} \hat{u} D^b_{1}+2 \hat{s} M_{\tilde{u}_s}^2
D^b_{1}
\right.\right.
}$

$\displaystyle{
\hspace{36pt}\left.\left.
+2 \hat{t} \hat{u} D^b_{12}-2 \hat{u}
M_{\neu{2}}^2 D^b_{12}-2 \hat{t} M_{\tilde{u}_s}^2
D^b_{12}+2 M_{\neu{2}}^2 M_{\tilde{u}_s}^2 D^b_{12}+2
\hat{t} \hat{u} D^b_{13}-2 \hat{u} M_{\neu{2}}^2
D^b_{13}
\right.\right.
}$

$\displaystyle{
\hspace{36pt}\left.\left.
-2 \hat{t} M_{\tilde{u}_s}^2 D^b_{13}+2
M_{\neu{2}}^2 M_{\tilde{u}_s}^2 D^b_{13}+2 \hat{u}^2
D^b_{2}+2 \hat{t} \hat{u} D^b_{2}-2 \hat{u}
M_{\chap{1}}^2 D^b_{2}-2 \hat{u} M_{\neu{2}}^2
D^b_{2}
\right.\right.
}$

$\displaystyle{
\hspace{36pt}\left.\left.
-2 \hat{t} M_{\tilde{u}_s}^2 D^b_{2}-2
\hat{u} M_{\tilde{u}_s}^2 D^b_{2}+2 M_{\chap{1}}^2
M_{\tilde{u}_s}^2 D^b_{2}+2 M_{\neu{2}}^2 M_{\tilde{u}_s}^2
D^b_{2}+2 \hat{t} \hat{u} D^b_{23}-2 \hat{u}
M_{\chap{1}}^2 D^b_{23}
\right.\right.
}$

$\displaystyle{
\hspace{36pt}\left.\left.
-2 \hat{u} M_{\neu{2}}^2
D^b_{23}-2 \hat{t} M_{\tilde{u}_s}^2 D^b_{23}+2
M_{\chap{1}}^2 M_{\tilde{u}_s}^2 D^b_{23}+2 M_{\neu{2}}^2
M_{\tilde{u}_s}^2 D^b_{23}+2 \hat{u}^2 D^b_{3}-2
\hat{s} \hat{u} D^b_{3}
\right.\right.
}$

$\displaystyle{
\hspace{36pt}\left.\left.
+2 \hat{t} \hat{u}
D^b_{3}-2 \hat{u} M_{\chap{1}}^2 D^b_{3}-2
\hat{u} M_{\neu{2}}^2 D^b_{3}+2 \hat{s}
M_{\tilde{u}_s}^2 D^b_{3}-2 \hat{t} M_{\tilde{u}_s}^2
D^b_{3}-2 \hat{u} M_{\tilde{u}_s}^2 D^b_{3}
\right.\right.
}$

$\displaystyle{
\hspace{36pt}\left.\left.
+2
M_{\chap{1}}^2 M_{\tilde{u}_s}^2 D^b_{3}+2 M_{\neu{2}}^2
M_{\tilde{u}_s}^2 D^b_{3}-2
(-\hat{t}+M_{\chap{1}}^2+M_{\neu{2}}^2)
(\hat{u}-M_{\tilde{u}_s}^2)
D^b_{33}\right)+A_0(M_{\tilde{u}_s}^2)\right)
}$

$\displaystyle{
f_{QCDV4}^{RL}=
-\sum_{s=1}^{2}
\frac{a^{\tilde{d}}_{s2} C_U^s
\alpha _s }{3 \pi
(\hat{t}-M_{\tilde{d}_s}^2)}
\left(B^a_{0}+2
(\hat{t}-M_{\chap{1}}^2) C^b_{0}+2
(\hat{t}-M_{\chap{1}}^2) C^b_{1}+(3
\hat{t}-M_{\chap{1}}^2) C^b_{2}
\right.}$

\hspace{36pt}$\displaystyle{\left.
-2
(\hat{t}-M_{\tilde{d}_s}^2) \left(-C^e_{0}+\hat{s}
D^a_{0}+\hat{s} D^a_{1}+(M_{\chap{1}}^2-\hat{u})
D^a_{13}
+(2 \hat{s}+\hat{t}+\hat{u}-M_{\chap{1}}^2-M_{\neu{2}}^2)
D^a_{2}
\right.\right.}$

\hspace{36pt}$\displaystyle{\left.\left.
+(\hat{s}+\hat{t}-M_{\chap{1}}^2)
D^a_{23}+2 \hat{s} D^a_{3}+(\hat{s}+\hat{t})
D^a_{33}\right)\right)
}$

$\displaystyle{
f_{QCDV4}^{LL}=
-\sum_{s=1}^{2}
\frac{b^{\tilde{d}}_{s2} C_U^s
\alpha _s }{3 \pi
(\hat{t}-M_{\tilde{d}_s}^2)}
\left(B^a_{0}+2
(\hat{t}-M_{\chap{1}}^2) C^b_{0}+2
(\hat{t}-M_{\chap{1}}^2) C^b_{1}+(3
\hat{t}-M_{\chap{1}}^2) C^b_{2}
\right.}$

\hspace{36pt}$\displaystyle{\left.
-2
(\hat{t}-M_{\tilde{d}_s}^2) \left(-C^e_{0}+\hat{s}
D^a_{0}+\hat{s} D^a_{1}+(M_{\chap{1}}^2-\hat{u})
D^a_{13}
+(2 \hat{s}+\hat{t}+\hat{u}-M_{\chap{1}}^2-M_{\neu{2}}^2)
D^a_{2}
\right.\right.}$

\hspace{36pt}$\displaystyle{\left.\left.
+(\hat{s}+\hat{t}-M_{\chap{1}}^2)
D^a_{23}+2 \hat{s} D^a_{3}+(\hat{s}+\hat{t})
D^a_{33}\right)\right)
}$

$\displaystyle{
f_{QCDV5}^{RR}=
\sum_{s=1}^{2}
\frac{2 a^{\tilde{u}}_{s2} M_{\neu{2}} C_V^s
\alpha _s
}{3 \pi }
\left(D^b_{12}+D^b_{13}+D^b_{23}+D^b_{3}
+D^b_{33}\right)
}$

$\displaystyle{
f_{QCDV5}^{RL}=
\sum_{s=1}^{2}
\frac{2 b^{\tilde{u}}_{s2} M_{\neu{2}} C_V^s
\alpha _s }{3 \pi }
\left(D^b_{12}+D^b_{13}+D^b_{23}+D^b_{3}
+D^b_{33}\right)
}$

$\displaystyle{
f_{QCDV6}^{RR}=
-\sum_{s=1}^{2}
\frac{2 b^{\tilde{u}}_{s2} C_V^s D^b_{12} \alpha _s
}{3 \pi }
}$

$\displaystyle{
f_{QCDV6}^{RL}=
-\sum_{s=1}^{2}
\frac{2 a^{\tilde{u}}_{s2} C_V^s D^b_{12} \alpha _s
}{3 \pi }
}$

$\displaystyle{
f_{QCDV7}^{RR}=
-
\sum_{s=1}^{2}
\frac{2 a^{\tilde{u}}_{s2} M_{\neu{2}} C_V^s D^b_{23} \alpha _s
}{3 \pi }
}$

$\displaystyle{
f_{QCDV7}^{RL}=
-
\sum_{s=1}^{2}
\frac{2 b^{\tilde{u}}_{s2} M_{\neu{2}} C_V^s D^b_{23} \alpha _s
}{3 \pi }
}$

$\displaystyle{
f_{QCDV8}^{RR}=
-\sum_{s=1}^{2}
\frac{2 a^{\tilde{d}}_{s2} M_{\chap{1}} C_U^s
\alpha _s }{3 \pi }
\left(D^a_{13}+D^a_{3}+D^a_{33}\right)
}$

$\displaystyle{
f_{QCDV8}^{LR}=
-\sum_{s=1}^{2}
\frac{2 b^{\tilde{d}}_{s2} M_{\chap{1}} C_U^s
\alpha _s }{3 \pi }
\left(D^a_{13}+D^a_{3}+D^a_{33}\right)
}$

$\displaystyle{
f_{QCDV9}^{RR}=
-\sum_{s=1}^{2}
\frac{2 b^{\tilde{u}}_{s2} M_{\chap{1}} C_V^s
\alpha _s }{3 \pi }
\left(D^b_{23}+D^b_{3}+D^b_{33}\right)
}$

$\displaystyle{
f_{QCDV9}^{LR}=
-\sum_{s=1}^{2}
\frac{2 a^{\tilde{u}}_{s2} M_{\chap{1}} C_V^s
\alpha _s }{3 \pi }
\left(D^b_{23}+D^b_{3}+D^b_{33}\right)
}$

$\displaystyle{
f_{QCDV10}^{LR}=
\sum_{s=1}^{2}
\frac{2 a^{\tilde{d}}_{s2} M_{\neu{2}} C_U^s
\alpha _s }{3 \pi }
\left(D^a_{23}+D^a_{3}+D^a_{33}\right)
}$

$\displaystyle{
f_{QCDV10}^{LL}=
\sum_{s=1}^{2}
\frac{2 b^{\tilde{d}}_{s2} M_{\neu{2}} C_U^s
\alpha _s
}{3 \pi }
\left(D^a_{23}+D^a_{3}+D^a_{33}\right)
}$

$\displaystyle{
f_{QCDV11}^{RR}=
\sum_{s=1}^{2}
\frac{b^{\tilde{u}}_{s2} C_V^s D^b_{12} \alpha _s
}{3 \pi }
}$

$\displaystyle{
f_{QCDV11}^{RL}=
\sum_{s=1}^{2}
\frac{a^{\tilde{u}}_{s2} C_V^s D^b_{12} \alpha _s
}{3 \pi }
}$

$\displaystyle{
f_{QCDV12}^{RR}=
\sum_{s=1}^{2}
\frac{a^{\tilde{u}}_{s2} M_{\neu{2}} C_V^s D^b_{23} \alpha _s
}{3 \pi }
}$

$\displaystyle{
f_{QCDV12}^{LR}=
\sum_{s=1}^{2}
\frac{b^{\tilde{u}}_{s2} M_{\neu{2}} C_V^s D^b_{23} \alpha _s
}{3 \pi }
}$

$\displaystyle{
f_{QCDV13}^{RR}=
-
\sum_{s=1}^{2}
\frac{a^{\tilde{d}}_{s2} M_{\chap{1}} C_U^s D^a_{23} \alpha _s
}{3 \pi }
}$

$\displaystyle{
f_{QCDV13}^{RL}=
-\sum_{s=1}^{2}
\frac{b^{\tilde{d}}_{s2} M_{\chap{1}} C_U^s D^a_{23} \alpha _s
}{3 \pi }
}$

$\displaystyle{
f_{QCDV14}^{RL}=
-\sum_{s=1}^{2}
\frac{a^{\tilde{d}}_{s2} C_U^s D^a_{12} \alpha _s
}{3 \pi }
}$

$\displaystyle{
f_{QCDV14}^{LL}=
-\sum_{s=1}^{2}
\frac{b^{\tilde{d}}_{s2} C_U^s D^a_{12} \alpha _s
}{3 \pi }
}$

$\displaystyle{
f_{QCDV15}^{RR}=
\sum_{s=1}^{2}
\frac{b^{\tilde{u}}_{s2} M_{\chap{1}} C_V^s D^b_{13} \alpha _s
}{3 \pi }
}$

$\displaystyle{
f_{QCDV15}^{RL}=
\sum_{s=1}^{2}
\frac{a^{\tilde{u}}_{s2} M_{\chap{1}} C_V^s D^b_{13} \alpha _s
}{3 \pi }
}$

$\displaystyle{
f_{QCDV16}^{RR}=
\sum_{s=1}^{2}
\frac{a^{\tilde{u}}_{s2} M_{\chap{1}} M_{\neu{2}} C_V^s D^b_{33}
\alpha _s
}{3 \pi }
}$

$\displaystyle{
f_{QCDV16}^{RL}=
\sum_{s=1}^{2}
\frac{b^{\tilde{u}}_{s2} M_{\chap{1}} M_{\neu{2}} C_V^s D^b_{33}
\alpha _s
}{3 \pi }
}$

$\displaystyle{
f_{QCDV17}^{RR}=
-
\sum_{s=1}^{2}
\frac{a^{\tilde{d}}_{s2} M_{\chap{1}} M_{\neu{2}} C_U^s D^a_{33}
\alpha _s
}{3 \pi }
}$

$\displaystyle{
f_{QCDV17}^{LR}=
-\sum_{s=1}^{2}
\frac{b^{\tilde{d}}_{s2} M_{\chap{1}} M_{\neu{2}} C_U^s D^a_{33}
\alpha _s
}{3 \pi }
}$

$\displaystyle{
f_{QCDV18}^{RL}=
-
\sum_{s=1}^{2}
\frac{a^{\tilde{d}}_{s2} M_{\neu{2}} C_U^s D^a_{13} \alpha _s
}{3 \pi }
}$

$\displaystyle{
f_{QCDV18}^{LL}=
-
\sum_{s=1}^{2}
\frac{b^{\tilde{d}}_{s2} M_{\neu{2}} C_U^s D^a_{13} \alpha _s
}{3 \pi }
}$

$\displaystyle{
f_{QCDV19}^{RR}=
-
\sum_{s=1}^{2}
\frac{b^{\tilde{u}}_{s2} C_V^s D^b_{00} \alpha _s
}{3 \pi }
}$

$\displaystyle{
f_{QCDV19}^{RL}=
-
\sum_{s=1}^{2}
\frac{a^{\tilde{u}}_{s2} C_V^s D^b_{00} \alpha _s
}{3 \pi }
}$

$\displaystyle{
f_{QCDV20}^{RL}=
\sum_{s=1}^{2}
\frac{a^{\tilde{d}}_{s2} C_U^s D^a_{00} \alpha _s
}{3 \pi }
}$

$\displaystyle{
f_{QCDV20}^{LL}=
\sum_{s=1}^{2}
\frac{b^{\tilde{d}}_{s2} C_U^s D^a_{00} \alpha _s
}{3 \pi }
}$

$\displaystyle{
f_{SUSYV1}^{RR}=
\sum_{m=1}^{2} \sum_{s=1}^{2}
\frac{2
\alpha _s
R^{\tilde{u}}_{s1}
R^{\tilde{d}}_{m1}}{3 \pi  (\hat{s}-M_W^2)}
(a^{\tilde{u}}_{s2} C_U^m D^c_{00}
(\hat{s}-M_W^2)-2 A_R C^i_{00} D_W^{m s})
}$

$\displaystyle{
f_{SUSYV1}^{LR}=
\sum_{m=1}^{2} \sum_{s=1}^{2}
\frac{2
\alpha _s
R^{\tilde{u}}_{s2}
R^{\tilde{d}}_{m2}}{3 \pi  (\hat{s}-M_W^2)}
(a^{\tilde{u}}_{s2} C_U^m D^c_{00}
(\hat{s}-M_W^2)-2 A_R C^i_{00} D_W^{m s})
}$

$\displaystyle{
f_{SUSYV1}^{RL}=
\sum_{m=1}^{2} \sum_{s=1}^{2}
-\frac{2
\alpha _s
R^{\tilde{u}}_{s1}
R^{\tilde{d}}_{m1}}{3 \pi  (\hat{s}-M_W^2)}
(2 A_L D_W^{m s} C^i_{00}+a^{\tilde{d}}_{m2} C_V^s
D^d_{00} (\hat{s}-M_W^2))
}$

$\displaystyle{
f_{SUSYV1}^{LL}=
-\sum_{m=1}^{2} \sum_{s=1}^{2}
\frac{2 \alpha _s R^{\tilde{u}}_{s2} R^{\tilde{d}}_{m2}}{3 \pi  (\hat{s}-M_W^2)}
(2 A_L D_W^{m s} C^i_{00}+a^{\tilde{d}}_{m2} C_V^s
D^d_{00} (\hat{s}-M_W^2))
}$

$\displaystyle{
f_{SUSYV2}^{LR}=
\sum_{m=1}^{2} \sum_{s=1}^{2}
\frac{
\alpha _s
}{3 \pi
(\hat{t}-M_{\tilde{d}_m}^2) (\hat{t}-M_{\tilde{d}_s}^2)}
\left\{2 M_{\tilde{g}}
a^{\tilde{d}}_{s2} M_{\chap{1}} (\hat{t}-M_{\tilde{d}_m}^2)
(C^h_{1}+C^h_{2}) R^{\tilde{u}}_{m1}
R^{\tilde{d}}_{s1} C_V^m\right.
}$

\hspace{36pt}$\displaystyle{
+C_U^s \left[2 b^{\tilde{d}}_{m2}
(\hat{t}-M_{\tilde{d}_m}^2) \left(B^j_{0}+(\hat{s}+2
\hat{t}+\hat{u}-M_{\chap{1}}^2-M_{\neu{2}}^2) C^m_{1}+
(2 \hat{s}+2 \hat{t}+2 \hat{u}-2 M_{\chap{1}}^2
\right.\right.}$

\hspace{36pt}$\displaystyle{\left.
-M_{\neu{2}}^2)
C^m_{2}\right) R^{\tilde{d}}_{m1}
R^{\tilde{d}}_{s2}+a^{\tilde{d}}_{m2} \left(4
(R^{\tilde{d}}_{m1} R^{\tilde{d}}_{s1}+R^{\tilde{d}}_{m2}
R^{\tilde{d}}_{s2}) (\hat{t}B^i_{1}+A_0(M_{\tilde{g}}^2) )
\right.-\sum_{n=1}^{2} D_{\tilde{d}}^{mnns} A_0(M_{\tilde{d}_n}^2)
}$

\hspace{36pt}$\displaystyle{\left.\left.\left.
+2 M_{\tilde{g}} M_{\neu{2}} (\hat{t}-M_{\tilde{d}_m}^2) (C^m_{1}+C^m_{2}) R^{\tilde{d}}_{m1} R^{\tilde{d}}_{s1}
\right)\right]\right\}
}$

$\displaystyle{
f_{SUSYV2}^{LL}=
\sum_{m=1}^{2} \sum_{s=1}^{2}
\frac{
\alpha _s
}{3 \pi
(\hat{t}-M_{\tilde{d}_m}^2) (\hat{t}-M_{\tilde{d}_s}^2)}
\left\{2 M_{\tilde{g}}
b^{\tilde{d}}_{s2} M_{\chap{1}} (\hat{t}-M_{\tilde{d}_m}^2)
(C^h_{1}+C^h_{2}) R^{\tilde{u}}_{m1}
R^{\tilde{d}}_{s1} C_V^m\right.
}$

\hspace{36pt}$\displaystyle{
+C_U^s \left[2 a^{\tilde{d}}_{m2}
(\hat{t}-M_{\tilde{d}_m}^2) \left(B^j_{0}+(\hat{s}+2
\hat{t}+\hat{u}-M_{\chap{1}}^2-M_{\neu{2}}^2) C^m_{1}+
(2 \hat{s}+2 \hat{t}+2 \hat{u}-2 M_{\chap{1}}^2
\right.\right.}$

\hspace{36pt}$\displaystyle{\left.
-M_{\neu{2}}^2)
C^m_{2}\right) R^{\tilde{d}}_{m1}
R^{\tilde{d}}_{s2}+b^{\tilde{d}}_{m2} \left(4
(R^{\tilde{d}}_{m1} R^{\tilde{d}}_{s1}+R^{\tilde{d}}_{m2}
R^{\tilde{d}}_{s2}) (\hat{t}B^i_{1}+A_0(M_{\tilde{g}}^2) )
\right.-\sum_{n=1}^{2} D_{\tilde{d}}^{mnns} A_0(M_{\tilde{d}_n}^2)
}$

\hspace{36pt}$\displaystyle{\left.\left.\left.
+2 M_{\tilde{g}} M_{\neu{2}} (\hat{t}-M_{\tilde{d}_m}^2) (C^m_{1}+C^m_{2}) R^{\tilde{d}}_{m1} R^{\tilde{d}}_{s1}
\right)\right]\right\}
}$

$\displaystyle{
f_{SUSYV2}^{RR}=
\sum_{m=1}^{2} \sum_{s=1}^{2}
\frac{2 a^{\tilde{d}}_{s2}
C_V^m \alpha _s
R^{\tilde{u}}_{m2}
R^{\tilde{d}}_{s1}}{3 \pi  (\hat{t}-M_{\tilde{d}_s}^2)}
(B^g_{0}+\hat{t}
C^h_{1}+M_{\chap{1}}^2 C^h_{2})
}$

$\displaystyle{
f_{SUSYV2}^{RL}=
\sum_{m=1}^{2} \sum_{s=1}^{2}
\frac{2 b^{\tilde{d}}_{s2}
C_V^m \alpha _s
R^{\tilde{u}}_{m2} R^{\tilde{d}}_{s1}}{3 \pi  (\hat{t}-M_{\tilde{d}_s}^2)}
(B^g_{0}+\hat{t} C^h_{1}+M_{\chap{1}}^2 C^h_{2})
}$

$\displaystyle{
f_{SUSYV3}^{RR}=
\sum_{m=1}^{2} \sum_{s=1}^{2}
\frac{\alpha _s
}{3 \pi  (\hat{u}-M_{\tilde{u}_m}^2) (\hat{u}-M_{\tilde{u}_s}^2)}
(
-2 M_{\tilde{g}} b^{\tilde{u}}_{m2} M_{\neu{2}} (\hat{u}-M_{\tilde{u}_m}^2)
(C^l_{0}+C^l_{1})
R^{\tilde{u}}_{m2} R^{\tilde{u}}_{s2} C_V^s
}$

\hspace{36pt}$\displaystyle{
+2 a^{\tilde{u}}_{m2}
R^{\tilde{u}}_{m2} R^{\tilde{u}}_{s1} C_V^s
(\hat{u}-M_{\tilde{u}_m}^2)
(B^m_{0}+(-\hat{s}-\hat{t}-\hat{u}+M_{\tilde{g}}^2
+M_{\chap{1}}^2+M_{\neu{2}}^2)
C^l_{0}
}$

\hspace{36pt}$\displaystyle{
+(-\hat{s}-\hat{t}+M_{\chap{1}}^2+M_{\neu{2}}^2)
C^l_{1})
+b^{\tilde{u}}_{s2} (C_V^m (4
(M_{\tilde{g}}^2 B^n_{0}+\hat{u} B^k_{1})
(R^{\tilde{u}}_{m1} R^{\tilde{u}}_{s1}+R^{\tilde{u}}_{m2}
R^{\tilde{u}}_{s2})
}$

\hspace{36pt}$\displaystyle{
-\sum_{n=1}^{2}
D_{\tilde{u}}^{mnns}
A_0(M_{\tilde{u}_n}^2))
-2 M_{\tilde{g}} M_{\chap{1}}
(\hat{u}-M_{\tilde{u}_m}^2)
(C^k_{0}+C^k_{1}) C_U^m R^{\tilde{d}}_{m1}
R^{\tilde{u}}_{s1}))
}$

$\displaystyle{
f_{SUSYV3}^{RL}=
\sum_{m=1}^{2} \sum_{s=1}^{2}
\frac{\alpha _s
}{3 \pi  (\hat{u}-M_{\tilde{u}_m}^2) (\hat{u}-M_{\tilde{u}_s}^2)}
(-2 M_{\tilde{g}}
a^{\tilde{u}}_{m2} M_{\neu{2}} (\hat{u}-M_{\tilde{u}_m}^2)
(C^l_{0}+C^l_{1}) R^{\tilde{u}}_{m1}
R^{\tilde{u}}_{s1} C_V^s
}$

\hspace{36pt}$\displaystyle{
+2 b^{\tilde{u}}_{m2}
R^{\tilde{u}}_{m1} R^{\tilde{u}}_{s2} C_V^s
(\hat{u}-M_{\tilde{u}_m}^2)
(B^m_{0}+(-\hat{s}-\hat{t}-\hat{u}+M_{\tilde{g}}^2
+M_{\chap{1}}^2+M_{\neu{2}}^2)
C^l_{0}
}$

\hspace{36pt}$\displaystyle{
+(-\hat{s}-\hat{t}+M_{\chap{1}}^2+M_{\neu{2}}^2)
C^l_{1})
+a^{\tilde{u}}_{s2} (C_V^m (4 (M_{\tilde{g}}^2
B^n_{0}+\hat{u} B^k_{1}) (R^{\tilde{u}}_{m1}
R^{\tilde{u}}_{s1}+R^{\tilde{u}}_{m2}
R^{\tilde{u}}_{s2})
}$

\hspace{36pt}$\displaystyle{
-\sum_{n=1}^{2}
D_{\tilde{u}}^{mnns}
A_0(M_{\tilde{u}_n}^2))
-2 M_{\tilde{g}} M_{\chap{1}}
(\hat{u}-M_{\tilde{u}_m}^2)
(C^k_{0}+C^k_{1}) C_U^m R^{\tilde{d}}_{m1}
R^{\tilde{u}}_{s1}))
}$

$\displaystyle{
f_{SUSYV3}^{LR}=
\sum_{m=1}^{2} \sum_{s=1}^{2}
\frac{2
b^{\tilde{u}}_{s2}
C_U^m \alpha _s
R^{\tilde{d}}_{m2}
R^{\tilde{u}}_{s1}}{3 \pi  (\hat{u}-M_{\tilde{u}_s}^2)}
(B^h_{0}+M_{\tilde{g}}^2
C^k_{0}+\hat{u} C^k_{1})
}$

$\displaystyle{
f_{SUSYV3}^{LL}=
\sum_{m=1}^{2} \sum_{s=1}^{2}
\frac{2 a^{\tilde{u}}_{s2}
C_U^m \alpha _s
R^{\tilde{d}}_{m2} R^{\tilde{u}}_{s1}}{3 \pi  (\hat{u}-M_{\tilde{u}_s}^2)}
(B^h_{0}+M_{\tilde{g}}^2
C^k_{0}+\hat{u} C^k_{1})
}$

$\displaystyle{
f_{SUSYV21}^{RR}=
\sum_{m=1}^{2} \sum_{s=1}^{2}
\frac{2 M_{\tilde{g}}
\alpha _s R^{\tilde{u}}_{s2} R^{\tilde{d}}_{m1}
}{3 \pi (\hat{s}-M_{H^-}^2) (\hat{s}-M_W^2)}
((-C_L^G (\hat{s}-M_{H^-}^2) D_G^{m
s}-C_L^H D_H^{m s} (\hat{s}-M_W^2))
C^i_{0}
}$

\hspace{36pt}$\displaystyle{
+(\hat{s}-M_{H^-}^2) (M_{\neu{2}}
A_L-M_{\chap{1}} A_R) (C^i_{0}+2
C^i_{2}) D_W^{m s}+(\hat{s}-M_{H^-}^2) a^{\tilde{u}}_{s2}
M_{\chap{1}} C_U^m D^c_{3}
(\hat{s}-M_W^2)
}$

\hspace{36pt}$\displaystyle{
+(\hat{s}-M_{H^-}^2) a^{\tilde{d}}_{m2}
M_{\neu{2}} C_V^s D^d_{3} (\hat{s}-M_W^2))
}$

$\displaystyle{
f_{SUSYV21}^{RL}=
\sum_{m=1}^{2} \sum_{s=1}^{2}
\frac{2 M_{\tilde{g}}
\alpha _s R^{\tilde{u}}_{s2} R^{\tilde{d}}_{m1}
}{3 \pi (\hat{s}-M_{H^-}^2)(\hat{s}-M_W^2)}
( (-C_R^G (\hat{s}-M_{H^-}^2)
D_G^{m s}-C_R^H D_H^{m s} (\hat{s}-M_W^2))
C^i_{0}
}$

\hspace{36pt}$\displaystyle{
-(\hat{s}-M_{H^-}^2) (M_{\chap{1}}
A_L-M_{\neu{2}} A_R) (C^i_{0}+2
C^i_{2}) D_W^{m s}+
a^{\tilde{d}}_{m2}
M_{\chap{1}} C_V^s (D^d_{1}+D^d_{2})
}$

\hspace{36pt}$\displaystyle{
\times (\hat{s}-M_{H^-}^2) (\hat{s}-M_W^2)
-a^{\tilde{u}}_{s2} M_{\neu{2}} C_U^m (D^c_{0}+D^c_{1}+D^c_{3})
(\hat{s}-M_{H^-}^2)(\hat{s}-M_W^2) )
}$

$\displaystyle{
f_{SUSYV21}^{LR}=
\sum_{m=1}^{2} \sum_{s=1}^{2}
\frac{2 M_{\tilde{g}}
\alpha _s R^{\tilde{u}}_{s1} R^{\tilde{d}}_{m2}
}{3 (\hat{s}-M_{H^-}^2) \pi (\hat{s}-M_W^2)}
((-C_L^G (\hat{s}-M_{H^-}^2)
D_G^{m s}-C_L^H D_H^{m s} (\hat{s}-M_W^2)) C^i_{0}
}$

\hspace{36pt}$\displaystyle{
+(\hat{s}-M_{H^-}^2) (M_{\neu{2}}
A_L-M_{\chap{1}} A_R) (C^i_{0}+2
C^i_{2}) D_W^{m s}+(\hat{s}-M_{H^-}^2) a^{\tilde{u}}_{s2}
M_{\chap{1}} C_U^m D^c_{3}
(\hat{s}-M_W^2)
}$

\hspace{36pt}$\displaystyle{
+(\hat{s}-M_{H^-}^2) a^{\tilde{d}}_{m2}
M_{\neu{2}} C_V^s D^d_{3} (\hat{s}-M_W^2))
}$

$\displaystyle{
f_{SUSYV21}^{LL}=
\sum_{m=1}^{2} \sum_{s=1}^{2}
\frac{2 M_{\tilde{g}}
\alpha _s R^{\tilde{u}}_{s1} R^{\tilde{d}}_{m2}}{3 \pi
(\hat{s}-M_{H^-}^2) (\hat{s}-M_W^2)
 }
(
(-C_R^G (\hat{s}-M_{H^-}^2)
D_G^{m s}-C_R^H D_H^{m s} (\hat{s}-M_W^2))
C^i_{0}
}$

\hspace{36pt}$\displaystyle{
-(\hat{s}-M_{H^-}^2) (M_{\chap{1}}
A_L-M_{\neu{2}} A_R) (C^i_{0}+2
C^i_{2}) D_W^{m s}+
a^{\tilde{d}}_{m2} M_{\chap{1}} C_V^s (D^d_{1}+D^d_{2})
}$

\hspace{36pt}$\displaystyle{
\times(\hat{s}-M_{H^-}^2) (\hat{s}-M_W^2)
-a^{\tilde{u}}_{s2} M_{\neu{2}} C_U^m
(D^c_{0}+D^c_{1}+D^c_{3})
(\hat{s}-M_{H^-}^2) (\hat{s}-M_W^2)
)
}$

$\displaystyle{
f_{SUSYV22}^{RR}=
\sum_{m=1}^{2} \sum_{s=1}^{2}
\frac{2 M_{\tilde{g}}
\alpha _s
R^{\tilde{u}}_{s2} R^{\tilde{d}}_{m1}
}{3 \pi  (\hat{s}-M_W^2)}
(a^{\tilde{u}}_{s2} C_U^m
(D^c_{0}+D^c_{1}+D^c_{2}+D^c_{3}
) (\hat{s}-M_W^2)
}$

\hspace{36pt}$\displaystyle{
-2 A_R (C^i_{0}+C^i_{1}+C^i_{2}) D_W^{m s}
)
}$

$\displaystyle{
f_{SUSYV22}^{LR}=
\sum_{m=1}^{2} \sum_{s=1}^{2}
\frac{2 M_{\tilde{g}}
\alpha _s R^{\tilde{u}}_{s1} R^{\tilde{d}}_{m2}}{3 \pi (\hat{s}-M_W^2)}
(a^{\tilde{u}}_{s2} C_U^m
(D^c_{0}+D^c_{1}+D^c_{2}+D^c_{3}
) (\hat{s}-M_W^2)
}$

\hspace{36pt}$\displaystyle{
-2 A_R
(C^i_{0}+C^i_{1}+C^i_{2})
D_W^{m s})
}$

$\displaystyle{
f_{SUSYV22}^{RL}=
\sum_{m=1}^{2} \sum_{s=1}^{2}
\frac{2 M_{\tilde{g}}
\alpha _s R^{\tilde{u}}_{s2} R^{\tilde{d}}_{m1}}{3 \pi (\hat{s}-M_W^2)}
(a^{\tilde{d}}_{m2} C_V^s D^d_{1}
(\hat{s}-M_W^2)-2 A_L
(C^i_{0}+C^i_{1}+C^i_{2})
D_W^{m s})
}$

$\displaystyle{
f_{SUSYV22}^{LL}=
\sum_{m=1}^{2} \sum_{s=1}^{2}
\frac{2 M_{\tilde{g}} \alpha _s
R^{\tilde{u}}_{s1} R^{\tilde{d}}_{m2}}{3 \pi
(\hat{s}-M_W^2)}
(a^{\tilde{d}}_{m2} C_V^s D^d_{1}
(\hat{s}-M_W^2)-2 A_L
(C^i_{0}+C^i_{1}+C^i_{2})
D_W^{m s})
}$

$\displaystyle{
f_{SUSYV23}^{RR}=
\sum_{m=1}^{2} \sum_{s=1}^{2}
\frac{2 \alpha _s
R^{\tilde{u}}_{s1} R^{\tilde{d}}_{m1}}{3 \pi }
(a^{\tilde{u}}_{s2} M_{\chap{1}} C_U^m
D^c_{13}+a^{\tilde{d}}_{m2} M_{\neu{2}} C_V^s
(D^d_{13}+D^d_{23}+D^d_{3}+D^d_{33}
))
}$

$\displaystyle{
f_{SUSYV23}^{RL}=
\sum_{m=1}^{2} \sum_{s=1}^{2}
\frac{2 \alpha _s
R^{\tilde{u}}_{s2} R^{\tilde{d}}_{m2}}{3 \pi }
(a^{\tilde{u}}_{s2} M_{\chap{1}} C_U^m
D^c_{13}+a^{\tilde{d}}_{m2} M_{\neu{2}} C_V^s
(D^d_{13}+D^d_{23}+D^d_{3}+D^d_{33}
))
}$

$\displaystyle{
f_{SUSYV23}^{LR}=
-\sum_{m=1}^{2} \sum_{s=1}^{2}
\frac{2
\alpha _s
R^{\tilde{u}}_{s1} R^{\tilde{d}}_{m1}}{3 \pi }
(a^{\tilde{u}}_{s2} M_{\neu{2}} C_U^m
(D^c_{1}+D^c_{11}+D^c_{13})
}$

\hspace{36pt}$\displaystyle{
-a^{\tilde{d}}_{m2}
M_{\chap{1}} C_V^s (D^d_{1}+D^d_{11}+2
D^d_{12}+D^d_{13}+D^d_{2}+D^d_{22}
+D^d_{23}))
}$

$\displaystyle{
f_{SUSYV23}^{LL}=
-\sum_{m=1}^{2} \sum_{s=1}^{2}
\frac{2
\alpha _s
R^{\tilde{u}}_{s2} R^{\tilde{d}}_{m2}}{3 \pi }
(a^{\tilde{u}}_{s2} M_{\neu{2}}
C_U^m
(D^c_{1}+D^c_{11}+D^c_{13})
}$

\hspace{36pt}$\displaystyle{
-a^{\tilde{d}}_{m2} M_{\chap{1}} C_V^s (D^d_{1}+D^d_{11}+2
D^d_{12}+D^d_{13}+D^d_{2}+D^d_{22}
+D^d_{23}))
}$

$\displaystyle{
f_{SUSYV24}^{RR}=
\sum_{m=1}^{2} \sum_{s=1}^{2}
\frac{2 a^{\tilde{u}}_{s2} C_U^m
\alpha _s
R^{\tilde{u}}_{s1} R^{\tilde{d}}_{m1}}{3 \pi }
(D^c_{1}+D^c_{11}+D^c_{12}+D^c_{13}
)
}$

$\displaystyle{
f_{SUSYV24}^{RL}=
\sum_{m=1}^{2} \sum_{s=1}^{2}
\frac{2 a^{\tilde{d}}_{m2} C_V^s
\alpha _s
R^{\tilde{u}}_{s1} R^{\tilde{d}}_{m1}}{3 \pi }
(D^d_{1}+D^d_{11}+D^d_{12}+D^d_{13}
)
}$

$\displaystyle{
f_{SUSYV24}^{LR}=
\sum_{m=1}^{2} \sum_{s=1}^{2}
\frac{2 a^{\tilde{u}}_{s2} C_U^m
\alpha _s
R^{\tilde{u}}_{s2} R^{\tilde{d}}_{m2}}{3 \pi }
(D^c_{1}+D^c_{11}+D^c_{12}+D^c_{13}
)
}$

$\displaystyle{
f_{SUSYV24}^{LL}=
\sum_{m=1}^{2} \sum_{s=1}^{2}
\frac{2 a^{\tilde{d}}_{m2} C_V^s
\alpha _s
R^{\tilde{u}}_{s2} R^{\tilde{d}}_{m2}}{3 \pi }
(D^d_{1}+D^d_{11}+D^d_{12}+D^d_{13}
)
}$

$\displaystyle{
f^{LR}_{QCDC2}=
-\sum_{s=1}^2 \frac{4 M_{\tilde{d}_s}^2
C_U^s \alphas a^{\tilde{d}}_{s2} }{3 \pi (\hat{t}-M_{\tilde{d}_s}^2)^2}
(B_0^t+B_1^t)
}$

$\displaystyle{
f^{LL}_{QCDC2}=
-\sum_{s=1}^2 \frac{4 M_{\tilde{d}_s}^2
C_U^s \alphas b^{\tilde{d}}_{s2} }{3 \pi (\hat{t}-M_{\tilde{d}_s}^2)^2}
(B_0^t+B_1^t)
}$

$\displaystyle{
f^{RR}_{QCDC3}=
-\sum_{s=1}^2
\frac{
4M_{\tilde{u}_s}^2
b^{\tilde{u}}_{s2} \alphas C_V^s }{3 \pi
(\hat{u}-M_{\tilde{u}_s}^2)^2}
(B_0^u+B_1^u)
}$

$\displaystyle{
f^{RL}_{QCDC3}=
-\sum_{s=1}^2
\frac{
4M_{\tilde{u}_s}^2
a^{\tilde{u}}_{s2} \alphas C_V^s }{3 \pi
(\hat{u}-M_{\tilde{u}_s}^2)^2}
(B_0^u+B_1^u)
}$

$\displaystyle{
f^{RR}_{SUSYC1}=
\sum_{k=1}^2
\frac{A_R D_L \alpha _s
}{3 \pi  (\hat{s}-M_W^2)}
(R^{\tilde{u}2}_{1k}
B_0^p+R^{\tilde{d}2}_{1k} B_0^q+B_1^p R^{\tilde{u}2}_{1k}+B_1^q
R^{\tilde{d}2}_{1k})
}$

$\displaystyle{
f^{RL}_{SUSYC1}=
\sum_{k=1}^2
\frac{A_L D_L \alpha _s
}{3 \pi  (\hat{s}-M_W^2)}
(R^{\tilde{u}2}_{1k}
B_0^p+R^{\tilde{d}2}_{1k} B_0^q+B_1^p R^{\tilde{u}2}_{1k}+B_1^q
R^{\tilde{d}2}_{1k})
}$

$\displaystyle{
f^{LR}_{SUSYC2}=
\sum_{s=1}^2 \left\{
\sum_{k=1}^2 \sum_{m=1}^2
\frac{C_U^s \alphas a^{\tilde{d}}_{m2} }{3 \pi (\hat{t}-M_{\tilde{d}_m}^2) (\hat{t}-M_{\tilde{d}_s})^2}
(R^{\tilde{d}}_{k1} R^{\tilde{d}}_{m1}-R^{\tilde{d}}_{k2}
R^{\tilde{d}}_{m2}) (R^{\tilde{d}}_{k1}
R^{\tilde{d}}_{s1}-R^{\tilde{d}}_{k2} R^{\tilde{d}}_{s2})
A_0(M_{\tilde{d}_k}^2)
\right.}$

\hspace{36pt}$\displaystyle{\left.
-\frac{C_U^s \alphas a^{\tilde{d}}_{s2} }{3 \pi (\hat{t}-M_{\tilde{d}_s}^2)^2}
\right[
(\hat{t}-M_{\tilde{d}_s}^2)
\sum_{k=1}^2 ( R^{\tilde{u}2}_{1k} B_0^p + R^{\tilde{u}2}_{1k} B_1^p +
R^{\tilde{d}2}_{1k} B_0^q + R^{\tilde{d}2}_{1k} B_1^q)
+4 M_{\tilde{d}_s}^2  B_1^r
}$

\hspace{36pt}$\displaystyle{
\left.\left.+4 A_0(M_{\tilde{g}}^2)+A_0(M_{\tilde{d}_s}^2)
-\sum_{k=1}^2 (R^{\tilde{d}}_{k1} R^{\tilde{d}}_{s1}-R^{\tilde{d}}_{k2} R^{\tilde{d}}_{s2})^2 A_0(M_{\tilde{d}_k}^2)
\right]\right\}
}$

$\displaystyle{
f^{LL}_{SUSYC2}=
\sum_{s=1}^2 \left\{
\sum_{k=1}^2 \sum_{m=1}^2
\frac{C_U^s \alphas b^{\tilde{d}}_{m2} }{3 \pi (\hat{t}-M_{\tilde{d}_m}^2) (\hat{t}-M_{\tilde{d}_s})^2}
(R^{\tilde{d}}_{k1} R^{\tilde{d}}_{m1}-R^{\tilde{d}}_{k2}
R^{\tilde{d}}_{m2}) (R^{\tilde{d}}_{k1}
R^{\tilde{d}}_{s1}-R^{\tilde{d}}_{k2} R^{\tilde{d}}_{s2})
A_0(M_{\tilde{d}_k}^2)
\right.}$

\hspace{36pt}$\displaystyle{\left.
-\frac{C_U^s \alphas b^{\tilde{d}}_{s2} }{3 \pi (\hat{t}-M_{\tilde{d}_s}^2)^2}
\right[
(\hat{t}-M_{\tilde{d}_s}^2)
\sum_{k=1}^2 ( R^{\tilde{u}2}_{1k} B_0^p + R^{\tilde{u}2}_{1k} B_1^p +
R^{\tilde{d}2}_{1k} B_0^q + R^{\tilde{d}2}_{1k} B_1^q)
+4 M_{\tilde{d}_s}^2  B_1^r
}$

\hspace{36pt}$\displaystyle{
\left.\left.+4 A_0(M_{\tilde{g}}^2)+A_0(M_{\tilde{d}_s}^2)
-\sum_{k=1}^2 (R^{\tilde{d}}_{k1} R^{\tilde{d}}_{s1}-R^{\tilde{d}}_{k2} R^{\tilde{d}}_{s2})^2 A_0(M_{\tilde{d}_k}^2)
\right]\right\}
}$

$\displaystyle{
f^{RR}_{SUSYC3}=
\sum_{s=1}^2
\left\{
\frac{b^{\tilde{u}}_{s2} \alphas C_V^m}{3 \pi  (\hat{u}-M_{\tilde{u}_m}^2) (\hat{u}-M_{\tilde{u}_s}^2)}
\sum_{k=1}^2 \sum_{m=1}^2 (R^{\tilde{u}}_{k1}
R^{\tilde{u}}_{m1}-R^{\tilde{u}}_{k2} R^{\tilde{u}}_{m2})
(R^{\tilde{u}}_{k1} R^{\tilde{u}}_{s1}-R^{\tilde{u}}_{k2}
R^{\tilde{u}}_{s2}) A_0(M_{\tilde{u}_k}^2)
\right.}$

\hspace{36pt}$\displaystyle{
-\frac{b^{\tilde{u}}_{s2} \alphas C_V^s }{3 \pi
(\hat{u}-M_{\tilde{u}_s}^2)^2}
\left[
\sum_{k=1}^2 (\hat{u}-M_{\tilde{u}_s}^2) (B_0^p  R^{\tilde{u}2}_{1k} +B_0^q R^{\tilde{d}2}_{1k}
+ R^{\tilde{u}2}_{1k} B_1^p+ R^{\tilde{d}2}_{1k} B_1^q
) +4 A_0(M_{\tilde{g}}^2)
\right.}$

\hspace{36pt}$\displaystyle{
\left.\left.
-\sum_{k=1}^2 (R^{\tilde{u}}_{k1}
R^{\tilde{u}}_{s1}-R^{\tilde{u}}_{k2} R^{\tilde{u}}_{s2})^2
A_0(M_{\tilde{u}_k}^2) +A_0(M_{\tilde{u}_s}^2)
+4M_{\tilde{u}_s}^2 B_1^s  \right]
\right\}
}$

$\displaystyle{
f^{RL}_{SUSYC3}=
\sum_{s=1}^2
\left\{
\frac{a^{\tilde{u}}_{s2} \alphas C_V^m}{3 \pi  (\hat{u}-M_{\tilde{u}_m}^2) (\hat{u}-M_{\tilde{u}_s}^2)}
\sum_{k=1}^2 \sum_{m=1}^2 (R^{\tilde{u}}_{k1}
R^{\tilde{u}}_{m1}-R^{\tilde{u}}_{k2} R^{\tilde{u}}_{m2})
(R^{\tilde{u}}_{k1} R^{\tilde{u}}_{s1}-R^{\tilde{u}}_{k2}
R^{\tilde{u}}_{s2}) A_0(M_{\tilde{u}_k}^2)
\right.}$

\hspace{36pt}$\displaystyle{
-\frac{a^{\tilde{u}}_{s2} \alphas C_V^s }{3 \pi
(\hat{u}-M_{\tilde{u}_s}^2)^2}
\left[
\sum_{k=1}^2 (\hat{u}-M_{\tilde{u}_s}^2) (B_0^p  R^{\tilde{u}2}_{1k} +B_0^q R^{\tilde{d}2}_{1k}
+ R^{\tilde{u}2}_{1k} B_1^p+ R^{\tilde{d}2}_{1k} B_1^q
) +4 A_0(M_{\tilde{g}}^2)
\right.}$

\hspace{36pt}$\displaystyle{
\left.\left.
-\sum_{k=1}^2 (R^{\tilde{u}}_{k1}
R^{\tilde{u}}_{s1}-R^{\tilde{u}}_{k2} R^{\tilde{u}}_{s2})^2
A_0(M_{\tilde{u}_k}^2) +A_0(M_{\tilde{u}_s}^2)
+4M_{\tilde{u}_s}^2 B_1^s  \right]
\right\}
}$

The standard matrix elements in Eq.(\ref{RGEQ}) for the subprocess
\begin{equation}
u(p_1)+\bar{d}(p_2)\to \chap{1}(p_3) +\neu{2}(p_4)+g(p_5),
\end{equation}
are defined as follows:

$\displaystyle{
M_{G1}^{ab}=\bar{v}_1^m P_a v_3 \bar{v}_2^n P_b v_4 [\epsilon^x\cdot(p_1-p_3)] (T^x)_{mn},
}$

$\displaystyle{
M_{G2}^{ab}=\bar{v}_2^n P_a v_4 \bar{u}_3 P_b u_1^m [\epsilon^x\cdot(p_1-p_3)] (T^x)_{mn},
}$

$\displaystyle{
M_{G3}^{ab}=\bar{v}_2^n P_a v_3 \bar{u}_4 P_b u_1^m [\epsilon^x\cdot(p_1-p_3)] (T^x)_{mn},
}$

$\displaystyle{
M_{G4}^{ab}=\bar{u}_4 P_a v_3 \bar{v}_2^n P_b \es u_1^m (T^x)_{mn},
}$

$\displaystyle{
M_{G5}^{ab}=\bar{u}_4 P_a u_1^m \bar{v}_2^n P_b \es v_3 (T^x)_{mn},
}$

$\displaystyle{
M_{G6}^{ab}=\bar{v}_1^m P_a v_3 \bar{v}_2^m P_b \es v_4 (T^x)_{mn},
}$

$\displaystyle{
M_{G7}^{ab}=\bar{v}_2^n P_a v_4 \bar{u}_3 P_b \es u_1^m (T^x)_{mn},
}$

$\displaystyle{
M_{G8}^{ab}=\bar{v}_2^n P_a v_3 \bar{u}_4 P_b \es u_1^m (T^x)_{mn},
}$

$\displaystyle{
M_{G9}^{ab}=\bar{v}_2^n P_a \ps_3 u_1^m \bar{u}_4 P_b \es v_3 (T^x)_{mn},
}$

$\displaystyle{
M_{G10}^{ab}=\bar{v}_2^n P_a \ps_4 u_1^m \bar{u}_4 P_b \es v_3 (T^x)_{mn},
}$

$\displaystyle{
M_{G11}^{ab}=\bar{v}_2^n P_a \es u_1^m \bar{u}_4 P_b \ps_1 v_3 (T^x)_{mn},
}$

$\displaystyle{
M_{G12}^{ab}=\bar{v}_2^n P_a \es u_1^m \bar{u}_4 P_b \ps_2 v_3 (T^x)_{mn},
}$

$\displaystyle{
M_{G13}^{ab}=\bar{v}_2^n P_a \gamma^\mu u_1^m \bar{u}_4 P_b \gamma_\mu v_3
[\epsilon^x\cdot(p_1+p_2)]
(T^x)_{mn},
}$

$\displaystyle{
M_{G14}^{ab}=\bar{u}_4 P_a u_1^m \bar{v}_2^n P_b \es \ps_1 v_3 (T^x)_{mn},
}$

$\displaystyle{
M_{G15}^{ab}=\bar{v}_1^m P_a v_3 \bar{v}_2^n P_b \es \ps_1 v_4 (T^x)_{mn},
}$

$\displaystyle{
M_{G16}^{ab}=\bar{v}_1^m P_a v_3 \bar{v}_2^n P_b \es \ps_3 v_4 (T^x)_{mn},
}$

$\displaystyle{
M_{G17}^{ab}=\bar{u}_4 P_a u_1^m \bar{v}_2^n P_b \es \ps_4 v_3 (T^x)_{mn},
}$

$\displaystyle{
M_{G18}^{ab}=\bar{v}_2^n P_a v_4 \bar{u}_3 P_b \es \ps_2 u_1^m (T^x)_{mn},
}$

$\displaystyle{
M_{G19}^{ab}=\bar{v}_2^n P_a v_4 \bar{u}_3 P_b \es \ps_4 u_1^m (T^x)_{mn},
}$

where $x$, $m$, and $n$ are color indices for gluons, up quarks and down quarks, respectively.

The nonzero form factors in Eq.(\ref{RGEQ}) are the following:

$\displaystyle{
f_{RG1}^{LR}=
\sum_{s=1}^2
\frac{2
g_s  a^{\tilde{d}}_{s2}
C_U^s
}{(\hat{t}-M_{\tilde{d}_s}^2)
(\hat{t}_{24}-M_{\tilde{d}_s}^2)}
}$

$\displaystyle{
f_{RG1}^{LL}=
\sum_{s=1}^2
\frac{2 g_s  b^{\tilde{d}}_{s2} C_U^s
}{(\hat{t}-M_{\tilde{d}_s}^2)
(\hat{t}_{24}-M_{\tilde{d}_s}^2)}
}$

$\displaystyle{
f_{RG2}^{RL}=
\sum_{s=1}^2
\frac{2
g_s  a^{\tilde{d}}_{s2}
C_U^s
}{(\hat{s}+\hat{t}+\hat{t}_{14}-M_{\chap{1}}^2-M_{\neu{2}}^2) (\hat{t}_{24}-M_{\tilde{d}_s}^2)}
}$

$\displaystyle{
f_{RG2}^{LL}=
\sum_{s=1}^2
\frac{2
g_s  b^{\tilde{d}}_{s2}
C_U^s
}{(\hat{s}+\hat{t}+\hat{t}_{14}-M_{\chap{1}}^2-M_{\neu{2}}^2) (\hat{t}_{24}-M_{\tilde{d}_s}^2)}
}$

$\displaystyle{
f_{RG3}^{RR}=
\sum_{s=1}^2
\frac{2
g_s  b^{\tilde{u}}_{s2}
C_V^s
(\hat{s}+\hat{t}-M_{\chap{1}}^2-M_{\neu{2}}^2+M_{\tilde{u}_s}^2)
}{(\hat{s}+\hat{t}+\hat{t}_{14}-M_{\chap{1}}^2-M_{\neu{2}}^2) (\hat{t}_{14}-M_{\tilde{u}_s}^2)
(M_{\tilde{u}_s}^2-\hat{u})}
}$

$\displaystyle{
f_{RG3}^{RL}=
\sum_{s=1}^2
\frac{2
g_s  a^{\tilde{u}}_{s2}
C_V^s
(\hat{s}+\hat{t}-M_{\chap{1}}^2-M_{\neu{2}}^2+M_{\tilde{u}_s}^2)
}{(\hat{s}+\hat{t}+\hat{t}_{14}-M_{\chap{1}}^2-M_{\neu{2}}^2) (\hat{t}_{14}-M_{\tilde{u}_s}^2)
(M_{\tilde{u}_s}^2-\hat{u})}
}$

$\displaystyle{
f_{RG4}^{RR}=
\frac{2 D_L g_s  (A_R M_{\chap{1}}-A_L
M_{\neu{2}})}{(M_W^2-\hat{s}_{34})
(\hat{s}+\hat{t}+\hat{t}_{14}-M_{\chap{1}}^2-M_{\neu{2}}^2)}
}$

$\displaystyle{
f_{RG4}^{LR}=
\frac{2 D_L g_s  (A_L M_{\chap{1}}-A_R
M_{\neu{2}})}{(M_W^2-\hat{s}_{34})
(\hat{s}+\hat{t}+\hat{t}_{14}-M_{\chap{1}}^2-M_{\neu{2}}^2)}
}$

$\displaystyle{
f_{RG5}^{RR}=
-\sum_{s=1}^2
\frac{g_s  b^{\tilde{u}}_{s2}
C_V^s M_{\chap{1}}
}{(\hat{s}+\hat{u}+\hat{t}_{24}-M_{\chap{1}}^2-M_{\neu{2}}^2) (\hat{t}_{14}-M_{\tilde{u}_s}^2)}
}$

$\displaystyle{
f_{RG5}^{LR}=
-\sum_{s=1}^2
\frac{g_s  a^{\tilde{u}}_{s2}
C_V^s M_{\chap{1}}
}{(\hat{s}+\hat{u}+\hat{t}_{24}-M_{\chap{1}}^2-M_{\neu{2}}^2) (\hat{t}_{14}-M_{\tilde{u}_s}^2)}
}$

$\displaystyle{
f_{RG6}^{LR}=
-\sum_{s=1}^2
\frac{g_s  a^{\tilde{d}}_{s2}
C_U^s M_{\neu{2}}
}{(\hat{s}+\hat{u}+\hat{t}_{24}-M_{\chap{1}}^2-M_{\neu{2}}^2) (\hat{t}-M_{\tilde{d}_s}^2)}
}$

$\displaystyle{
f_{RG6}^{LL}=
-\sum_{s=1}^2
\frac{g_s  b^{\tilde{d}}_{s2}
C_U^s M_{\neu{2}}
}{(\hat{s}+\hat{u}+\hat{t}_{24}-M_{\chap{1}}^2-M_{\neu{2}}^2) (\hat{t}-M_{\tilde{d}_s}^2)}
}$

$\displaystyle{
f_{RG7}^{RR}=
\sum_{s=1}^2
\frac{g_s  a^{\tilde{d}}_{s2} C_U^s
M_{\chap{1}}
}{(\hat{s}+\hat{t}+\hat{t}_{14}-M_{\chap{1}}^2-M_{\neu{2}}^2) (\hat{t}_{24}-M_{\tilde{d}_s}^2)}
}$

$\displaystyle{
f_{RG7}^{LR}=
\sum_{s=1}^2
\frac{g_s  b^{\tilde{d}}_{s2} C_U^s
M_{\chap{1}}
}{(\hat{s}+\hat{t}+\hat{t}_{14}-M_{\chap{1}}^2-M_{\neu{2}}^2) (\hat{t}_{24}-M_{\tilde{d}_s}^2)}
}$

$\displaystyle{
f_{RG8}^{RR}=
-\sum_{s=1}^2
\frac{g_s  a^{\tilde{u}}_{s2}
C_V^s M_{\neu{2}}
}{(\hat{s}+\hat{t}+\hat{t}_{14}-M_{\chap{1}}^2-M_{\neu{2}}^2) (\hat{u}-M_{\tilde{u}_s}^2)}
}$

$\displaystyle{
f_{RG8}^{RL}=
-\sum_{s=1}^2
\frac{g_s  b^{\tilde{u}}_{s2}
C_V^s M_{\neu{2}}
}{(\hat{s}+\hat{t}+\hat{t}_{14}-M_{\chap{1}}^2-M_{\neu{2}}^2) (\hat{u}-M_{\tilde{u}_s}^2)}
}$

$\displaystyle{
f_{RG9}^{RR}=
\frac{2 A_R D_L g_s }{(M_W^2-\hat{s}_{34})
(\hat{s}+\hat{t}+\hat{t}_{14}-M_{\chap{1}}^2-M_{\neu{2}}^2)}
}$

$\displaystyle{
f_{RG9}^{RL}=
\frac{2 A_L D_L g_s }{(M_W^2-\hat{s}_{34})
(\hat{s}+\hat{t}+\hat{t}_{14}-M_{\chap{1}}^2-M_{\neu{2}}^2)}
}$

$\displaystyle{
f_{RG10}^{RR}=
\frac{2 A_R D_L g_s }{(M_W^2-\hat{s}_{34})
(\hat{s}+\hat{t}+\hat{t}_{14}-M_{\chap{1}}^2-M_{\neu{2}}^2)}
}$

$\displaystyle{
f_{RG10}^{RL}=
\frac{2 A_L D_L g_s }{(M_W^2-\hat{s}_{34})
(\hat{s}+\hat{t}+\hat{t}_{14}-M_{\chap{1}}^2-M_{\neu{2}}^2)}
}$

$\displaystyle{
f_{RG11}^{RR}=
-\frac{2 A_R D_L g_s }{(M_W^2-\hat{s}_{34})
(\hat{s}+\hat{u}+\hat{t}_{24}-M_{\chap{1}}^2-M_{\neu{2}}^2)}
}$

$\displaystyle{
f_{RG11}^{RL}=
-\frac{2 A_L D_L g_s }{(M_W^2-\hat{s}_{34})
(\hat{s}+\hat{u}+\hat{t}_{24}-M_{\chap{1}}^2-M_{\neu{2}}^2)}
}$

$\displaystyle{
f_{RG12}^{RR}=
\frac{2 A_R D_L g_s }{(M_W^2-\hat{s}_{34})
(\hat{s}+\hat{t}+\hat{t}_{14}-M_{\chap{1}}^2-M_{\neu{2}}^2)}
}$

$\displaystyle{
f_{RG12}^{RL}=
\frac{2 A_L D_L g_s }{(M_W^2-\hat{s}_{34})
(\hat{s}+\hat{t}+\hat{t}_{14}-M_{\chap{1}}^2-M_{\neu{2}}^2)}
}$

$\displaystyle{
f_{RG13}^{RR}=
-\frac{2
A_R D_L g_s
}{(M_W^2-\hat{s}_{34})
(\hat{s}+\hat{t}+\hat{t}_{14}-M_{\chap{1}}^2-M_{\neu{2}}^2)}
}$

$\displaystyle{
f_{RG13}^{RL}=
-\frac{2
A_L D_L g_s
}{(M_W^2-\hat{s}_{34})
(\hat{s}+\hat{t}+\hat{t}_{14}-M_{\chap{1}}^2-M_{\neu{2}}^2)}
}$

$\displaystyle{
f_{RG14}^{RR}=
-\sum_{s=1}^2
\frac{g_s  b^{\tilde{u}}_{s2}
C_V^s
}{(\hat{s}+\hat{u}+\hat{t}_{24}-M_{\chap{1}}^2-M_{\neu{2}}^2) (\hat{t}_{14}-M_{\tilde{u}_s}^2)}
}$

$\displaystyle{
f_{RG14}^{LR}=
-\sum_{s=1}^2
\frac{g_s  a^{\tilde{u}}_{s2}
C_V^s
}{(\hat{s}+\hat{u}+\hat{t}_{24}-M_{\chap{1}}^2-M_{\neu{2}}^2) (\hat{t}_{14}-M_{\tilde{u}_s}^2)}
}$

$\displaystyle{
f_{RG15}^{LR}=
-\sum_{s=1}^2
\frac{g_s  a^{\tilde{d}}_{s2}
C_U^s
}{(\hat{s}+\hat{u}+\hat{t}_{24}-M_{\chap{1}}^2-M_{\neu{2}}^2) (\hat{t}-M_{\tilde{d}_s}^2)}
}$

$\displaystyle{
f_{RG15}^{LL}=
-\sum_{s=1}^2
\frac{g_s  b^{\tilde{d}}_{s2}
C_U^s
}{(\hat{s}+\hat{u}+\hat{t}_{24}-M_{\chap{1}}^2-M_{\neu{2}}^2) (\hat{t}-M_{\tilde{d}_s}^2)}
}$

$\displaystyle{
f_{RG16}^{LR}=
\sum_{s=1}^2
\frac{g_s  a^{\tilde{d}}_{s2} C_U^s
}{(\hat{s}+\hat{u}+\hat{t}_{24}-M_{\chap{1}}^2-M_{\neu{2}}^2) (\hat{t}-M_{\tilde{d}_s}^2)}
}$

$\displaystyle{
f_{RG16}^{LL}=
\sum_{s=1}^2
\frac{g_s  b^{\tilde{d}}_{s2} C_U^s
}{(\hat{s}+\hat{u}+\hat{t}_{24}-M_{\chap{1}}^2-M_{\neu{2}}^2) (\hat{t}-M_{\tilde{d}_s}^2)}
}$

$\displaystyle{
f_{RG17}^{RR}=
\sum_{s=1}^2
\frac{g_s  b^{\tilde{u}}_{s2} C_V^s
}{(\hat{s}+\hat{u}+\hat{t}_{24}-M_{\chap{1}}^2-M_{\neu{2}}^2) (\hat{t}_{14}-M_{\tilde{u}_s}^2)}
}$

$\displaystyle{
f_{RG17}^{LR}=
\sum_{s=1}^2
\frac{g_s  a^{\tilde{u}}_{s2} C_V^s
}{(\hat{s}+\hat{u}+\hat{t}_{24}-M_{\chap{1}}^2-M_{\neu{2}}^2) (\hat{t}_{14}-M_{\tilde{u}_s}^2)}
}$

$\displaystyle{
f_{RG18}^{RL}=
\sum_{s=1}^2
\frac{g_s  a^{\tilde{d}}_{s2} C_U^s
}{(\hat{s}+\hat{t}+\hat{t}_{14}-M_{\chap{1}}^2-M_{\neu{2}}^2) (\hat{t}_{24}-M_{\tilde{d}_s}^2)}
}$

$\displaystyle{
f_{RG18}^{LL}=
\sum_{s=1}^2
\frac{g_s  b^{\tilde{d}}_{s2} C_U^s
}{(\hat{s}+\hat{t}+\hat{t}_{14}-M_{\chap{1}}^2-M_{\neu{2}}^2) (\hat{t}_{24}-M_{\tilde{d}_s}^2)}
}$

$\displaystyle{
f_{RG19}^{RL}=
-\sum_{s=1}^2
\frac{g_s  a^{\tilde{d}}_{s2}
C_U^s
}{(\hat{s}+\hat{t}+\hat{t}_{14}-M_{\chap{1}}^2-M_{\neu{2}}^2) (\hat{t}_{24}-M_{\tilde{d}_s}^2)}
}$

$\displaystyle{
f_{RG19}^{LL}=
-\sum_{s=1}^2
\frac{g_s  b^{\tilde{d}}_{s2}
C_U^s
}{(\hat{s}+\hat{t}+\hat{t}_{14}-M_{\chap{1}}^2-M_{\neu{2}}^2) (\hat{t}_{24}-M_{\tilde{d}_s}^2)}
}$

$\displaystyle{
f_{RG20}^{RR}=
-\sum_{s=1}^2
\frac{g_s  b^{\tilde{u}}_{s2}
C_V^s
}{(\hat{s}+\hat{t}+\hat{t}_{14}-M_{\chap{1}}^2-M_{\neu{2}}^2) (\hat{u}-M_{\tilde{u}_s}^2)}
}$

$\displaystyle{
f_{RG20}^{RL}=
-\sum_{s=1}^2
\frac{g_s  a^{\tilde{u}}_{s2}
C_V^s
}{(\hat{s}+\hat{t}+\hat{t}_{14}-M_{\chap{1}}^2-M_{\neu{2}}^2) (\hat{u}-M_{\tilde{u}_s}^2)}
}$

$\displaystyle{
f_{RG21}^{RR}=
\sum_{s=1}^2
\frac{g_s  b^{\tilde{u}}_{s2} C_V^s
}{(\hat{s}+\hat{t}+\hat{t}_{14}-M_{\chap{1}}^2-M_{\neu{2}}^2) (\hat{u}-M_{\tilde{u}_s}^2)}
}$

$\displaystyle{
f_{RG21}^{RL}=
\sum_{s=1}^2
\frac{g_s  a^{\tilde{u}}_{s2} C_V^s
}{(\hat{s}+\hat{t}+\hat{t}_{14}-M_{\chap{1}}^2-M_{\neu{2}}^2) (\hat{u}-M_{\tilde{u}_s}^2)}
}$

$\displaystyle{
f_{RG22}^{RR}=
\frac{A_R D_L g_s  (2
\hat{s}+\hat{t}+\hat{u}+\hat{t}_{14}+\hat{t}_{24}-2
(M_{\chap{1}}^2+M_{\neu{2}}^2))}{(M_W^2-\hat{s}_{34})
(\hat{s}+\hat{t}+\hat{t}_{14}-M_{\chap{1}}^2-M_{\neu{2}}^2)
(\hat{s}+\hat{u}+\hat{t}_{24}-M_{\chap{1}}^2-M_{\neu{2}}^2)}
}$

$\displaystyle{
f_{RG22}^{LR}=
\frac{A_L D_L g_s  (2
\hat{s}+\hat{t}+\hat{u}+\hat{t}_{14}+\hat{t}_{24}-2
(M_{\chap{1}}^2+M_{\neu{2}}^2))}{(M_W^2-\hat{s}_{34})
(\hat{s}+\hat{t}+\hat{t}_{14}-M_{\chap{1}}^2-M_{\neu{2}}^2)
(\hat{s}+\hat{u}+\hat{t}_{24}-M_{\chap{1}}^2-M_{\neu{2}}^2)}
}$

$\displaystyle{
f_{RG23}^{RR}=
\frac{A_R D_L g_s  (2
\hat{s}+\hat{t}+\hat{u}+\hat{t}_{14}+\hat{t}_{24}-2
(M_{\chap{1}}^2+M_{\neu{2}}^2))}{(M_W^2-\hat{s}_{34})
(\hat{s}+\hat{t}+\hat{t}_{14}-M_{\chap{1}}^2-M_{\neu{2}}^2)
(\hat{s}+\hat{u}+\hat{t}_{24}-M_{\chap{1}}^2-M_{\neu{2}}^2)}
}$

$\displaystyle{
f_{RG23}^{LR}=
\frac{A_L D_L g_s  (2
\hat{s}+\hat{t}+\hat{u}+\hat{t}_{14}+\hat{t}_{24}-2
(M_{\chap{1}}^2+M_{\neu{2}}^2))}{(M_W^2-\hat{s}_{34})
(\hat{s}+\hat{t}+\hat{t}_{14}-M_{\chap{1}}^2-M_{\neu{2}}^2)
(\hat{s}+\hat{u}+\hat{t}_{24}-M_{\chap{1}}^2-M_{\neu{2}}^2)}
}$

The standard matrix elements in Eq.(\ref{RQEQ}) for the subprocesses
\begin{equation}
u(p_1)+g(p_2)\to\chap{1}(p_3)+\neu{2}(p_4)+d(p_5),
\end{equation}
and
\begin{equation}
\bar{d}(p_1)+g(p_2)\to\chap{1}(p_3)+\neu{2}(p_4)+\bar{u}(p_5),
\end{equation}
are defined as follows:

$\displaystyle{
M_{Q1}^{ab}=\bar{u}_4 P_a u_1^m \bar{u}_5^n P_b v_3 (T^x)_{mn},
}$

$\displaystyle{
M_{Q2}^{ab}=\bar{v}_1^m P_a v_3 \bar{u}_5^n P_b v_4 (T^x)_{mn},
}$

$\displaystyle{
M_{Q3}^{ab}=\bar{u}_3 P_a u_1^m \bar{u}_5^n P_b v_4 (T^x)_{mn},
}$

$\displaystyle{
M_{Q4}^{ab}=\bar{u}_4 P_a \ps_2 v_3 \bar{u}_5^n P_b \es u_1^m (T^x)_{mn},
}$

$\displaystyle{
M_{Q5}^{ab}=\bar{u}_4 P_a \es v_3 \bar{u}_5^n P_b \ps_2 u_1^m (T^x)_{mn},
}$

$\displaystyle{
M_{Q6}^{ab}=\bar{u}_4 P_a \gamma^\mu v_3 \bar{u}_5^n P_b \gamma_\mu u_1^m (T^x)_{mn},
}$

$\displaystyle{
M_{Q7}^{ab}=\bar{u}_5^n P_a v_4 \bar{u}_3 P_b \es \ps_2 u_1^m (T^x)_{mn},
}$

$\displaystyle{
M_{Q8}^{ab}=\bar{u}_5^n P_a v_3 \bar{u}_4 P_b \es \ps_2 u_1^m (T^x)_{mn},
}$

$\displaystyle{
M_{Q9}^{ab}=\bar{u}_4 P_a u_1^m \bar{u}_5^n P_b \es \ps_2 v_3 (T^x)_{mn},
}$

$\displaystyle{
M_{Q10}^{ab}=\bar{v}_1^m P_a v_3 \bar{u}_5^n P_b \es \ps_2 v_4 (T^x)_{mn},
}$

$\displaystyle{
M_{Q11}^{ab}=\bar{u}_4 P_a \gamma^\mu v_3 \bar{u}_5^n P_b \es \ps_2 \gamma_\mu u_1^m (T^x)_{mn},
}$

$\displaystyle{
M_{Q12}^{ab}=\bar{v}_1^n P_a v_4 \bar{u}_3 P_b v_5^m (T^x)_{mn},
}$

$\displaystyle{
M_{Q13}^{ab}=\bar{v}_1^n P_a v_3 \bar{u}_4 P_b v_5^m (T^x)_{mn},
}$

$\displaystyle{
M_{Q14}^{ab}=\bar{v}_1^n P_a \ps_2 v_5^m \bar{u}_4 P_b \es v_3 (T^x)_{mn},
}$

$\displaystyle{
M_{Q15}^{ab}=\bar{v}_1^n P_a \es v_5^m \bar{u}_4 P_b \ps_2 v_3 (T^x)_{mn},
}$

$\displaystyle{
M_{Q16}^{ab}=\bar{v}_1^n P_a \gamma^\mu v_5^m \bar{u}_4 P_b \gamma_\mu v_3 (T^x)_{mn},
}$

$\displaystyle{
M_{Q17}^{ab}=\bar{u}_4 P_a v_5^m \bar{v}_1^n P_b \es \ps_2 v_3 (T^x)_{mn},
}$

$\displaystyle{
M_{Q18}^{ab}=\bar{u}_3 P_a v_5^m \bar{v}_1^n P_b \es \ps_2 v_4 (T^x)_{mn},
}$

$\displaystyle{
M_{Q19}^{ab}=\bar{v}_1^n P_a v_4 \bar{u}_3 P_b \es \ps_2 v_5^m (T^x)_{mn},
}$

$\displaystyle{
M_{Q20}^{ab}=\bar{v}_1^n P_a v_3 \bar{u}_4 P_b \es \ps_2 v_5^m (T^x)_{mn},
}$

$\displaystyle{
M_{Q21}^{ab}=\bar{u}_4 P_a \gamma^\mu v_3 \bar{v}_1^n P_b \es \ps_2 \gamma_\mu v_5^m (T^x)_{mn},
}$

The nonzero form factors in Eq.(\ref{RQEQ}) are the following:

$\displaystyle{
f_{RQ1}^{RR}=
\sum_{s=1}^2
\frac{2 g_s  b^{\tilde{u}}_{s2} C_V^s
}{\hat{t}_{14}-M_{\tilde{u}_s}^2}
\left(\frac{-[\epsilon^x\cdot(p_1-p_4)] \hat{s}-(\epsilon^x\cdot p_1)
\hat{t}_{14}+(\epsilon^x\cdot p_1) M_{\tilde{u}_s}^2}{\hat{s}
(\hat{t}+\hat{u}+\hat{s}_{34}-2
M_{\chap{1}}^2-M_{\neu{2}}^2+M_{\tilde{u}_s}^2)}
-\frac{[\epsilon^x\cdot(p_1-p_3-p_4)]}{\hat{s}+\hat{u}+\hat{t}_{24}
-M_{\chap{1}}^2-M_{\neu{2}}^2}\right)
}$

$\displaystyle{
f_{RQ1}^{LR}=
\sum_{s=1}^2
\frac{2 g_s  a^{\tilde{u}}_{s2} C_V^s
}{\hat{t}_{14}-M_{\tilde{u}_s}^2}
\left(\frac{-[\epsilon^x\cdot(p_1-p_4)] \hat{s}-(\epsilon^x\cdot p_1)
\hat{t}_{14}+(\epsilon^x\cdot p_1) M_{\tilde{u}_s}^2}{\hat{s}
(\hat{t}+\hat{u}+\hat{s}_{34}-2
M_{\chap{1}}^2-M_{\neu{2}}^2+M_{\tilde{u}_s}^2)}
-\frac{[\epsilon^x\cdot(p_1-p_3-p_4)]}{\hat{s}+\hat{u}+\hat{t}_{24}
-M_{\chap{1}}^2-M_{\neu{2}}^2}\right)
}$

$\displaystyle{
f_{RQ2}^{LR}=
\sum_{s=1}^2
\frac{2 g_s  a^{\tilde{d}}_{s2} C_U^s
}{\hat{t}-M_{\tilde{d}_s}^2}
\left(-\frac{[\epsilon^x\cdot(p_1-p_3-p_4)]}{\hat{s}+\hat{u}+\hat{t}_{24}
-M_{\chap{1}}^2-M_{\neu{2}}^2}-\frac{[\epsilon^x\cdot(p_1-p_3)]}{\hat{s}_{34}
+\hat{t}_{14}+\hat{t}_{24}-M_{\chap{1}}^2
-2 M_{\neu{2}}^2+M_{\tilde{d}_s}^2}\right)
}$

$\displaystyle{
f_{RQ2}^{LL}=
\sum_{s=1}^2
\frac{2 g_s  b^{\tilde{d}}_{s2} C_U^s
}{\hat{t}-M_{\tilde{d}_s}^2}
\left(-\frac{[\epsilon^x\cdot(p_1-p_3-p_4)]}{\hat{s}+\hat{u}+\hat{t}_{24}
-M_{\chap{1}}^2-M_{\neu{2}}^2}-\frac{[\epsilon^x\cdot(p_1-p_3)]}{\hat{s}_{34}
+\hat{t}_{14}+\hat{t}_{24}-M_{\chap{1}}^2-2
M_{\neu{2}}^2+M_{\tilde{d}_s}^2}\right)
}$

$\displaystyle{
f_{RQ3}^{LR}=
\sum_{s=1}^2
\frac{2 g_s (\epsilon^x\cdot p_1)
a^{\tilde{d}}_{s2} C_U^s
}{\hat{s}
(\hat{s}_{34}+\hat{t}_{14}+\hat{t}_{24}-M_{\chap{1}}^2-2
M_{\neu{2}}^2+M_{\tilde{d}_s}^2)}
}$

$\displaystyle{
f_{RQ3}^{LL}=
\sum_{s=1}^2
\frac{2 g_s (\epsilon^x\cdot p_1)
b^{\tilde{d}}_{s2} C_U^s
}{\hat{s}
(\hat{s}_{34}+\hat{t}_{14}+\hat{t}_{24}-M_{\chap{1}}^2-2
M_{\neu{2}}^2+M_{\tilde{d}_s}^2)}
}$

$\displaystyle{
f_{RQ4}^{RR}=
\frac{2 A_R D_L g_s }{\hat{s}
(M_W^2-\hat{s}_{34})
}
}$

$\displaystyle{
f_{RQ4}^{LR}=
\frac{2 A_L D_L g_s }{\hat{s}
(M_W^2-\hat{s}_{34})
}
}$

$\displaystyle{
f_{RQ5}^{RR}=
\frac{2 A_R D_L g_s }{\hat{s}
(\hat{s}_{34}-M_W^2)
}
}$

$\displaystyle{
f_{RQ5}^{LR}=
\frac{2 A_L D_L g_s }{\hat{s}
(\hat{s}_{34}-M_W^2)
}
}$

$\displaystyle{
f_{RQ6}^{RR}=
\frac{2 A_R D_L g_s
}{\hat{s}_{34}-M_W^2}
\left(\frac{[\epsilon^x\cdot(p_1-p_3-p_4)]}{\hat{s}+\hat{u}+\hat{t}_{24}
-M_{\chap{1}}^2-M_{\neu{2}}^2}
-\frac{(\epsilon^x\cdot p_1)
}{\hat{s}}\right)
}$

$\displaystyle{
f_{RQ6}^{LR}=
\frac{2 A_L D_L g_s
}{\hat{s}_{34}-M_W^2}
\left(\frac{[\epsilon^x\cdot(p_1-p_3-p_4)]}{\hat{s}+\hat{u}+\hat{t}_{24}
-M_{\chap{1}}^2-M_{\neu{2}}^2}
-\frac{(\epsilon^x\cdot p_1)}{\hat{s}}\right)
}$

$\displaystyle{
f_{RQ7}^{RL}=
-\sum_{s=1}^2
\frac{g_s  a^{\tilde{d}}_{s2}
C_U^s }{\hat{s}
(\hat{s}_{34}+\hat{t}_{14}+\hat{t}_{24}-M_{\chap{1}}^2-2
M_{\neu{2}}^2+M_{\tilde{d}_s}^2)}
}$

$\displaystyle{
f_{RQ7}^{LL}=
-\sum_{s=1}^2
\frac{g_s  b^{\tilde{d}}_{s2}
C_U^s
}{\hat{s}
(\hat{s}_{34}+\hat{t}_{14}+\hat{t}_{24}-M_{\chap{1}}^2-2
M_{\neu{2}}^2+M_{\tilde{d}_s}^2)}
}$

$\displaystyle{
f_{RQ8}^{RR}=
\sum_{s=1}^2
\frac{g_s  b^{\tilde{u}}_{s2} C_V^s
}{\hat{s}
(\hat{t}+\hat{u}+\hat{s}_{34}-2
M_{\chap{1}}^2-M_{\neu{2}}^2+M_{\tilde{u}_s}^2)}
}$

$\displaystyle{
f_{RQ8}^{RL}=
\sum_{s=1}^2
\frac{g_s  a^{\tilde{u}}_{s2} C_V^s }{\hat{s}
(\hat{t}+\hat{u}+\hat{s}_{34}-2
M_{\chap{1}}^2-M_{\neu{2}}^2+M_{\tilde{u}_s}^2)}
}$

$\displaystyle{
f_{RQ9}^{RR}=
\sum_{s=1}^2
\frac{g_s  b^{\tilde{u}}_{s2}
C_V^s
}{(\hat{s}+\hat{u}+\hat{t}_{24}
-M_{\chap{1}}^2-M_{\neu{2}}^2)
(\hat{t}_{14}-M_{\tilde{u}_s}^2)}
}$

$\displaystyle{
f_{RQ9}^{LR}=
\sum_{s=1}^2
\frac{g_s  a^{\tilde{u}}_{s2}
C_V^s }{(\hat{s}+\hat{u}+\hat{t}_{24}
-M_{\chap{1}}^2-M_{\neu{2}}^2)
(\hat{t}_{14}-M_{\tilde{u}_s}^2)}
}$

$\displaystyle{
f_{RQ10}^{LR}=
\sum_{s=1}^2
\frac{g_s  a^{\tilde{d}}_{s2}
C_U^s
}{(\hat{s}+\hat{u}+\hat{t}_{24}
-M_{\chap{1}}^2-M_{\neu{2}}^2)
(\hat{t}-M_{\tilde{d}_s}^2)}
}$

$\displaystyle{
f_{RQ10}^{LL}=
\sum_{s=1}^2
\frac{g_s  b^{\tilde{d}}_{s2}
C_U^s
}{(\hat{s}+\hat{u}+\hat{t}_{24}
-M_{\chap{1}}^2-M_{\neu{2}}^2)
(\hat{t}-M_{\tilde{d}_s}^2)}
}$

$\displaystyle{
f_{RQ11}^{RR}=
-\frac{A_R D_L g_s
(\hat{u}+\hat{t}_{24}-M_{\chap{1}}^2-M_{\neu{2}}^2)
}{\hat{s} (M_W^2-\hat{s}_{34})
(\hat{s}+\hat{u}+\hat{t}_{24}-M_{\chap{1}}^2
-M_{\neu{2}}^2)}
}$

$\displaystyle{
f_{RQ11}^{LR}=
-\frac{A_L D_L g_s
(\hat{u}+\hat{t}_{24}-M_{\chap{1}}^2-M_{\neu{2}}^2)
}{\hat{s} (M_W^2-\hat{s}_{34})
(\hat{s}+\hat{u}+\hat{t}_{24}-M_{\chap{1}}^2
-M_{\neu{2}}^2)}
}$

$\displaystyle{
f_{RQ12}^{RL}=
\sum_{s=1}^2
\frac{2 g_s  a^{\tilde{d}}_{s2} C_U^s
}{\hat{t}_{14}-M_{\tilde{d}_s}^2}
\left(\frac{[\epsilon^x\cdot(p_1-p_3-p_4)]}{\hat{s}+\hat{u}+\hat{t}_{24}
-M_{\chap{1}}^2-M_{\neu{2}}^2}+\frac{[\epsilon^x\cdot(p_1-p_4)] \hat{s}+(\epsilon^x\cdot p_1)
\hat{t}_{14}-(\epsilon^x\cdot p_1) M_{\tilde{d}_s}^2}{\hat{s}
(\hat{t}+\hat{u}+\hat{s}_{34}-2
M_{\chap{1}}^2-M_{\neu{2}}^2+M_{\tilde{d}_s}^2)}\right)
}$

$\displaystyle{
f_{RQ12}^{LL}=
\sum_{s=1}^2
\frac{2 g_s  b^{\tilde{d}}_{s2} C_U^s
}{\hat{t}_{14}-M_{\tilde{d}_s}^2}
\left(\frac{[\epsilon^x\cdot(p_1-p_3-p_4)]}{\hat{s}+\hat{u}+\hat{t}_{24}
-M_{\chap{1}}^2-M_{\neu{2}}^2}+\frac{[\epsilon^x\cdot(p_1-p_4)] \hat{s}+(\epsilon^x\cdot p_1)
\hat{t}_{14}-(\epsilon^x\cdot p_1) M_{\tilde{d}_s}^2}{\hat{s}
(\hat{t}+\hat{u}+\hat{s}_{34}-2
M_{\chap{1}}^2-M_{\neu{2}}^2+M_{\tilde{d}_s}^2)}\right)
}$

$\displaystyle{
f_{RQ13}^{RR}=
\sum_{s=1}^2
\frac{2 g_s  b^{\tilde{u}}_{s2} C_V^s
}{\hat{t}-M_{\tilde{u}_s}^2}
\left(\frac{-[\epsilon^x\cdot(p_1-p_3)] \hat{s}-\hat{t}
(\epsilon^x\cdot p_1)+(\epsilon^x\cdot p_1) M_{\tilde{u}_s}^2}{\hat{s}
(\hat{s}_{34}+\hat{t}_{14}+\hat{t}_{24}-M_{\chap{1}}^2-2
M_{\neu{2}}^2+M_{\tilde{u}_s}^2)}-\frac{[\epsilon^x\cdot(p_1-p_3-p_4)]
}{\hat{s}+\hat{u}+\hat{t}_{24}-M_{\chap{1}}^2-M_{\neu{2}}^2}\right)
}$

$\displaystyle{
f_{RQ13}^{RL}=
\sum_{s=1}^2
\frac{2 g_s  a^{\tilde{u}}_{s2} C_V^s
}{\hat{t}-M_{\tilde{u}_s}^2}
\left(\frac{-[\epsilon^x\cdot(p_1-p_3)] \hat{s}-\hat{t}
(\epsilon^x\cdot p_1)+(\epsilon^x\cdot p_1) M_{\tilde{u}_s}^2}{\hat{s}
(\hat{s}_{34}+\hat{t}_{14}+\hat{t}_{24}-M_{\chap{1}}^2-2
M_{\neu{2}}^2+M_{\tilde{u}_s}^2)}-\frac{[\epsilon^x\cdot(p_1-p_3-p_4)]
}{\hat{s}+\hat{u}+\hat{t}_{24}-M_{\chap{1}}^2-M_{\neu{2}}^2}\right)
}$

$\displaystyle{
f_{RQ14}^{RR}=
-\frac{2 A_R D_L g_s }{(M_W^2-\hat{s}_{34})
(\hat{s}+\hat{u}+\hat{t}_{24}-M_{\chap{1}}^2-M_{\neu{2}}^2)}
}$

$\displaystyle{
f_{RQ14}^{RL}=
-\frac{2 A_L D_L g_s }{(M_W^2-\hat{s}_{34})
(\hat{s}+\hat{u}+\hat{t}_{24}-M_{\chap{1}}^2-M_{\neu{2}}^2)}
}$

$\displaystyle{
f_{RQ15}^{RR}=
\frac{2 A_R D_L g_s }{(M_W^2-\hat{s}_{34})
(\hat{s}+\hat{u}+\hat{t}_{24}-M_{\chap{1}}^2-M_{\neu{2}}^2)}
}$

$\displaystyle{
f_{RQ15}^{RL}=
\frac{2 A_L D_L g_s }{(M_W^2-\hat{s}_{34})
(\hat{s}+\hat{u}+\hat{t}_{24}-M_{\chap{1}}^2-M_{\neu{2}}^2)}
}$

$\displaystyle{
f_{RQ16}^{RR}=
\frac{2 A_R D_L g_s
}{M_W^2-\hat{s}_{34}}
\left(\frac{(\epsilon^x\cdot p_1)}{\hat{s}}-\frac{[\epsilon^x\cdot(p_1-p_3-p_4)]}{\hat{s}
+\hat{u}+\hat{t}_{24}-M_{\chap{1}}^2-M_{\neu{2}}^2}\right)
}$

$\displaystyle{
f_{RQ16}^{RL}=
\frac{2 A_L D_L g_s
}{M_W^2-\hat{s}_{34}}
\left(\frac{(\epsilon^x\cdot p_1)}{\hat{s}}-\frac{[\epsilon^x\cdot(p_1-p_3-p_4)]}{\hat{s}
+\hat{u}+\hat{t}_{24}-M_{\chap{1}}^2-M_{\neu{2}}^2}\right)
}$

$\displaystyle{
f_{RQ17}^{RR}=
-\sum_{s=1}^2
\frac{g_s  b^{\tilde{u}}_{s2} C_V^s }{\hat{s}
(\hat{s}_{34}+\hat{t}_{14}+\hat{t}_{24}-M_{\chap{1}}^2-2
M_{\neu{2}}^2+M_{\tilde{u}_s}^2)}
}$

$\displaystyle{
f_{RQ17}^{LR}=
-\sum_{s=1}^2
\frac{g_s  a^{\tilde{u}}_{s2} C_V^s }{\hat{s}
(\hat{s}_{34}+\hat{t}_{14}+\hat{t}_{24}-M_{\chap{1}}^2-2
M_{\neu{2}}^2+M_{\tilde{u}_s}^2)}
}$

$\displaystyle{
f_{RQ18}^{LR}=
\sum_{s=1}^2
\frac{g_s  a^{\tilde{d}}_{s2} C_U^s
}{\hat{s}
(\hat{t}+\hat{u}+\hat{s}_{34}-2
M_{\chap{1}}^2-M_{\neu{2}}^2+M_{\tilde{d}_s}^2)}
}$

$\displaystyle{
f_{RQ18}^{LL}=
\sum_{s=1}^2
\frac{g_s  b^{\tilde{d}}_{s2} C_U^s
}{\hat{s}
(\hat{t}+\hat{u}+\hat{s}_{34}-2
M_{\chap{1}}^2-M_{\neu{2}}^2+M_{\tilde{d}_s}^2)}
}$

$\displaystyle{
f_{RQ19}^{RL}=
\sum_{s=1}^2
\frac{g_s  a^{\tilde{d}}_{s2} C_U^s
}{(\hat{s}+\hat{u}+\hat{t}_{24}
-M_{\chap{1}}^2-M_{\neu{2}}^2) (\hat{t}_{14}-M_{\tilde{d}_s}^2)}
}$

$\displaystyle{
f_{RQ19}^{LL}=
\sum_{s=1}^2
\frac{g_s  b^{\tilde{d}}_{s2} C_U^s
}{(\hat{s}+\hat{u}+\hat{t}_{24}
-M_{\chap{1}}^2-M_{\neu{2}}^2) (\hat{t}_{14}-M_{\tilde{d}_s}^2)}
}$

$\displaystyle{
f_{RQ20}^{RR}=
-\sum_{s=1}^2
\frac{g_s  b^{\tilde{u}}_{s2}
C_V^s
}{(\hat{s}+\hat{u}+\hat{t}_{24}
-M_{\chap{1}}^2-M_{\neu{2}}^2) (\hat{t}-M_{\tilde{u}_s}^2)}
}$

$\displaystyle{
f_{RQ20}^{RL}=
-\sum_{s=1}^2
\frac{g_s  a^{\tilde{u}}_{s2}
C_V^s
}{(\hat{s}+\hat{u}+\hat{t}_{24}
-M_{\chap{1}}^2-M_{\neu{2}}^2) (\hat{t}-M_{\tilde{u}_s}^2)}
}$

$\displaystyle{
f_{RQ21}^{RR}=
\frac{A_R D_L g_s
(\hat{u}+\hat{t}_{24}-M_{\chap{1}}^2-M_{\neu{2}}^2)}{\hat{s}
(M_W^2-\hat{s}_{34})
(\hat{s}+\hat{u}+\hat{t}_{24}-M_{\chap{1}}^2-M_{\neu{2}}^2)}
}$

$\displaystyle{
f_{RQ21}^{LR}=
\frac{A_L D_L g_s
(\hat{u}+\hat{t}_{24}-M_{\chap{1}}^2-M_{\neu{2}}^2)}{\hat{s}
(M_W^2-\hat{s}_{34})
(\hat{s}+\hat{u}+\hat{t}_{24}-M_{\chap{1}}^2-M_{\neu{2}}^2)}
}$

\bibliography{gaugino}

\begin{figure}[!hp]
\includegraphics[width=0.8\textwidth]{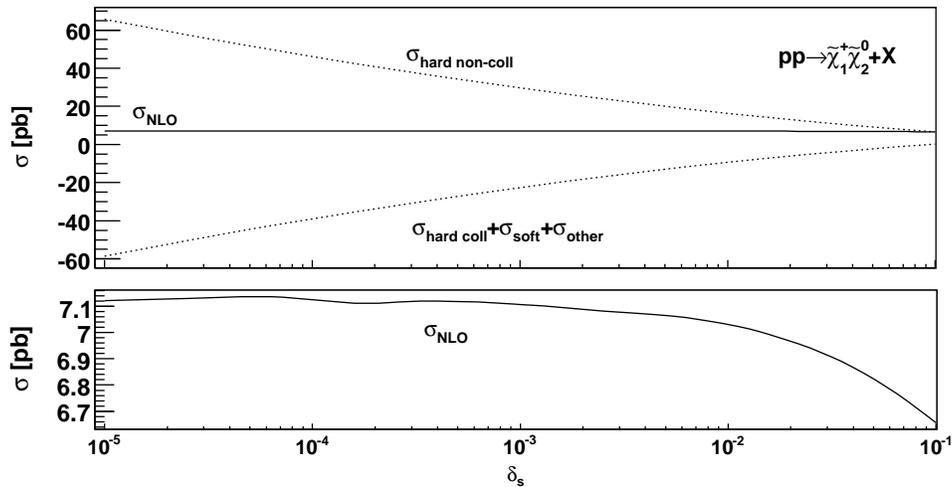}
\caption{The dependence of the total cross sections for the associated
production of $\chap{1}\neu{2}$ at the LHC on the cutoff $\delta_s$,
assuming $m_0=200$GeV, $m_{1/2}=150$GeV, $\tan\beta=5$ and
$\delta_c=\delta_s/100$. } \label{CUTDEP}
\end{figure}

\begin{figure}[!hp]
\includegraphics[width=0.8\textwidth]{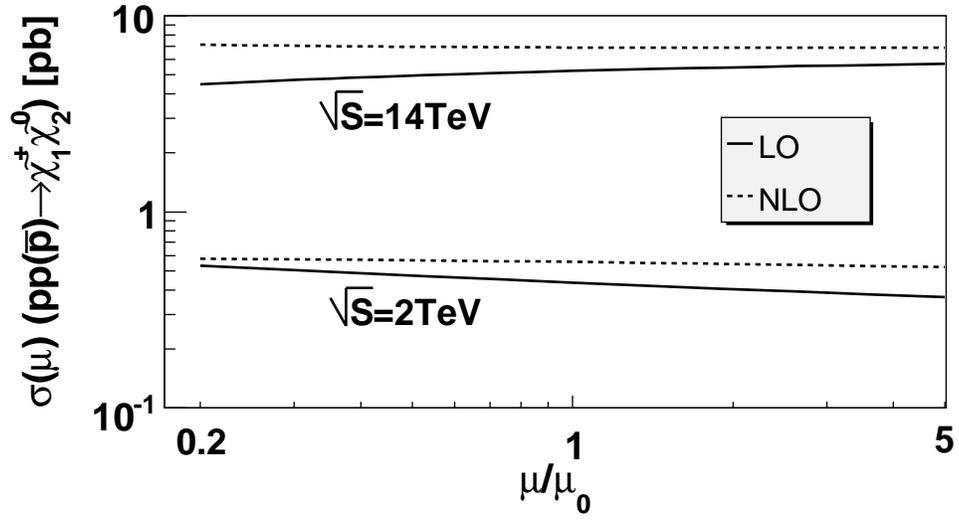}
\caption{The dependence of the total cross sections on the
renormalization/factorization scale with the same parameters chosen
as in Fig.2 of Ref.\cite{Beenakker:1999xh}. } \label{PLEHN}
\end{figure}

\begin{figure}[!hp]
\includegraphics[width=0.8\textwidth]{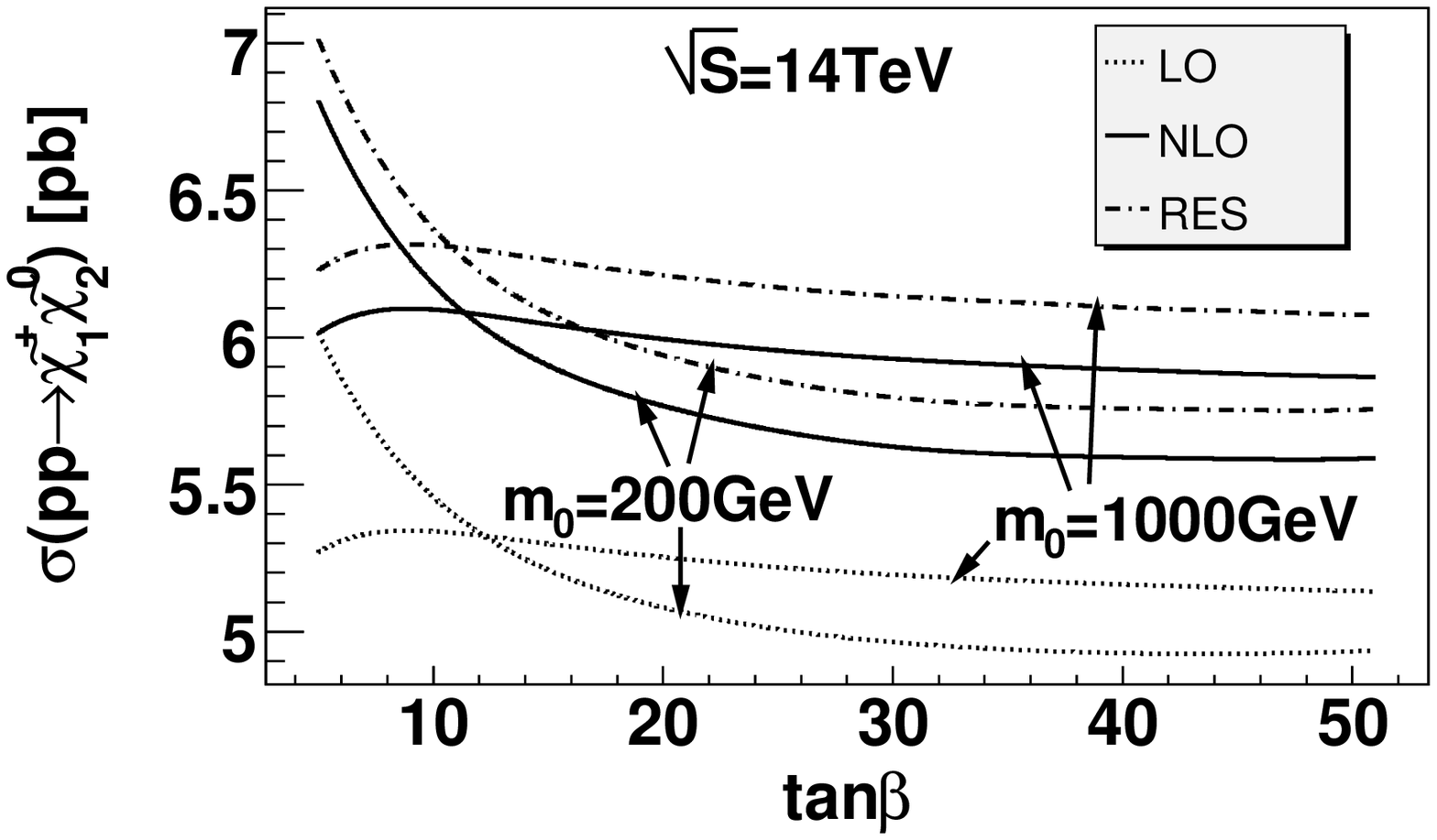}
\includegraphics[width=0.8\textwidth]{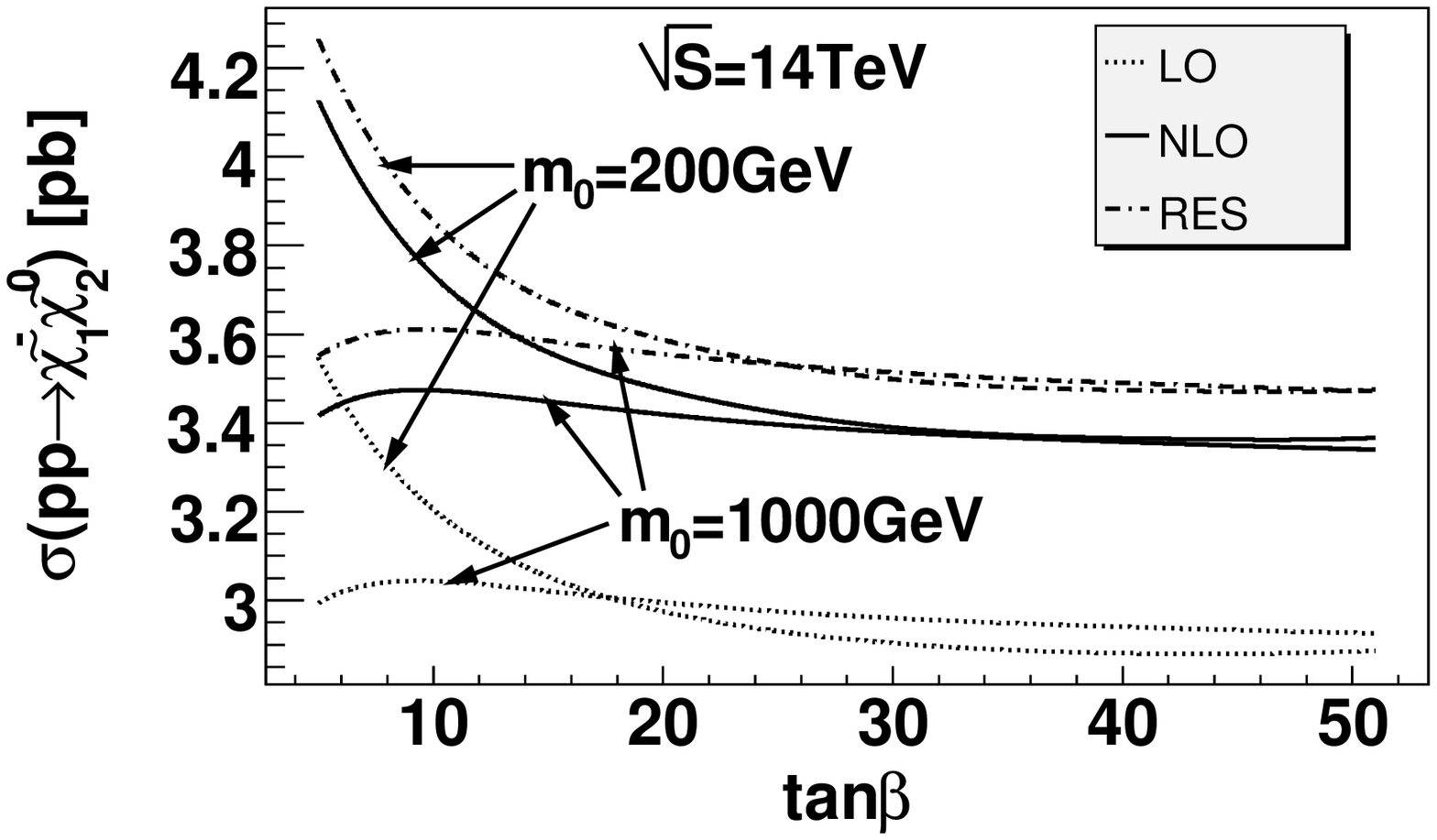}
\includegraphics[width=0.8\textwidth]{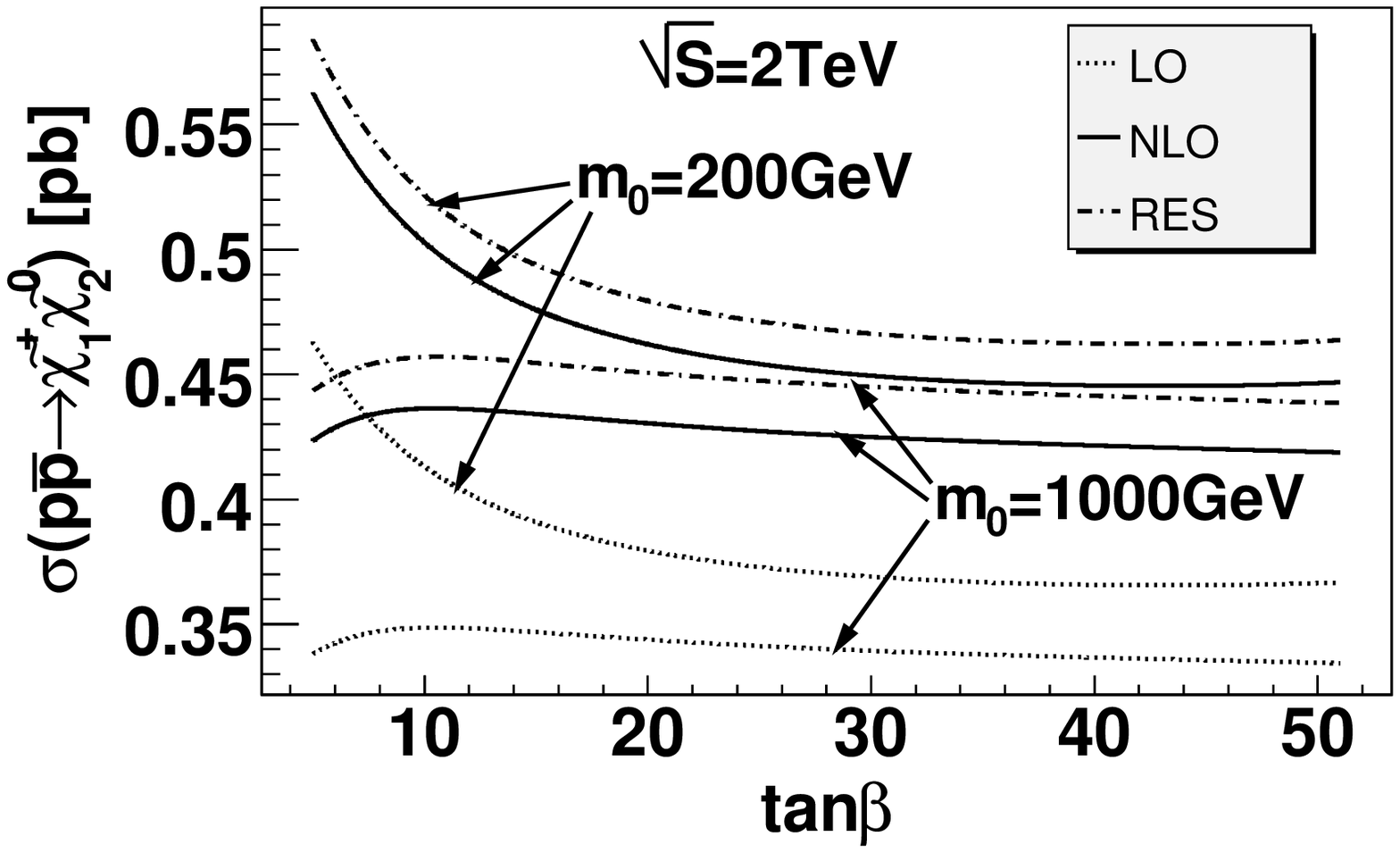}
\caption{The total cross sections as a function of $\tan\beta$ for
the associated production of $\cha{1}\neu{2}$ at the two colliders
assuming $m_{1/2}=150$GeV, $m_0=200$GeV and $1000$GeV, $A_0=0$ and
$\mu>0$.} \label{TBDEP}
\end{figure}

\begin{figure}[ht!]
\includegraphics[width=0.8\textwidth]{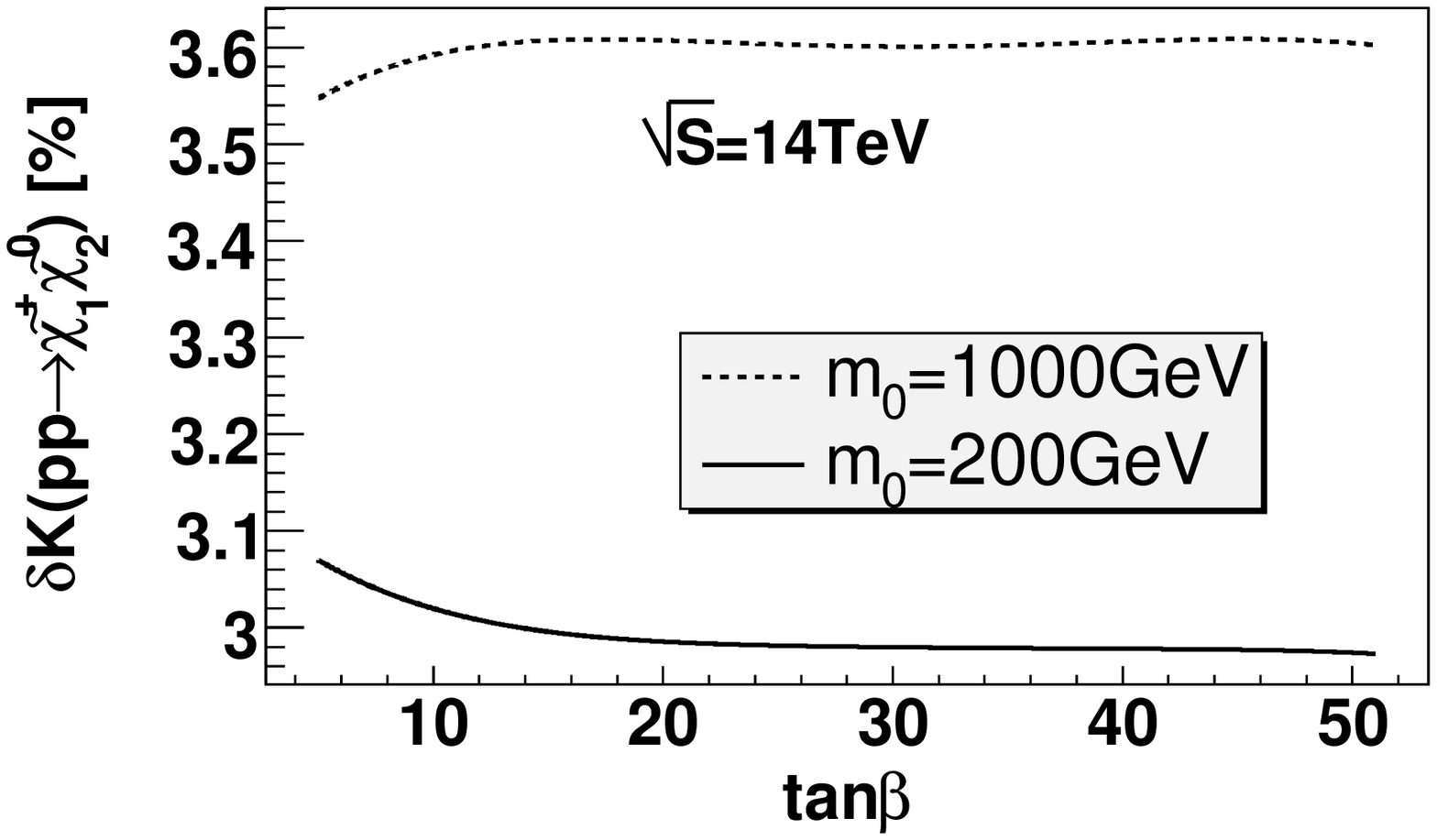}
\includegraphics[width=0.8\textwidth]{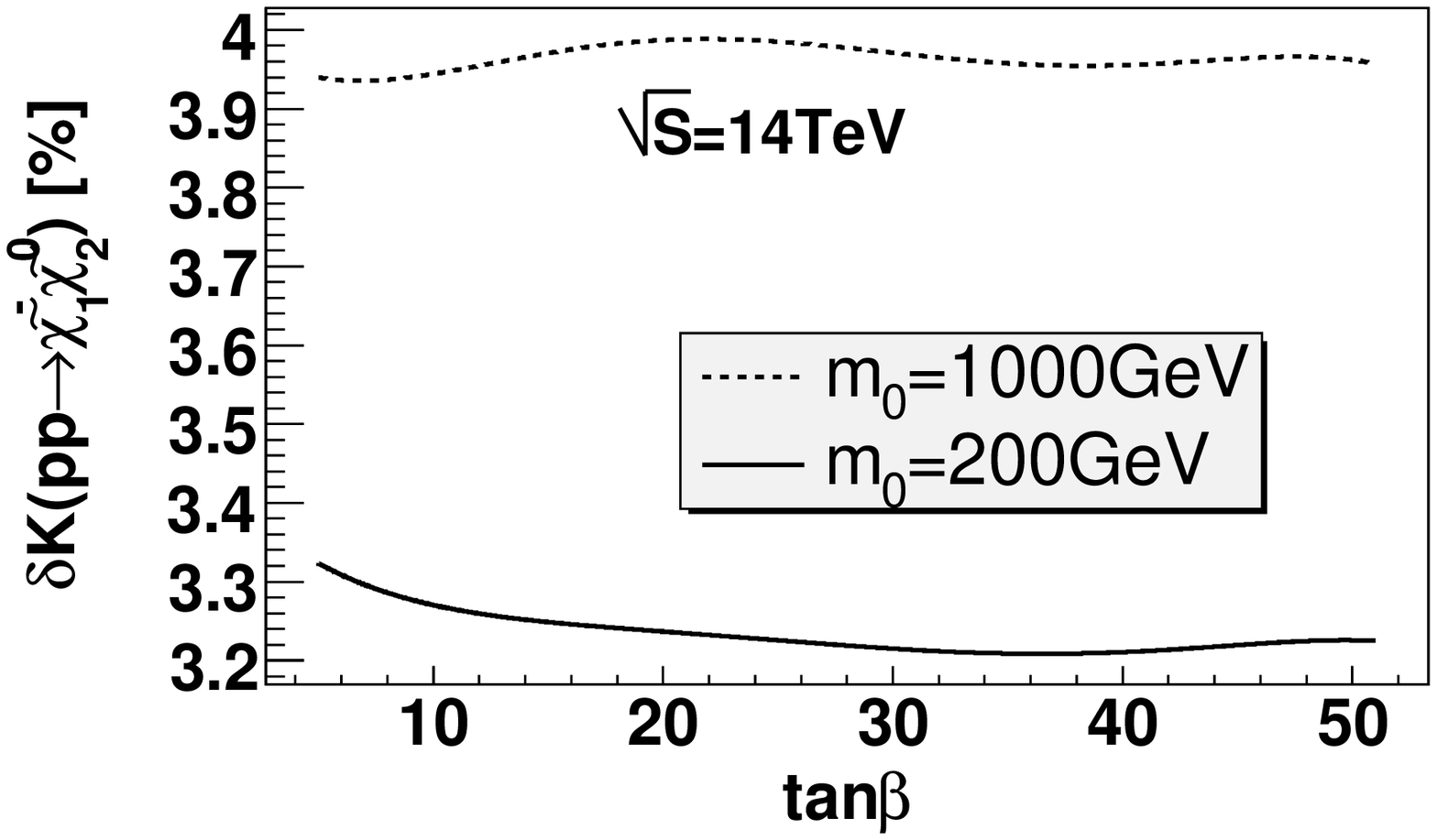}
\includegraphics[width=0.8\textwidth]{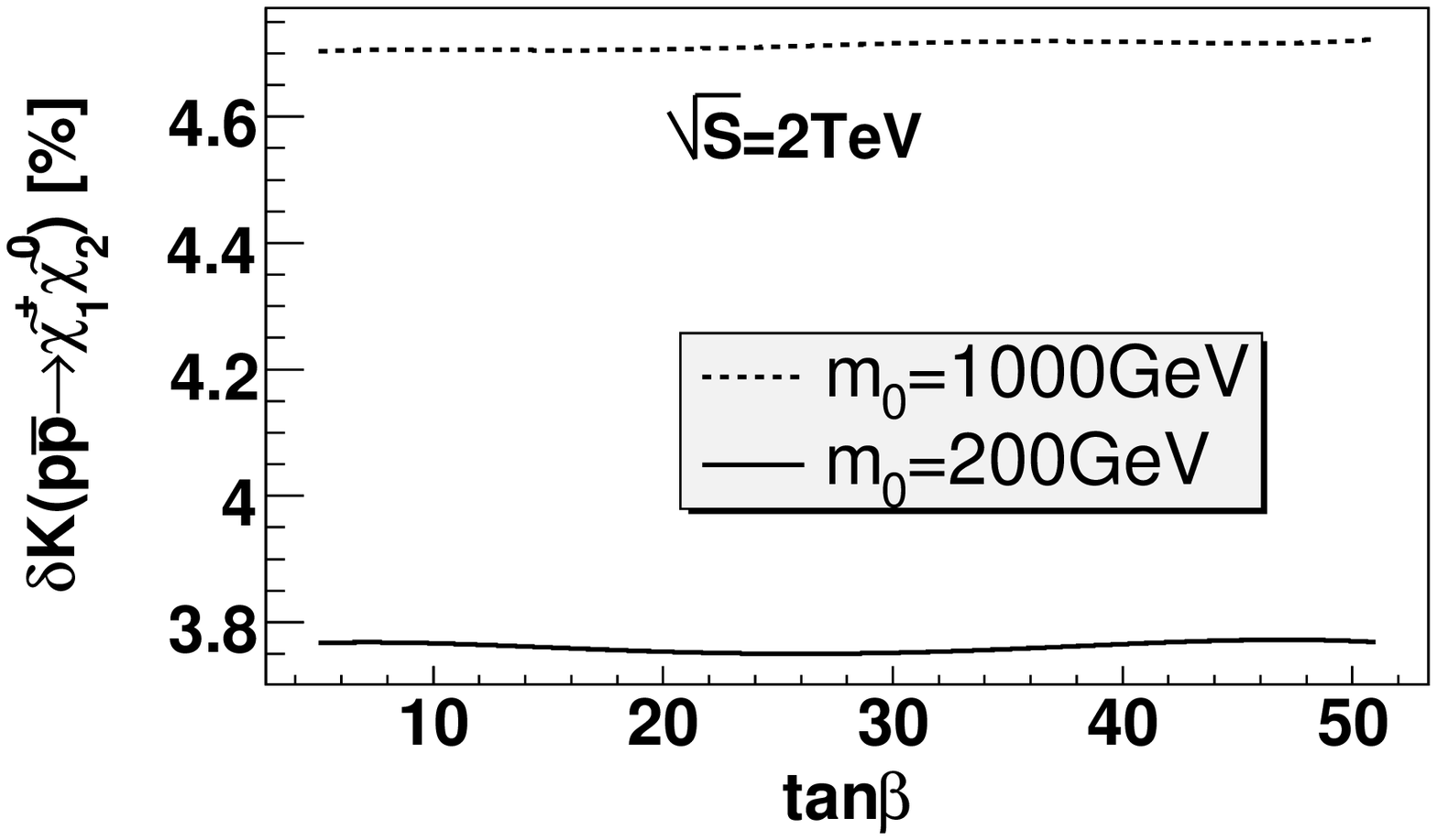}
\caption{$\delta K$, defined as $\delta
K=(\sigma^{RES}-\sigma^{NLO})/\sigma^{NLO}$, as a function of
$\tan\beta$ for the associated production of $\cha{1}\neu{2}$ at the
two colliders assuming $m_{1/2}=150$GeV, $m_0=200$GeV and
$1000$GeV, $A_0=0$ and $\mu>0$.} \label{RELTBDEP}
\end{figure}

\begin{figure}[ht!]
\includegraphics[width=0.8\textwidth]{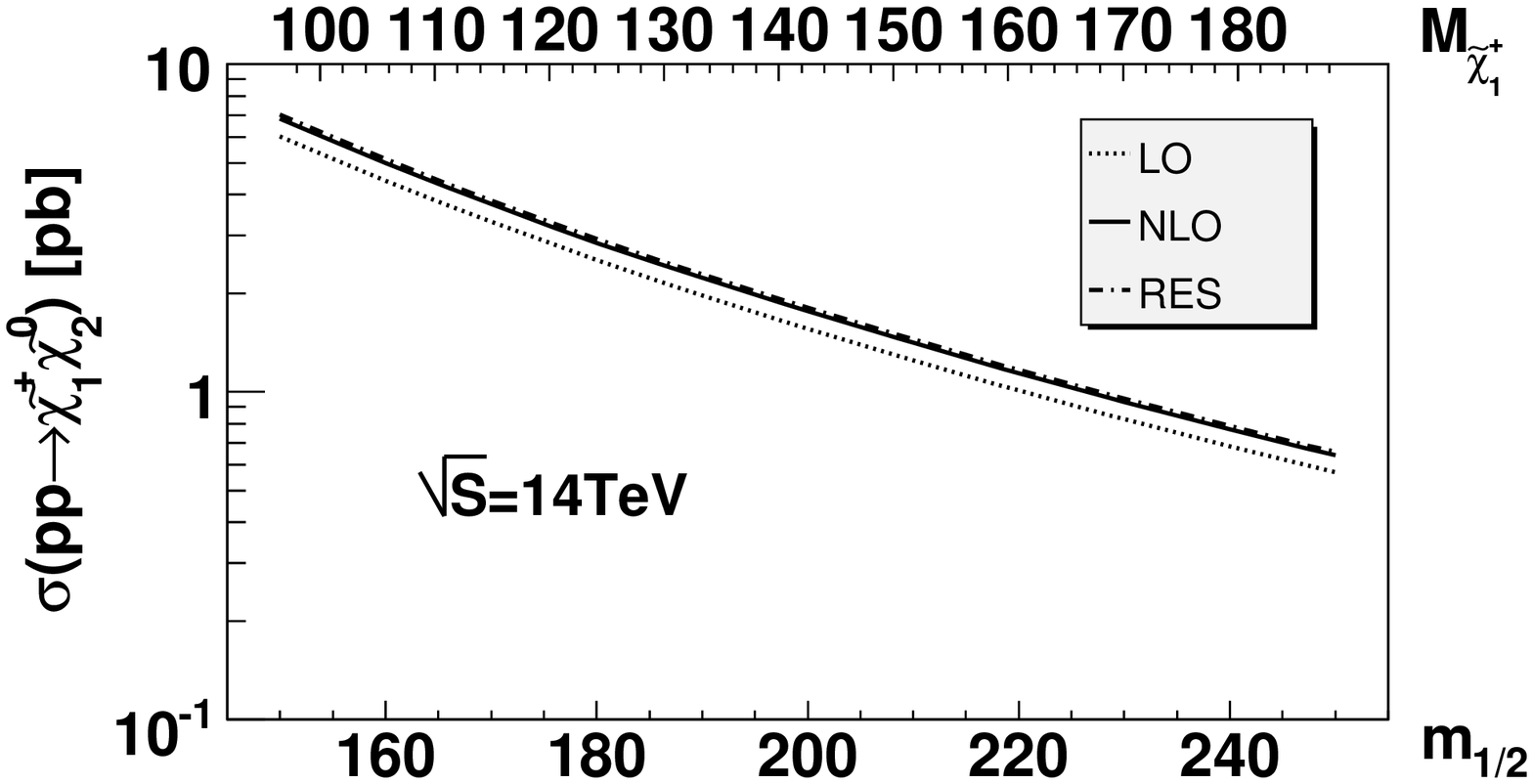}
\includegraphics[width=0.8\textwidth]{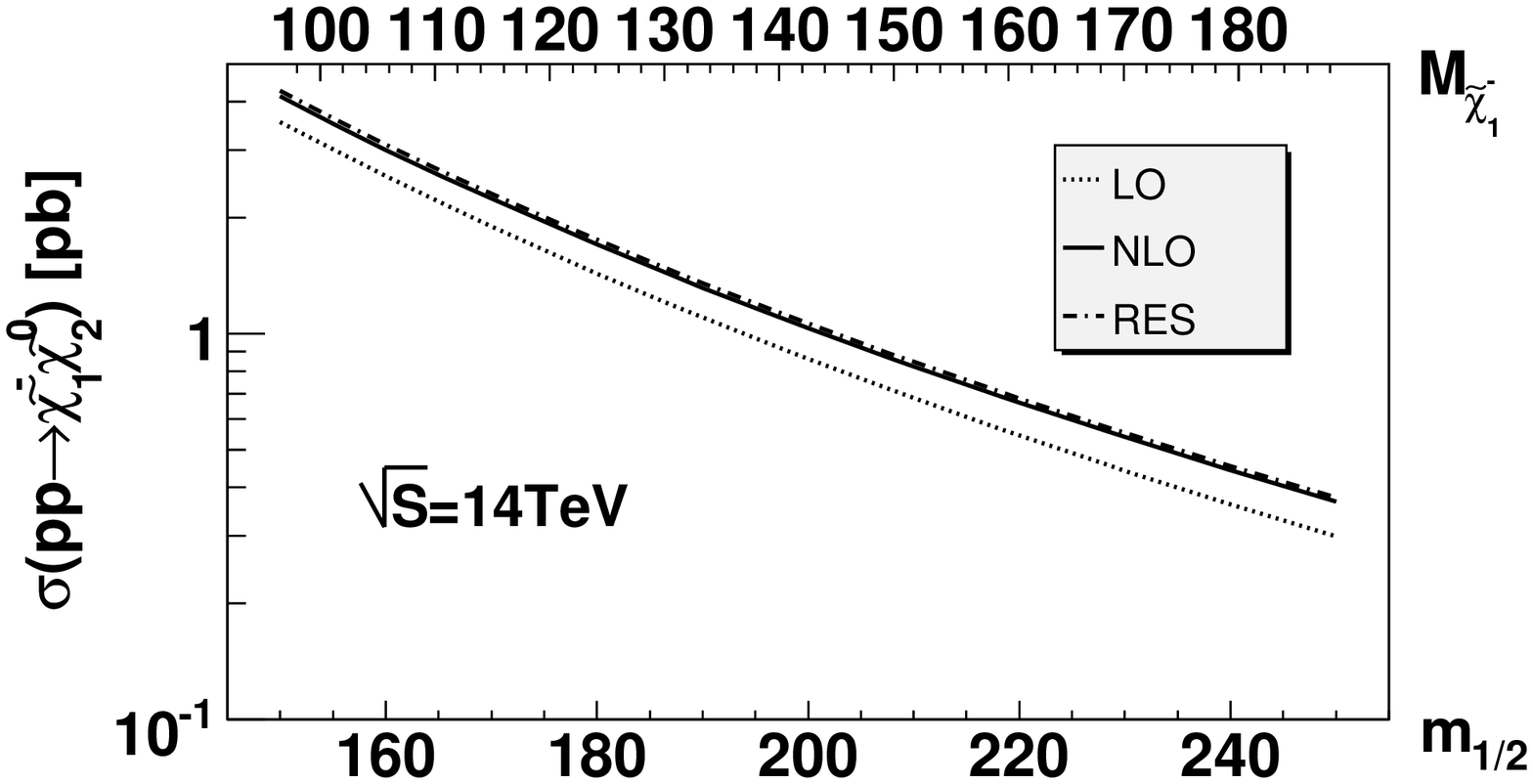}
\includegraphics[width=0.8\textwidth]{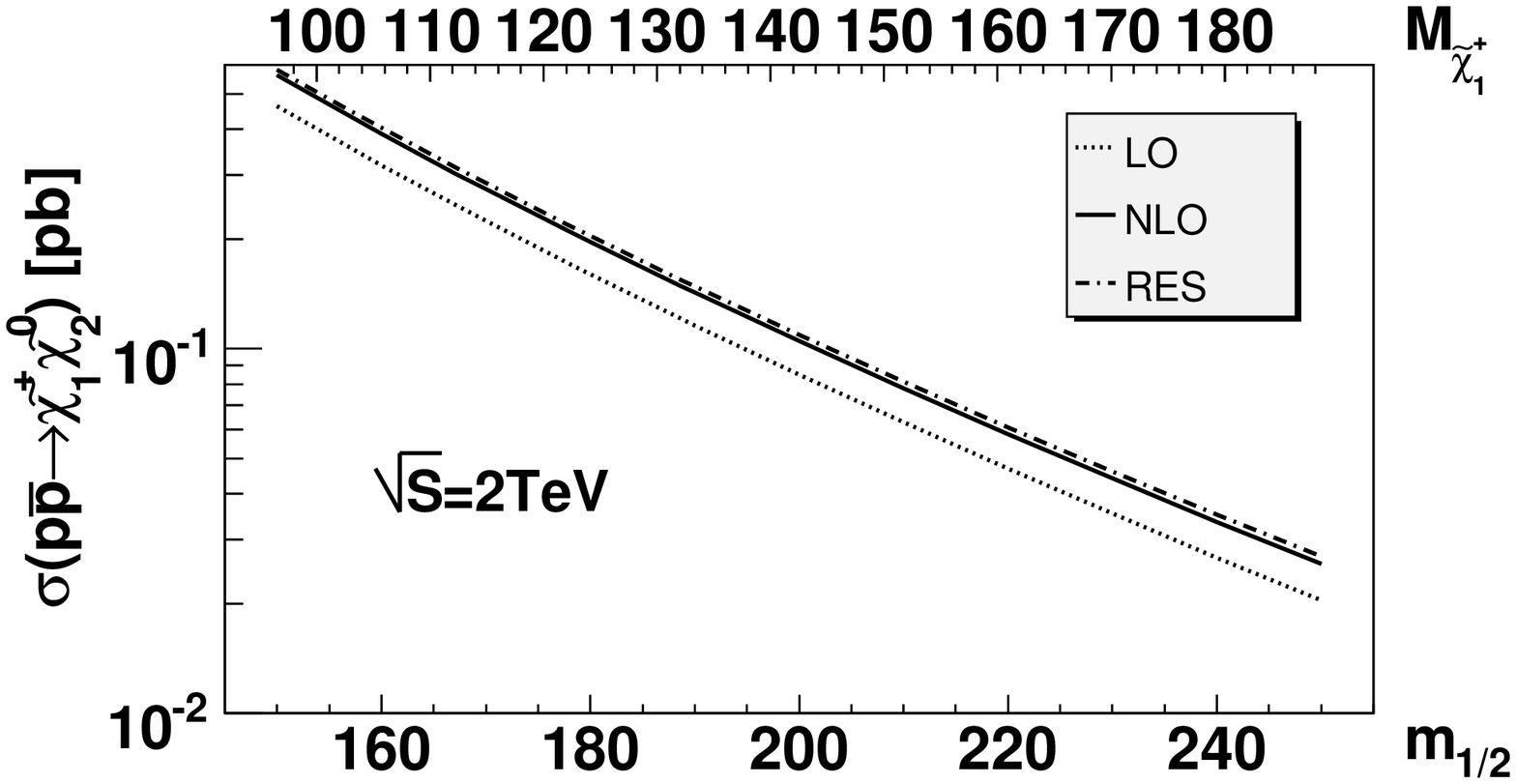}
\caption{The total cross sections as a function of $m_{1/2}$ for the
associated production of $\cha{1}\neu{2}$ at the two colliders
assuming $m_0=200$GeV, $\tan\beta=5$, $A_0=0$ and $\mu>0$.}
\label{MHFDEP}
\end{figure}

\begin{figure}[ht!]
\includegraphics[width=0.8\textwidth]{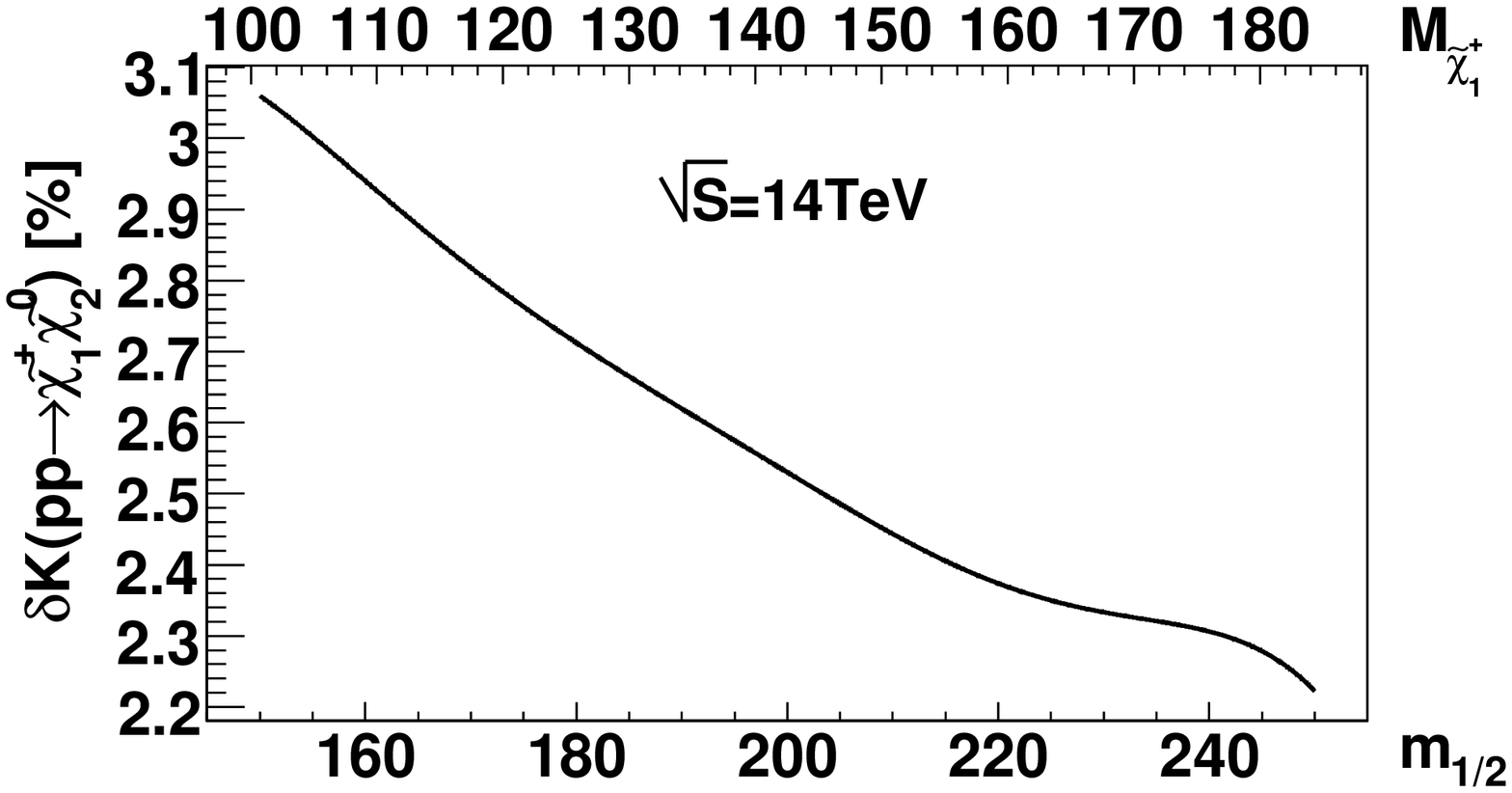}
\includegraphics[width=0.8\textwidth]{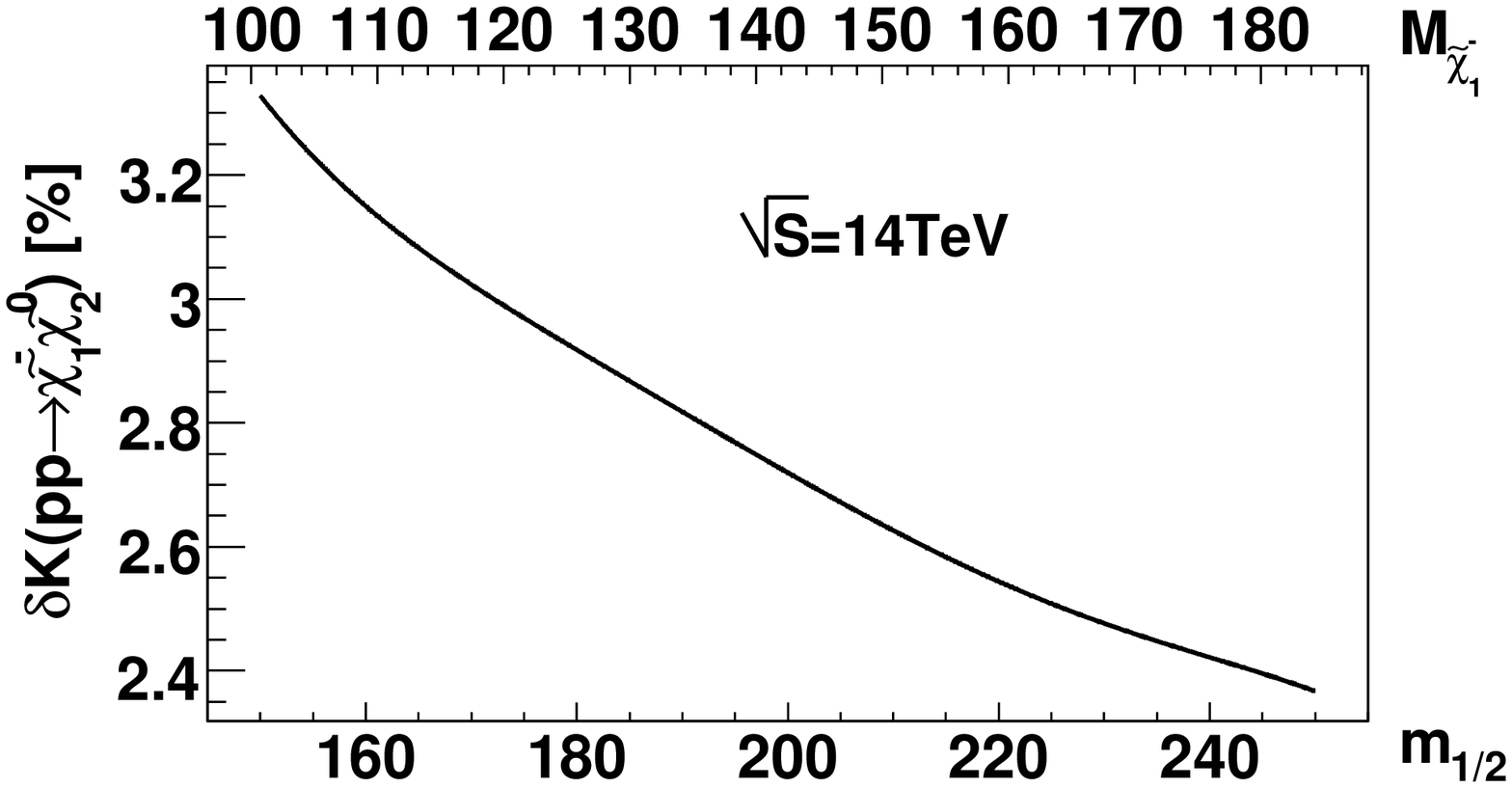}
\includegraphics[width=0.8\textwidth]{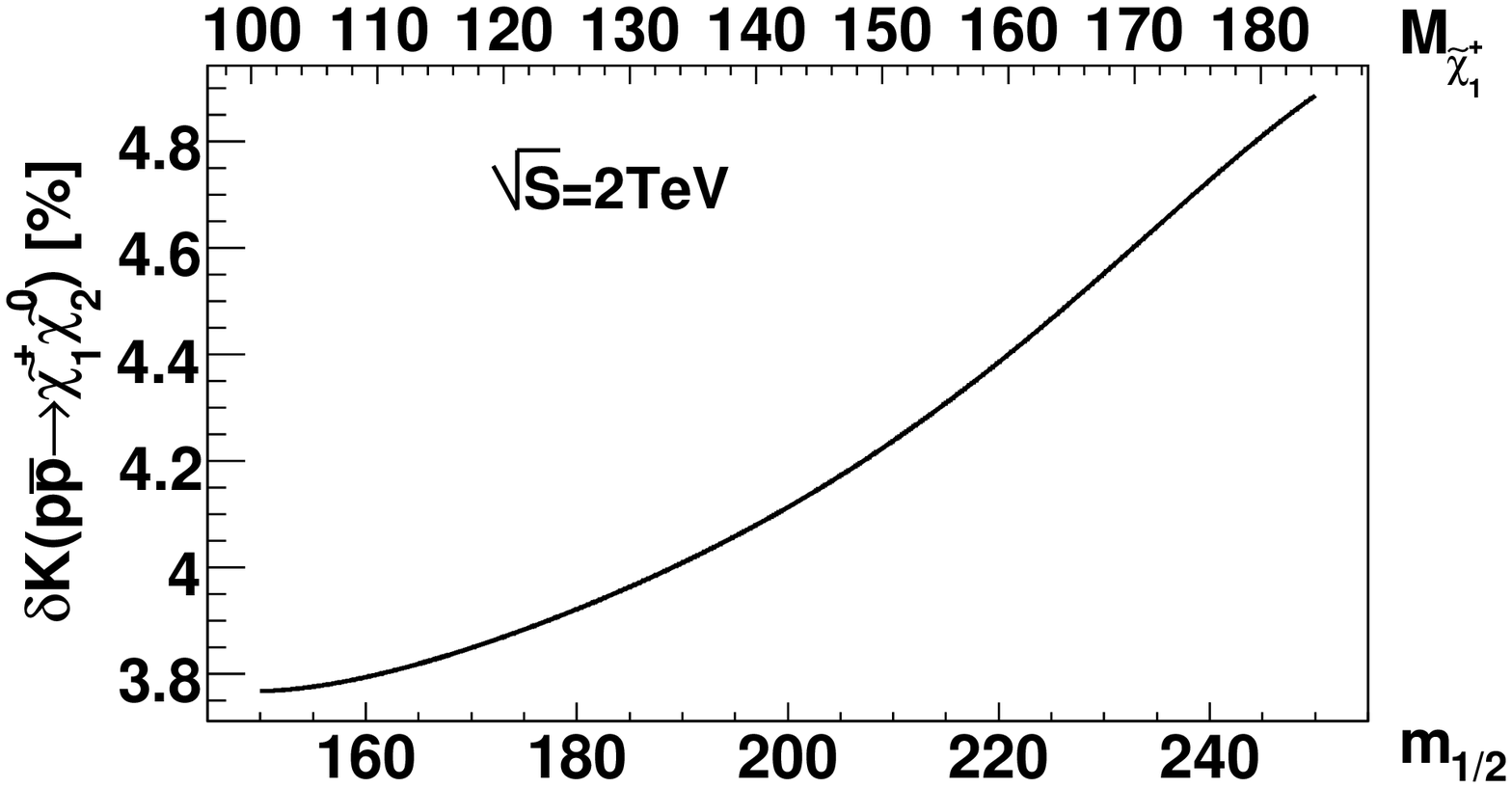}
\caption{$\delta K$, defined as $\delta
K=(\sigma^{RES}-\sigma^{NLO})/\sigma^{NLO}$, as a function of
$m_{1/2}$ for the associated production of $\cha{1}\neu{2}$ at the
two colliders assuming $m_0=200$GeV, $\tan\beta=5$, $A_0=0$ and
$\mu>0$.} \label{RELMHFDEP}
\end{figure}

\begin{figure}[ht!]
\includegraphics[width=0.8\textwidth]{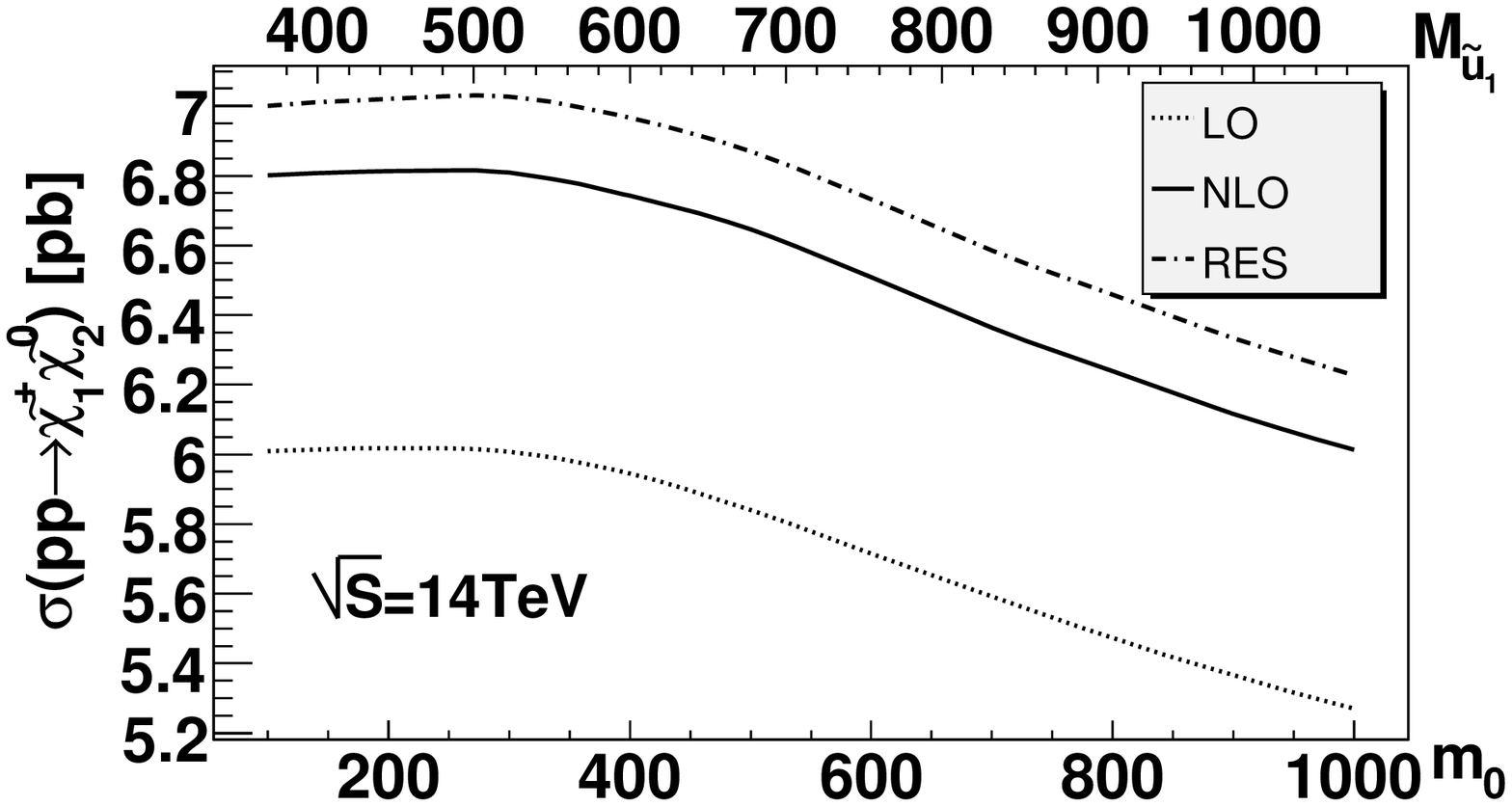}
\includegraphics[width=0.8\textwidth]{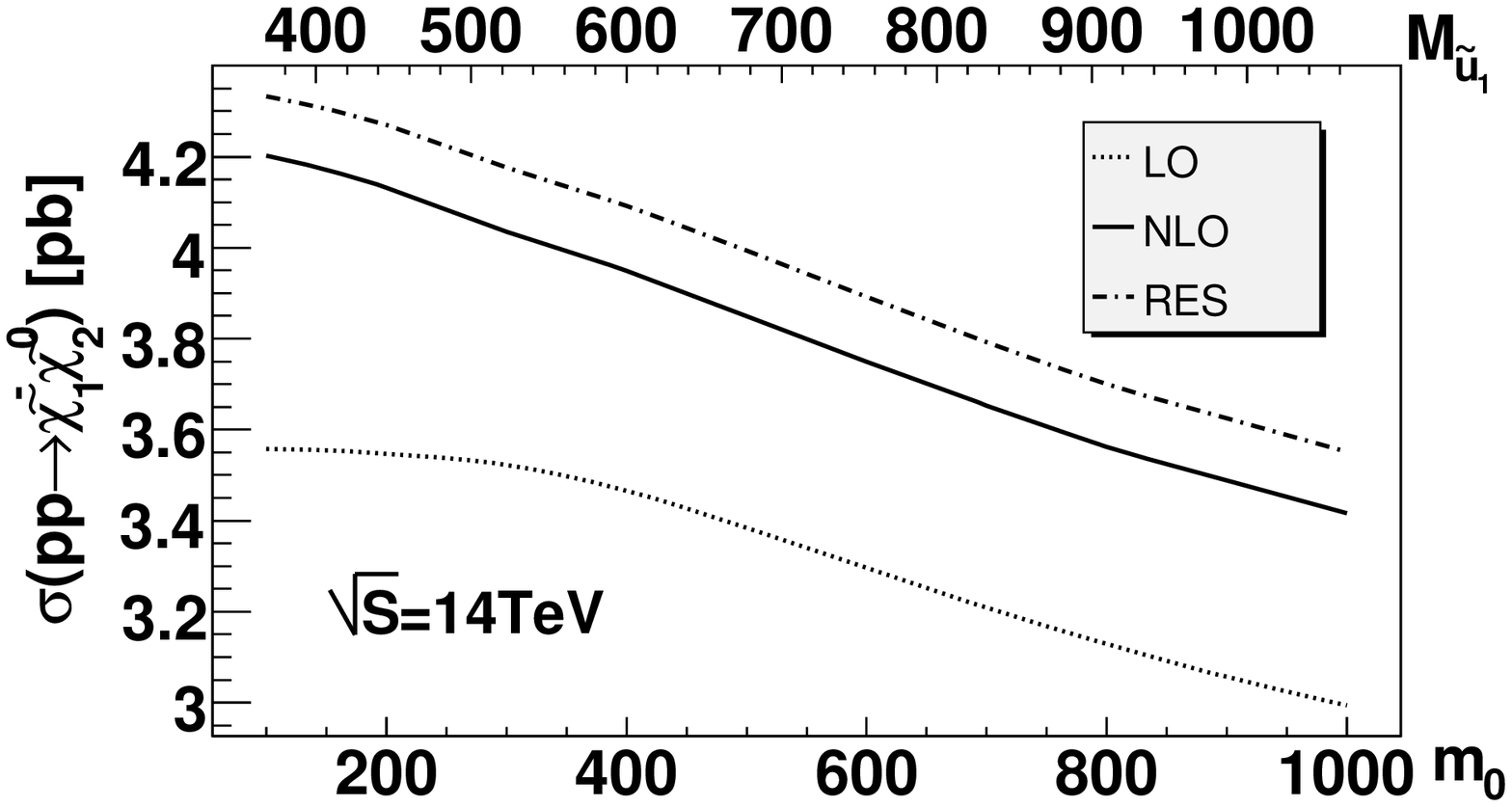}
\includegraphics[width=0.8\textwidth]{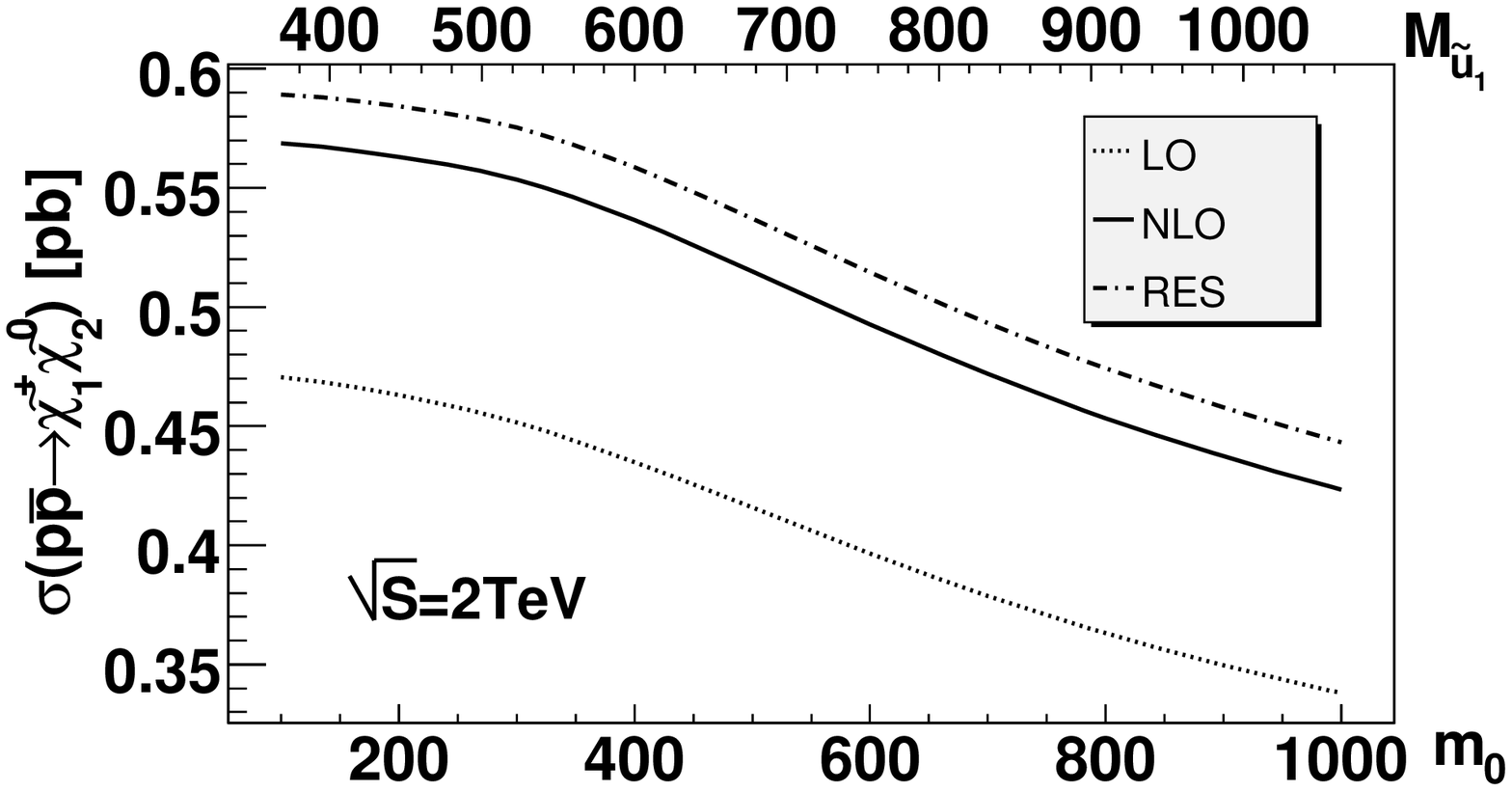}
\caption{The total cross sections as a function of $m_{0}$ for the
associated production of $\cha{1}\neu{2}$ at the two colliders
assuming $m_{1/2}=150$GeV, $\tan\beta=5$, $A_0=0$ and $\mu>0$.}
\label{M0DEP}
\end{figure}

\begin{figure}[ht!]
\includegraphics[width=0.8\textwidth]{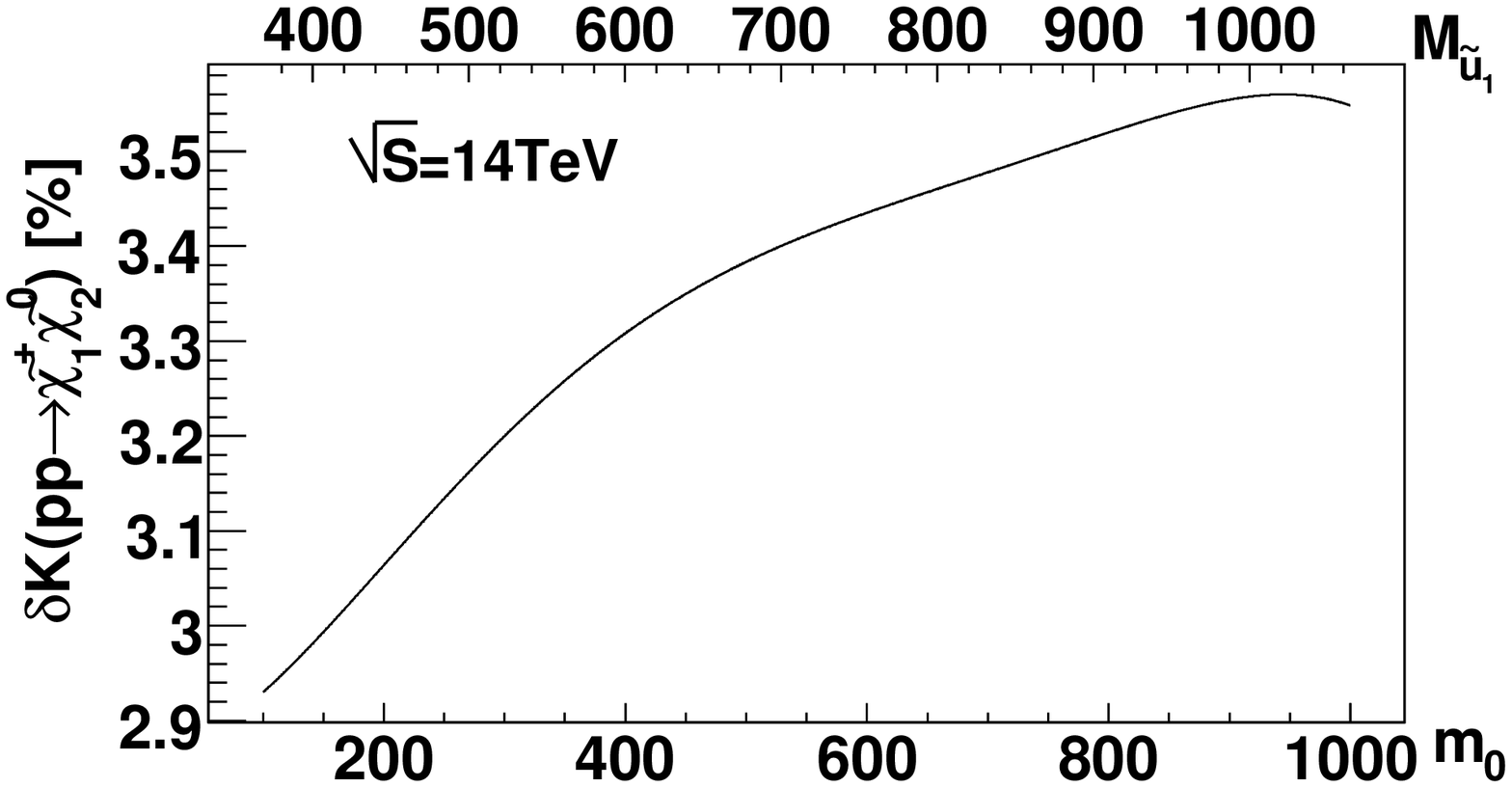}
\includegraphics[width=0.8\textwidth]{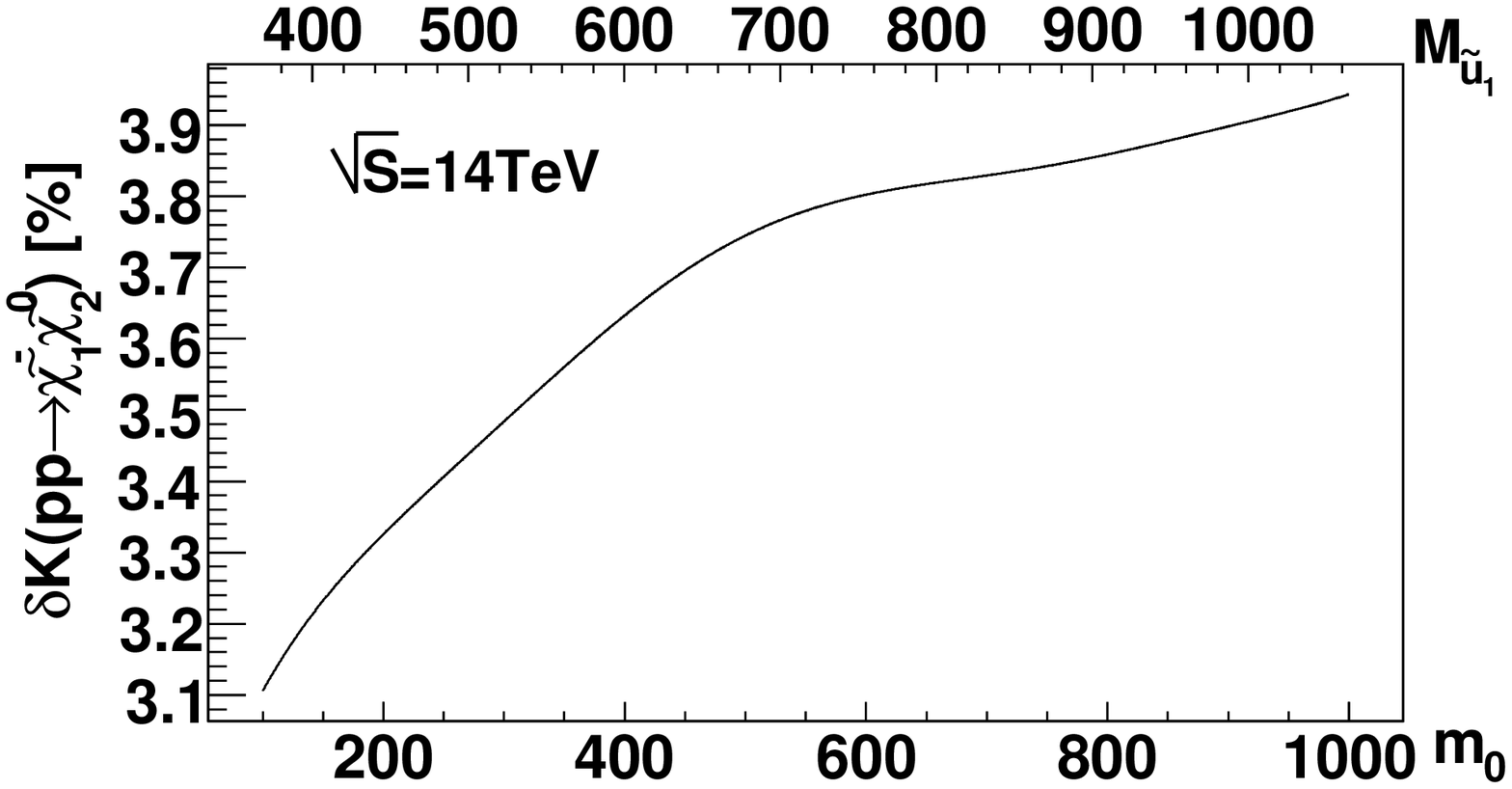}
\includegraphics[width=0.8\textwidth]{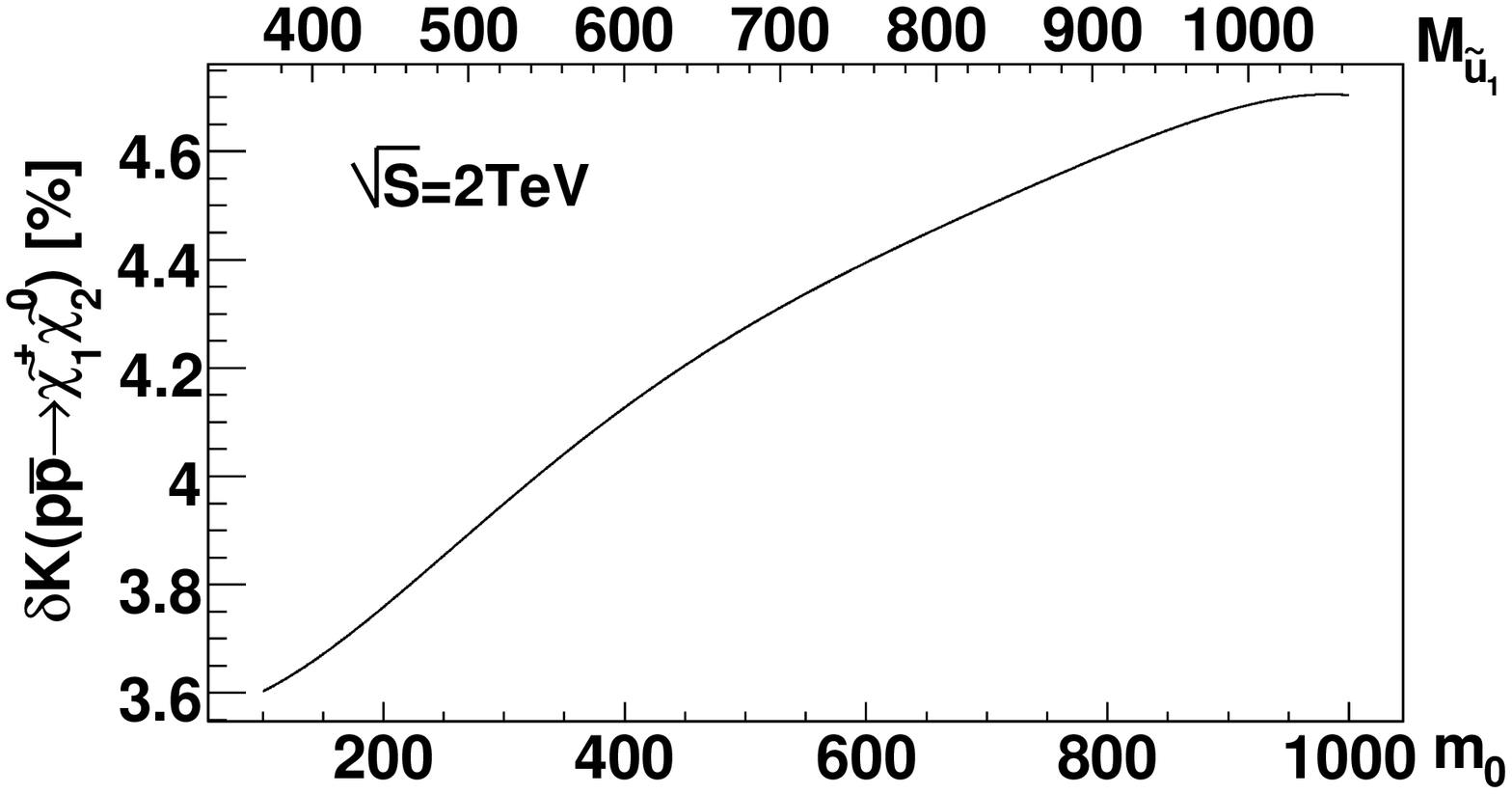}
\caption{$\delta K$, defined as $\delta
K=(\sigma^{RES}-\sigma^{NLO})/\sigma^{NLO}$, as a function of
$m_{0}$ for the associated production of $\cha{1}\neu{2}$ at the two
colliders assuming $m_{1/2}=150$GeV, $\tan\beta=5$, $A_0=0$ and
$\mu>0$.} \label{RELM0DEP}
\end{figure}

\begin{figure}[ht!]
\includegraphics[width=0.9\textwidth]{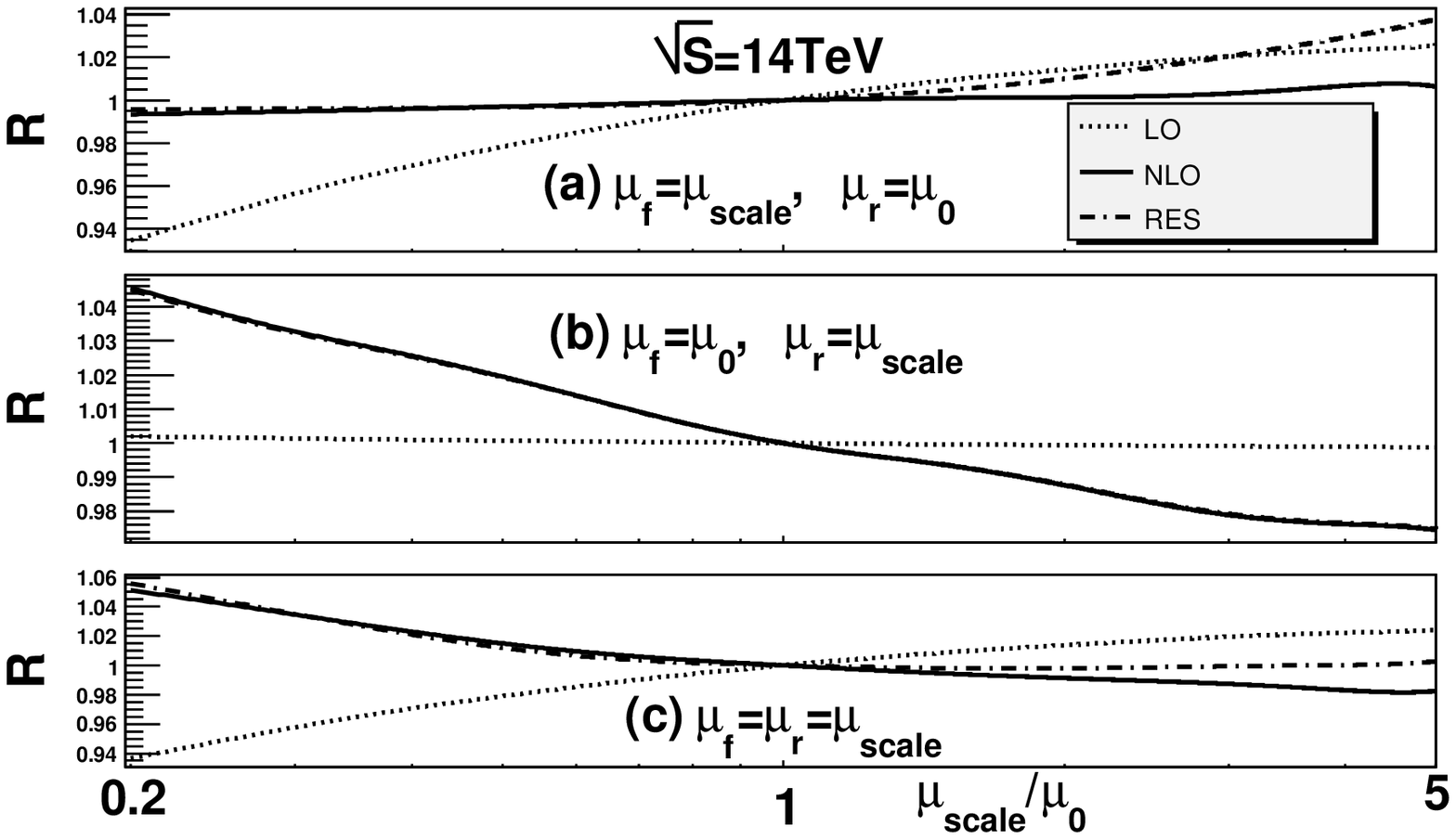}
\caption{The dependence of the total cross sections for $\chap{1}\neu{2}$ production on the
factorization scale(a), the renormalization scale(b), and both scales equal(c) 
at the LHC assuming
$m_{1/2}=200$GeV, $m_0=200$GeV, $\tan\beta=5$, $A_0=0$ and $\mu>0$.
$\mu_0=(m_{\cha{1}}+m_{\neu{2}})/2$.
$R=\sigma(\mu_{scale})/\sigma(\mu_0)$.
} \label{LHCSCALEDEP}
\end{figure}

\begin{figure}[ht!]
\includegraphics[width=0.9\textwidth]{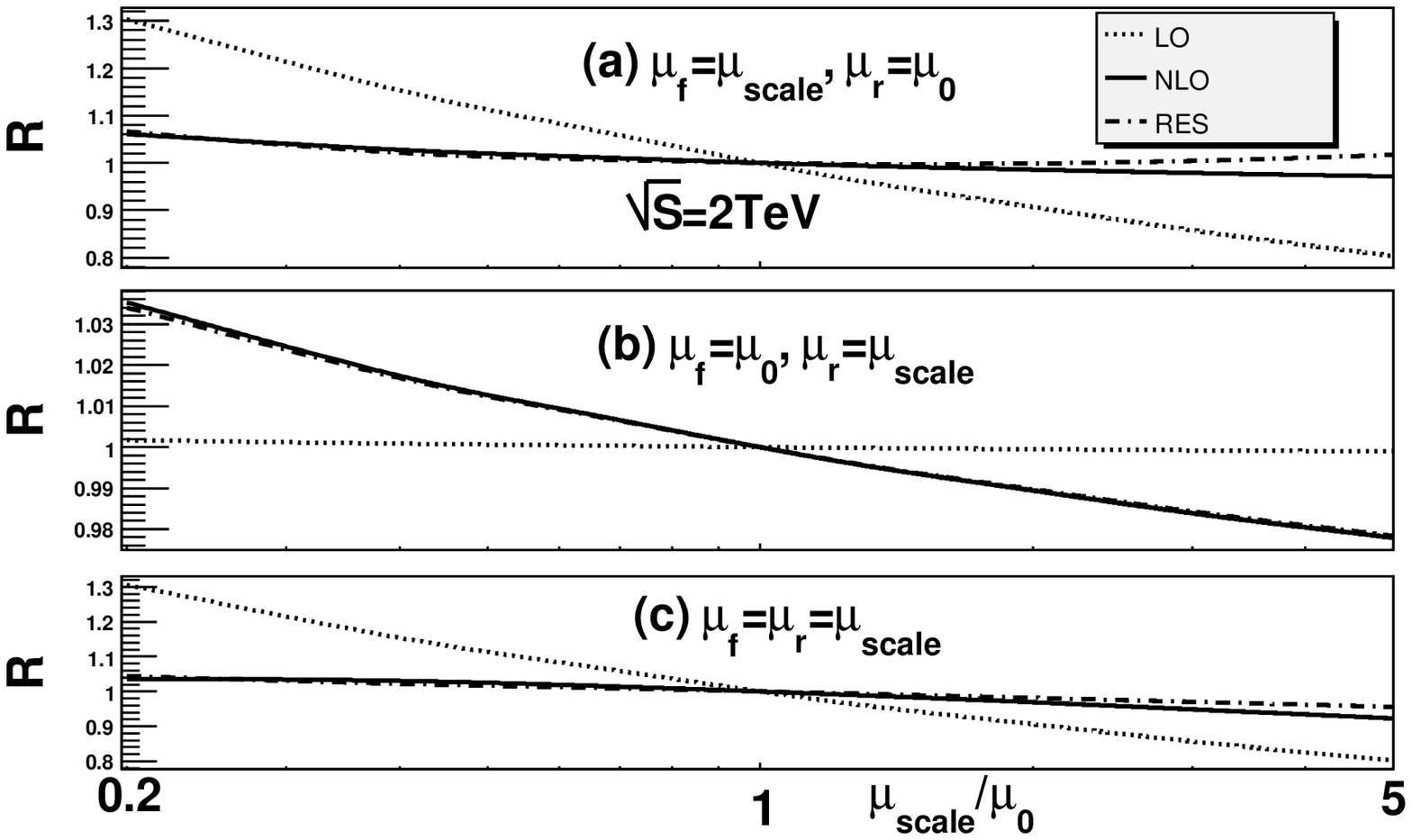}
\caption{The dependence of the total cross sections for $\chap{1}\neu{2}$ production on the
factorization scale(a), the renormalization scale(b), and both scales equal(c) at the Tevatron assuming
$m_{1/2}=200$GeV, $m_0=200$GeV, $\tan\beta=5$, $A_0=0$ and $\mu>0$.
Here $\mu_0=(m_{\cha{1}}+m_{\neu{2}})/2$ and 
$R=\sigma(\mu_{scale})/\sigma(\mu_0)$.
} \label{TEVSCALEDEP}
\end{figure}

\begin{figure}[ht!]
\includegraphics[width=0.8\textwidth]{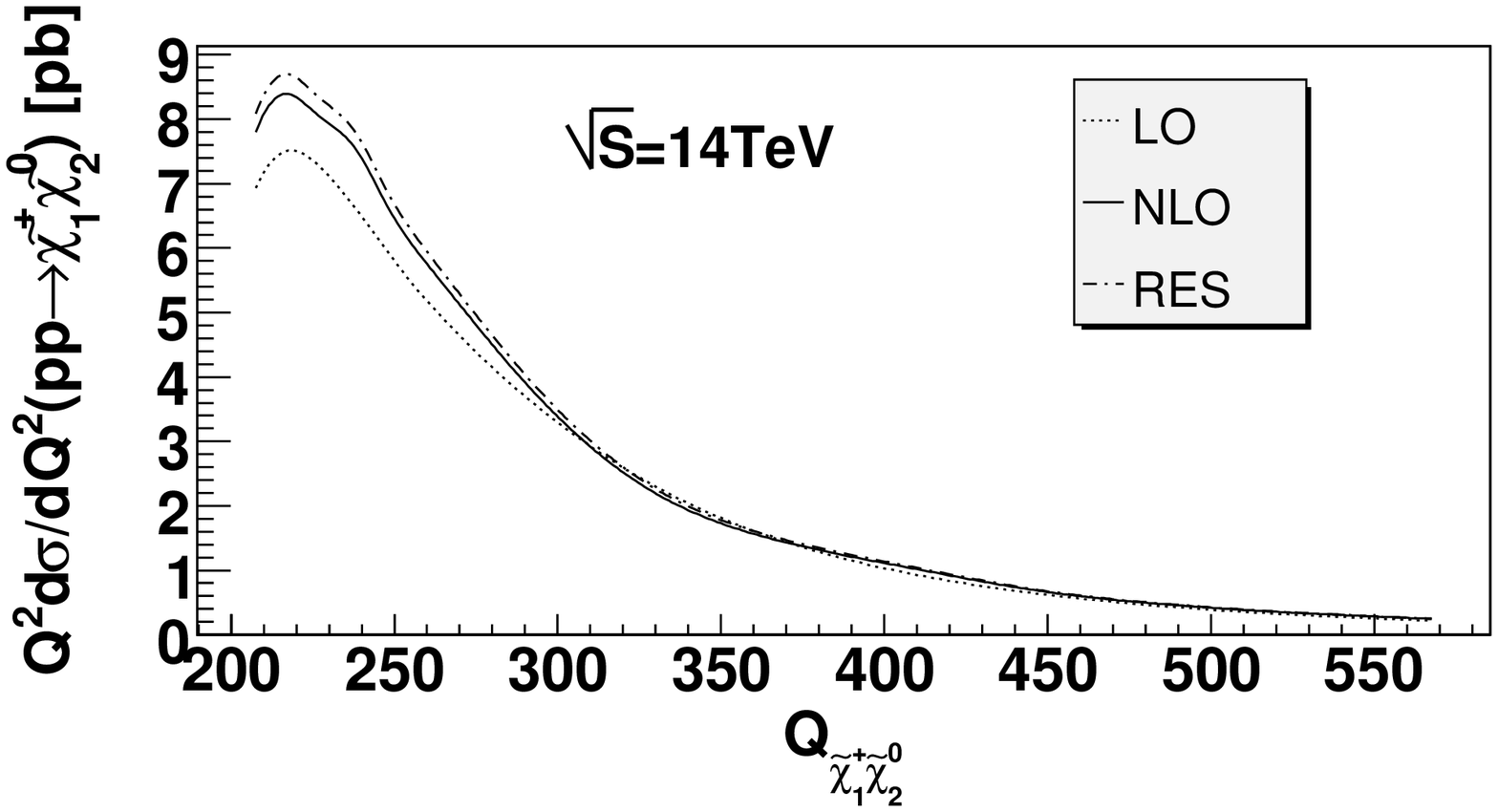}
\includegraphics[width=0.8\textwidth]{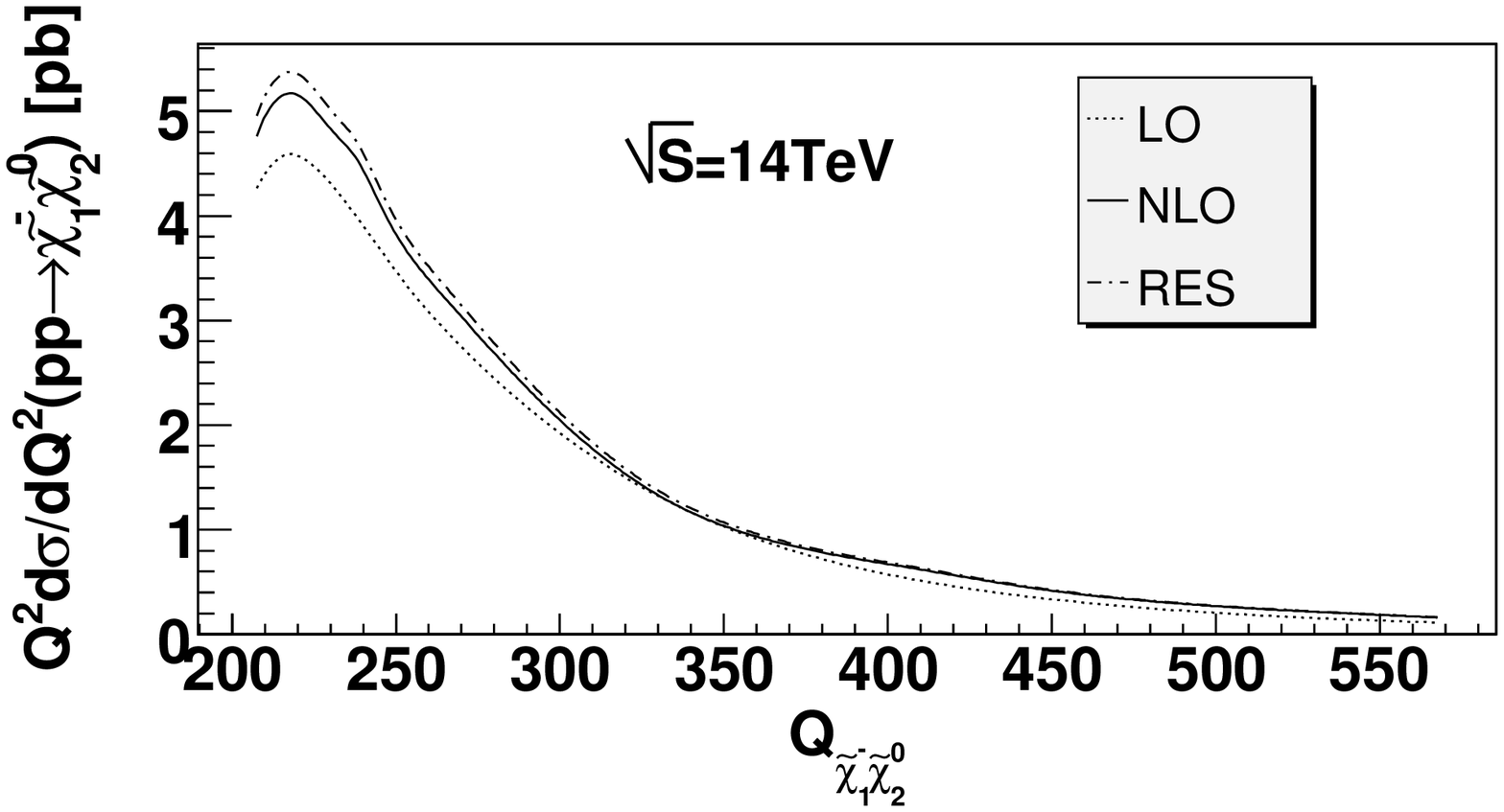}
\includegraphics[width=0.8\textwidth]{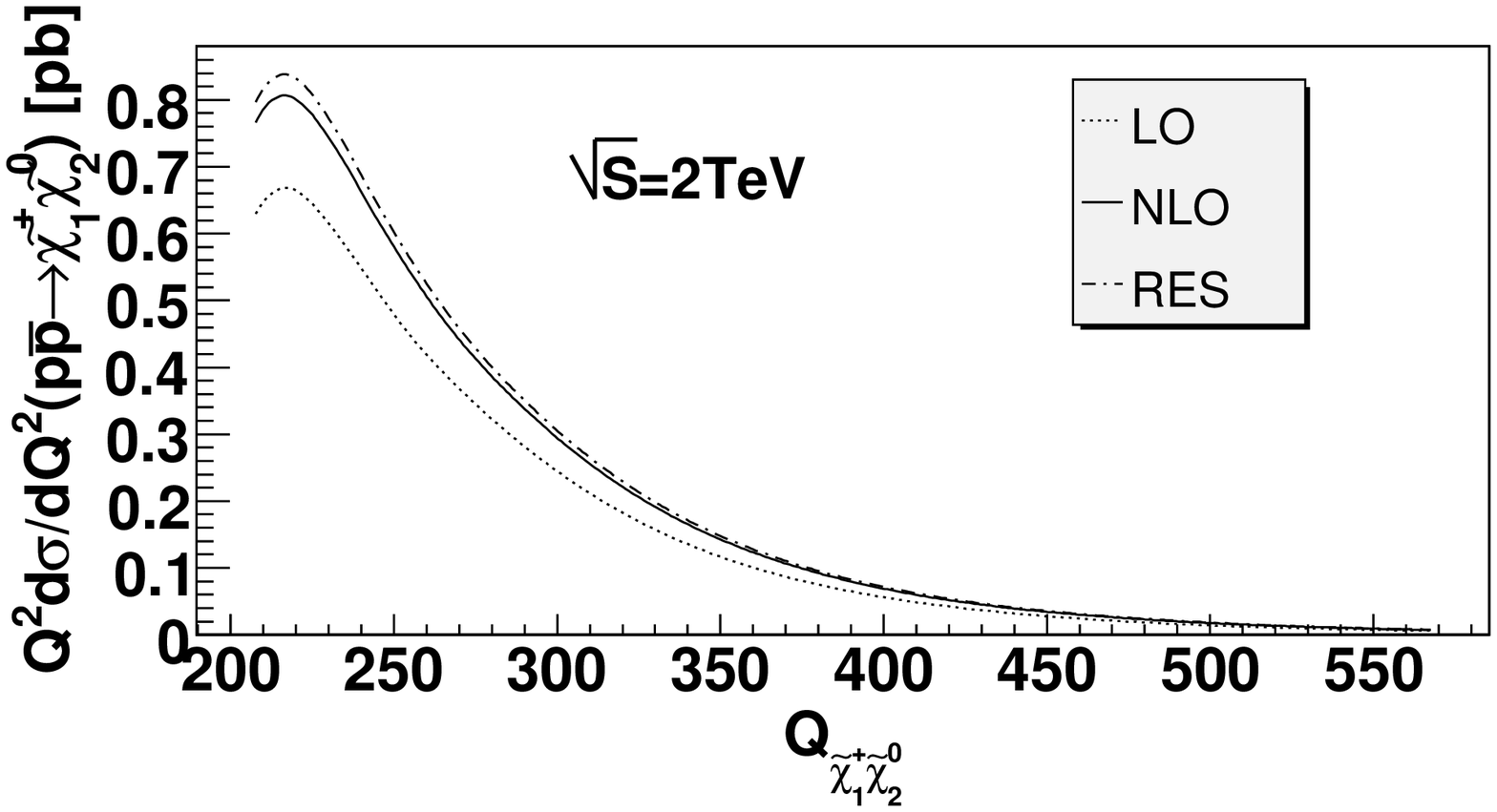}
\caption{The invariant mass differential cross sections for the
associated production of $\cha{1}\neu{2}$ at the two colliders
assuming $m_{1/2}=150$GeV, $\tan\beta=5$, $m_0=200$GeV, $A_0=0$ and
$\mu>0$ .} \label{INV200}
\end{figure}

\begin{figure}[ht!]
\includegraphics[width=0.8\textwidth]{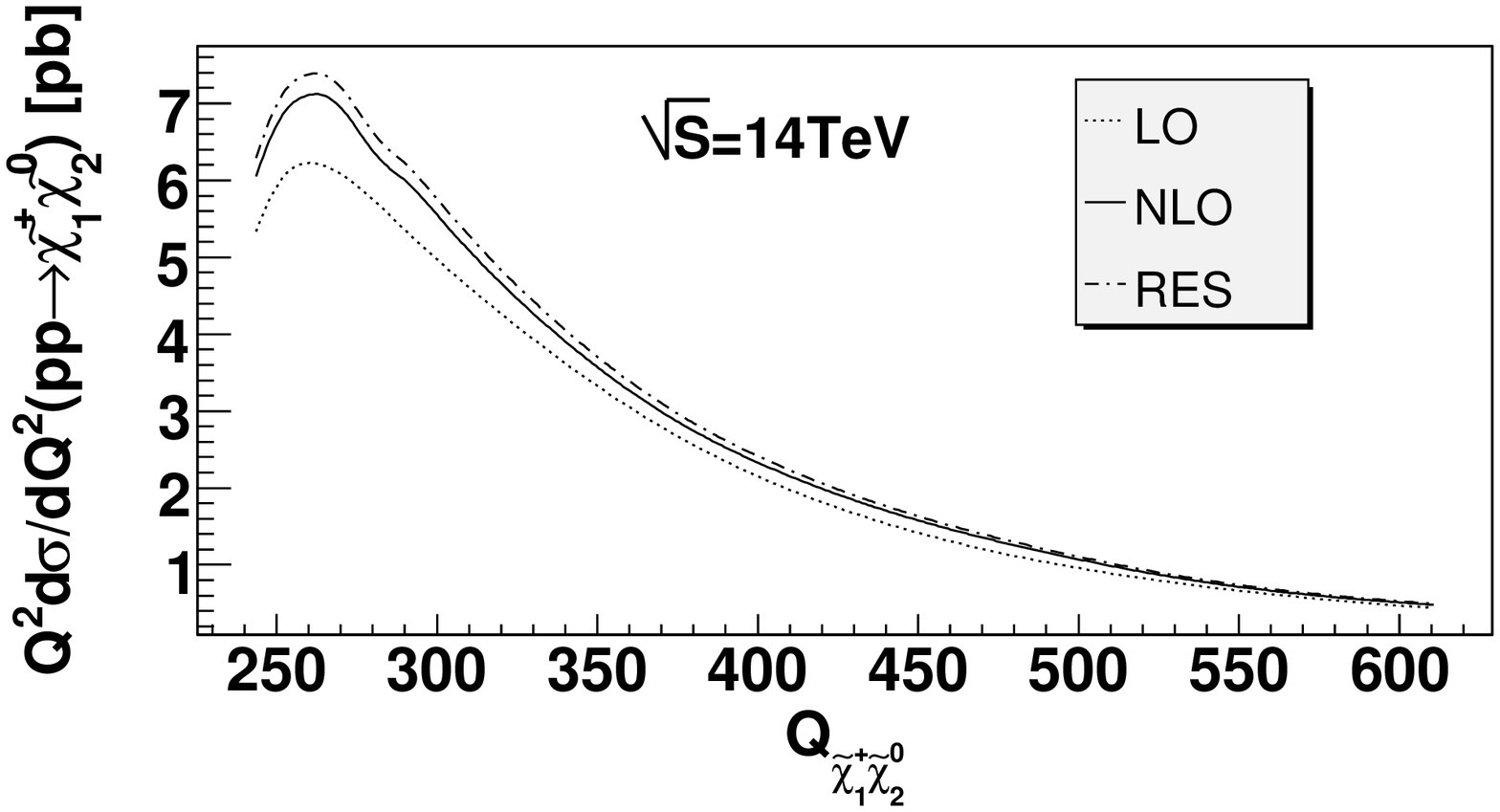}
\includegraphics[width=0.8\textwidth]{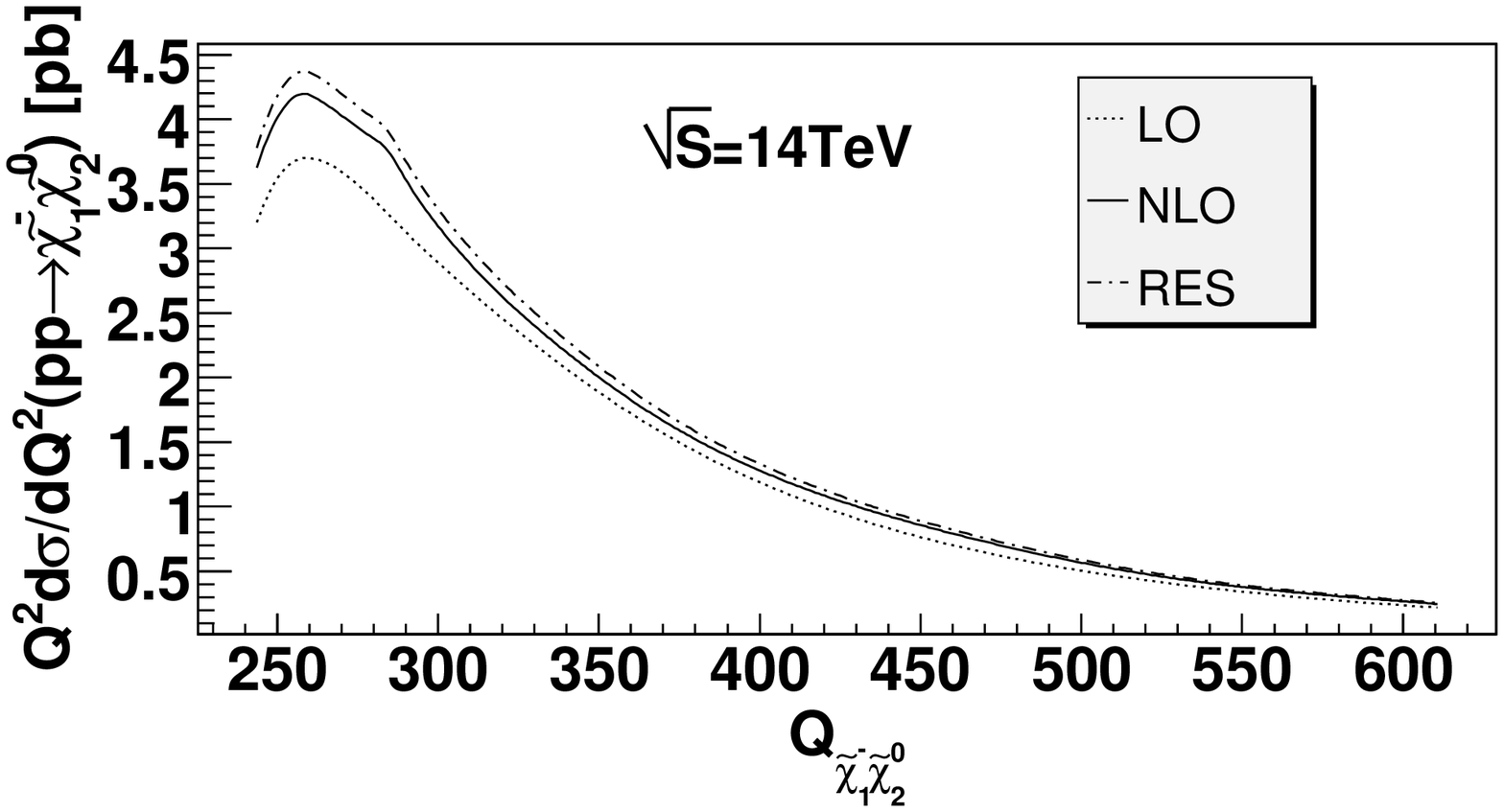}
\includegraphics[width=0.8\textwidth]{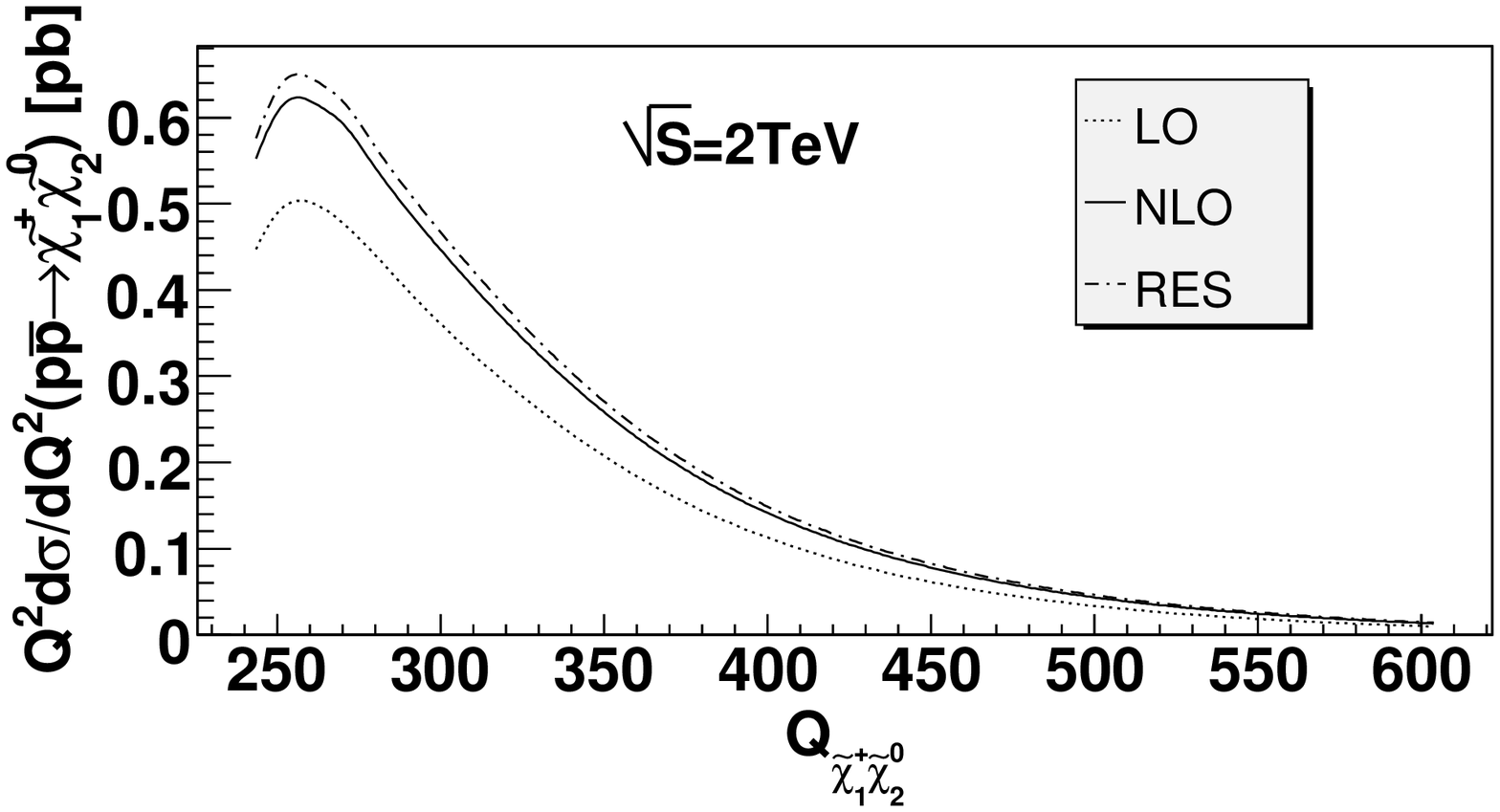}
\caption{The invariant mass differential cross sections for the
associated production of $\cha{1}\neu{2}$ at the two colliders
assuming $m_{1/2}=150$GeV, $\tan\beta=5$, $m_0=1000$GeV, $A_0=0$
and $\mu>0$ .} \label{INV1000}
\end{figure}

\begin{figure}[ht!]
\includegraphics[width=0.8\textwidth]{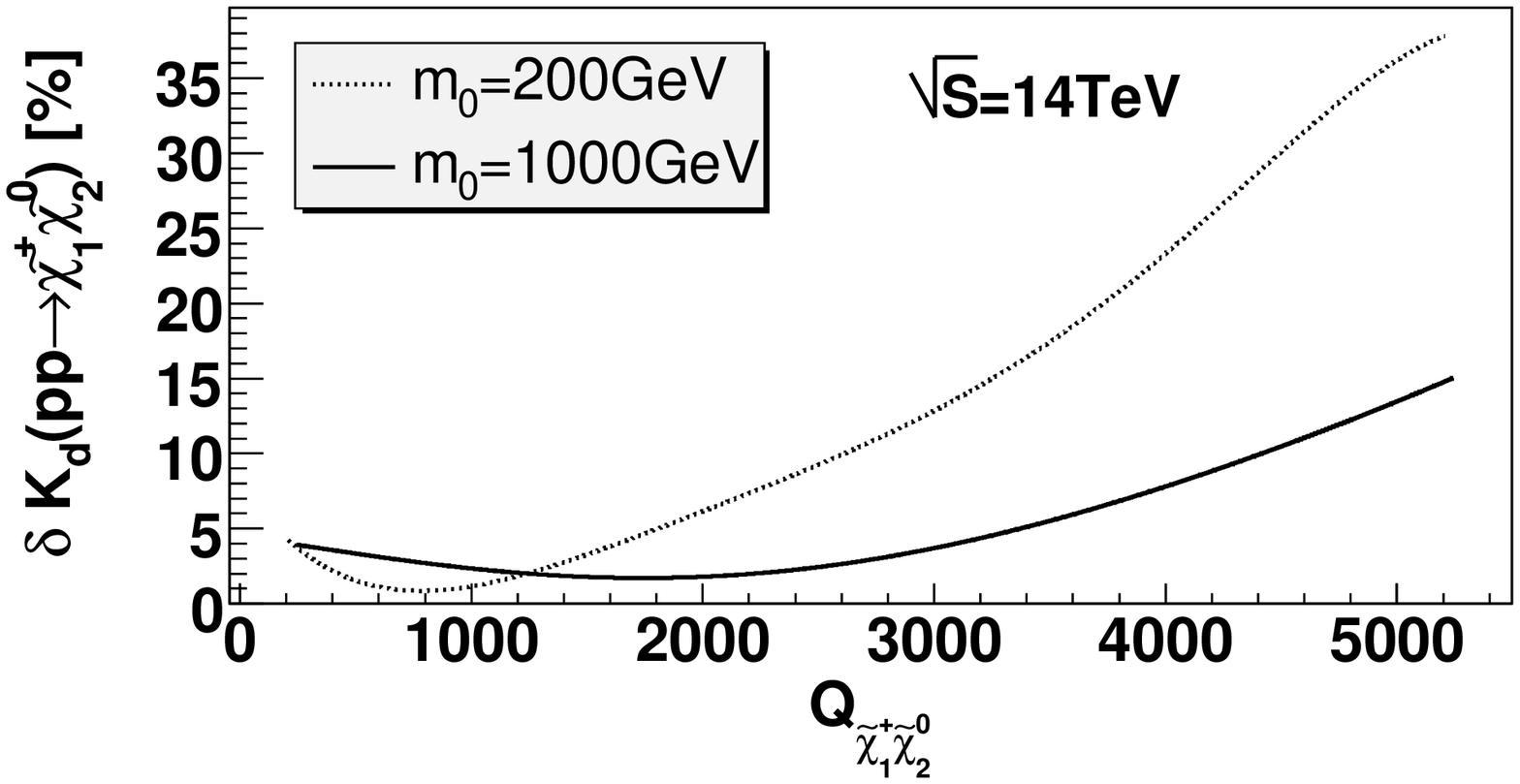}
\includegraphics[width=0.8\textwidth]{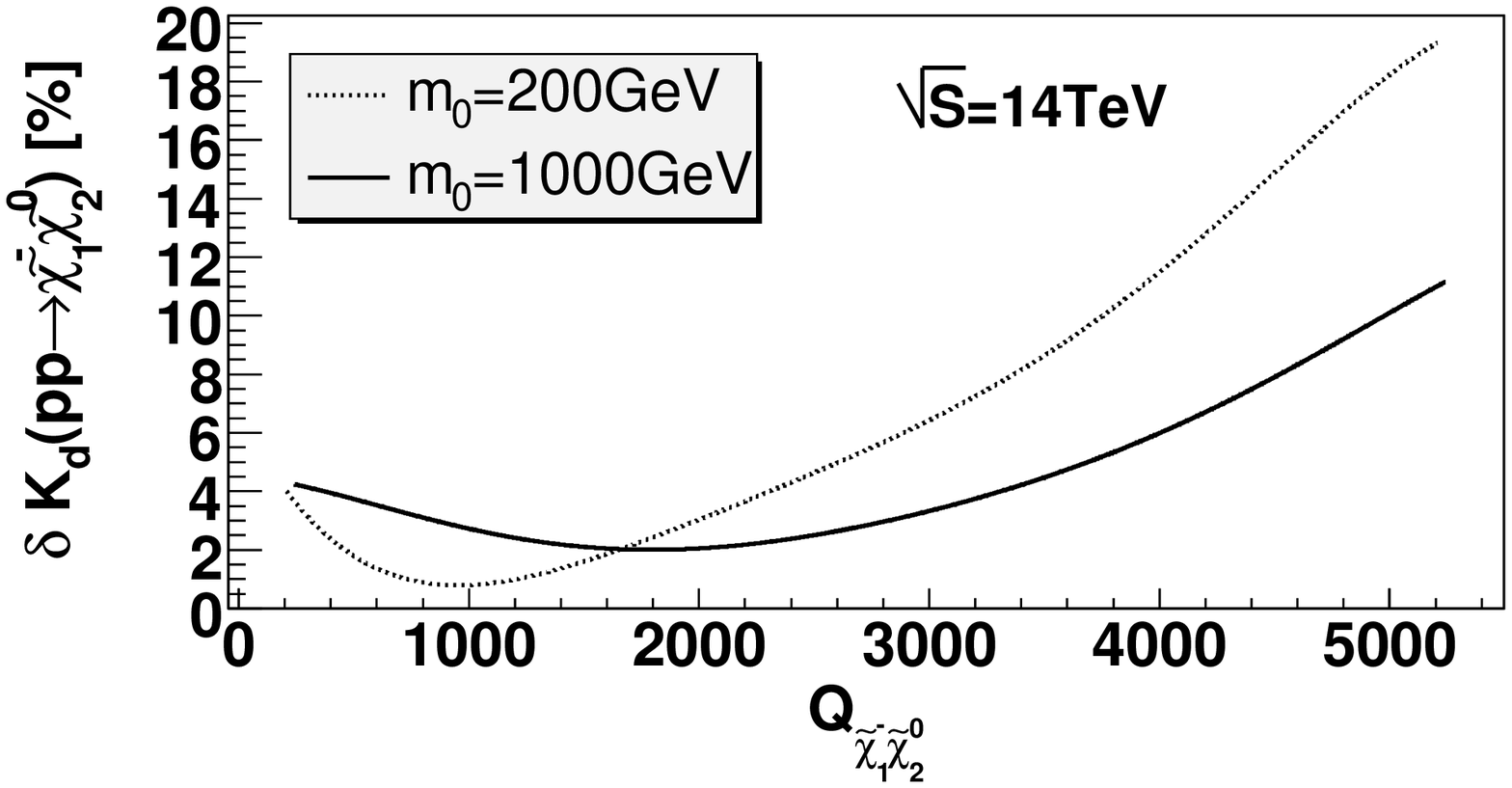}
\includegraphics[width=0.8\textwidth]{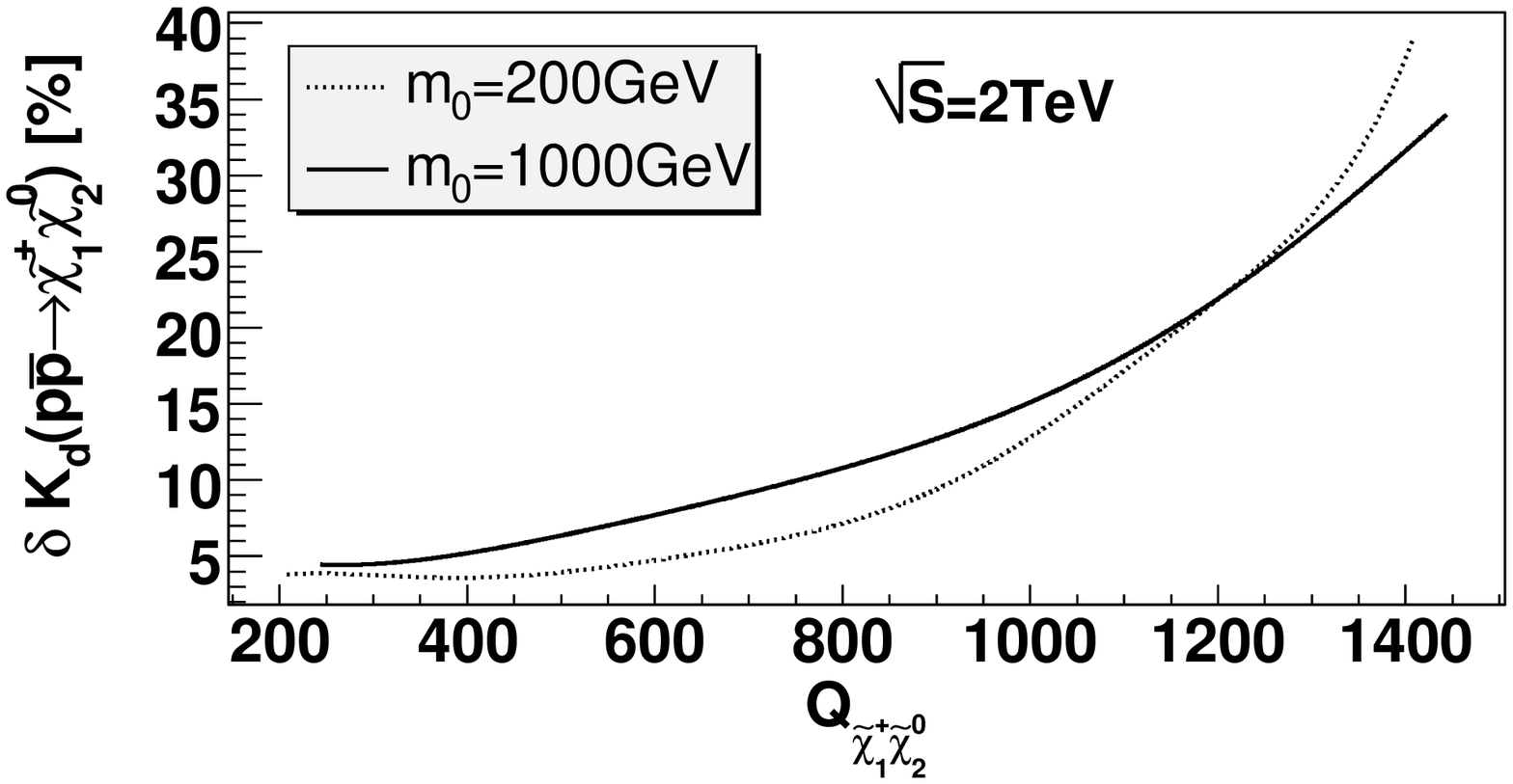}
\caption{The dependence of $\delta K_{d}$, defined as $\delta
K_d=(d\sigma^{RES}-d\sigma^{NLO})/d\sigma^{NLO}$, on the invariant
mass for the associated production of $\cha{1}\neu{2}$ at the two
colliders assuming $m_{1/2}=150$GeV, $\tan\beta=5$, $m_0=200$GeV
and $1000$GeV, $A_0=0$ and $\mu>0$ .} \label{RELINV}
\end{figure}

\end{document}